\documentclass{aa} 
\usepackage{color}
\usepackage{float}
\usepackage{graphicx}
\usepackage{txfonts}
\newcommand{\vc}{V746~Cas }
\newcommand{\ve}{V746~Cas}

\newcommand{\spefo}{{\tt SPEFO} }
\newcommand{\spefoe}{{\tt SPEFO}}
\newcommand{\phoebe}{{\tt PHOEBE} }

\newcommand{\fotel}{{\tt FOTEL} }
\newcommand{\fotele}{{\tt FOTEL}}
\newcommand{\korel}{{\tt KOREL} }
\newcommand{\korele}{{\tt KOREL}}
\newcommand{\pyt}{{\tt PYTERPOL} }
\newcommand{\pyte}{{\tt PYTERPOL}}
\newcommand{\kms}{km~s$^{-1}$ }
\newcommand{\ks}{km~s$^{-1}$}
\newcommand{\vsin}{$v$~sin~$i$ }

\newcommand{\tef}{$T_{\rm eff}$ }
\newcommand{\teff}{$T_{\rm eff}$}
\newcommand{\lgg}{{\rm log}~$g$ }
\newcommand{\ms}{M$_{\odot}$}
\newcommand{\rs}{R$_{\odot}$}
\newcommand{\cd}{c$\,$d$^{-1}$}

\newcommand{\ubv}{\hbox{$U\!B{}V$}}
\newcommand{\bv}{\hbox{$B\!-\!V$}}
\newcommand{\ub}{\hbox{$U\!-\!B$}}
\newcommand{\uvby}{\hbox{$uvby$}}
\newcommand{\hp}{\hbox{$H_{\rm p}$}}
\newcommand{\oc}{\hbox{$O\!-\!C$}}
\newcommand{\m}{\ifmmode.\kern-2.3pt^{\rm m}\else$.\kern-2.3pt^{\rm m}$\fi}

\newcommand{\Ame}{\AA~mm$^{-1}$}
\newcommand{\ha}{H$\alpha$ }
\newcommand{\hae}{H$\alpha$}

\newcommand{\he}{\ion{He}{i}~6678~\AA\ }

\begin{document}

   \title{Improved model of the triple system V746 Cas that has a bipolar magnetic field associated with the tertiary
\thanks{Based on observations from the Ond\v{r}ejov, Haute Provence, Bernard
Lyot, ESO Hipparcos and Mercator Observatories.}
}

   \titlerunning{Multiple system \ve}

\author{
        P. Harmanec\inst{1}
        \and
        M.~Bro\v{z}\inst{1}
        \and
        P. Mayer\inst{1}
        \and
        P.~Zasche\inst{1}
        \and
        L.~Kotkov\'a\inst{2}
        \and
        J.A. Nemravov\'{a}\inst{1}
        \and
        R.J. Dukes\inst{3}
        \and
        D.~Kor\v{c}\'akov\'a\inst{1}
        \and
        M.~\v{S}lechta\inst{2}
        \and
        E.~K\i ran\inst{4,1}
        \and
        R.~K\v{r}\'\i\v{c}ek\inst{1}
        \and
        J.~Jury\v{s}ek\inst{1,5}
}
\institute{
   Astronomical Institute of the Charles University,
   Faculty of Mathematics and Physics,\\
   V~Hole\v{s}ovi\v{c}k\'ach~2, CZ-180 00 Praha~8,
   Czech Republic
   \and
   Astronomical Institute, Czech Academy of Sciences,
   CZ-251 65 Ond\v{r}ejov, Czech Republic
\and
   Department of Physics and Astronomy, The College of Charleston,
   Charleston, SC 29424, USA
\and
   University of Ege, Department of Astronomy \& Space Sciences,
              35 100 Bornova - {\.I}zmir, Turkey
\and
   Institute of Physics, Czech Academy of Sciences, Na~Slovance 1999/2,
   CZ-182~21 Praha~8, Czech Republic
   }

\authorrunning{P. Harmanec et al.}

\date{Received \today; accepted}

\abstract{
V746 Cas is known to be a triple system composed of a close binary
with an alternatively reported period of either 25\fd4 or 27\fd8
and a~distant third component in a~170~yr (62000~d) orbit.
The object was also reported to exhibit multiperiodic
light variations with periods from 0\fd83 to 2\fd50, on the basis of which it
was classified as a~slowly pulsating B star. Interest in further investigation
of this system was raised by the recent detection of a variable magnetic
field. Analysing spectra from four instruments, earlier published
radial velocities, and several sets of photometric observations,
we arrived at the following conclusions:
(1)~The optical spectrum is dominated by the lines of the B-type primary
($T_{\rm eff\,1}\sim 16500(100)\,{\rm K}$), contributing 70\,\% of the light
in the optical region, and a~slightly cooler B tertiary
($T_{\rm eff\,3}\sim 13620(150)\,{\rm K}$).
The lines of the low-mass secondary are below our detection threshold;
we estimate that it could be a~normal A or F star.
(2)~We resolved the ambiguity in the value of the inner binary period
and arrived at a linear ephemeris of
$T_{\rm super.\,conj.}={\rm HJD}~2443838.78(81)+25\fd41569(42)\times E$.
(3)~The intensity of the magnetic field undergoes a~sinusoidal variation
in phase with one of the known photometric periods, namely 2\fd503867(19),
which we identify with the rotational period of the tertiary.
(4)~The second dominant photometric 1\fd0649524(40) period
is tentatively identified with the rotational period of the broad-lined
B-type primary, but this interpretation is much less certain and needs
further verification.
(5)~If our interpretation of photometric periods is confirmed, the
classification of the object as a slowly pulsating B star should
be revised.
(6)~Applying an N-body model to different types of available observational
data, we can constrain the orbital inclination of the inner orbit
to $\sim60^\circ < i_1 < 85^\circ$ even in the absence of binary eclipses,
and we estimate the probable properties of the triple system and its
components.}
\maketitle
%


\section{Introduction}

\begin{figure}
\includegraphics[width=9.0cm]{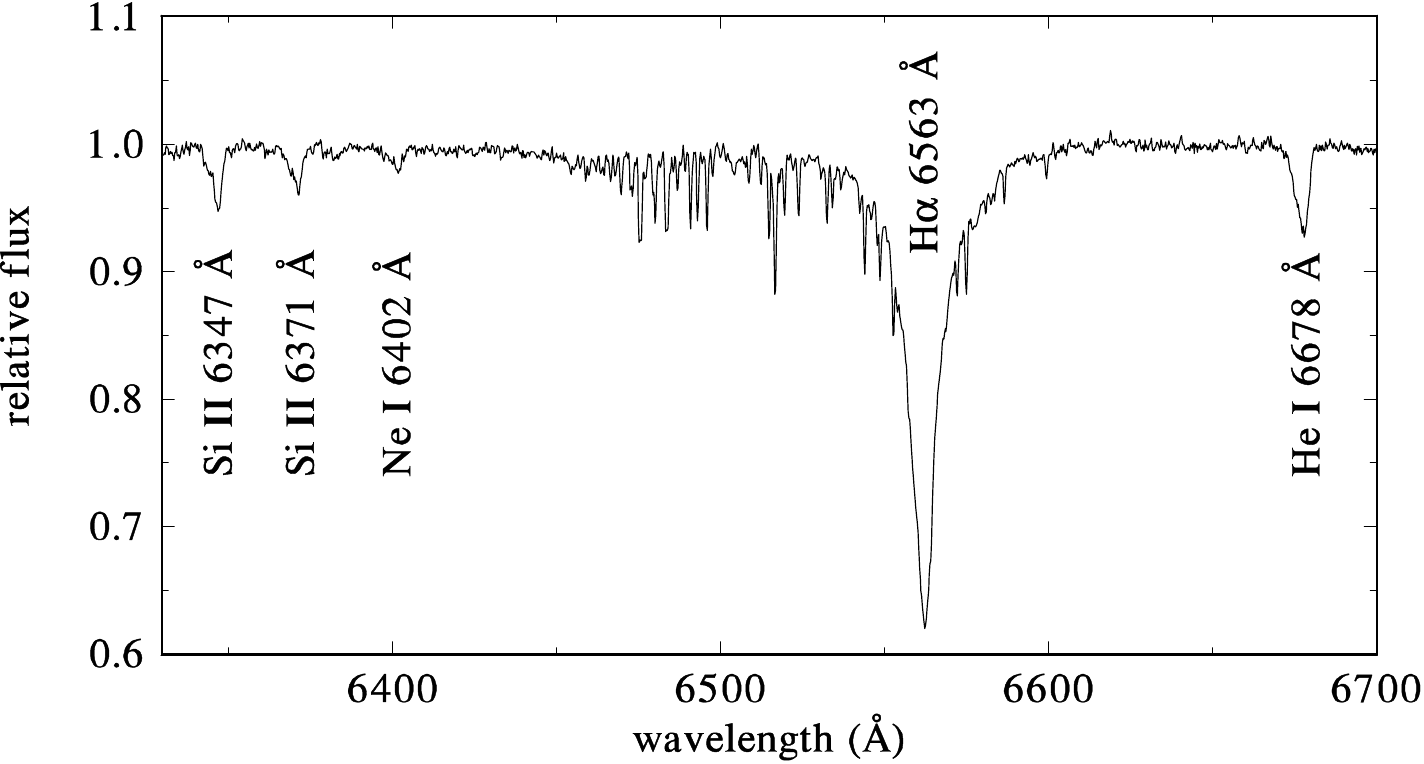}
\includegraphics[width=9.0cm]{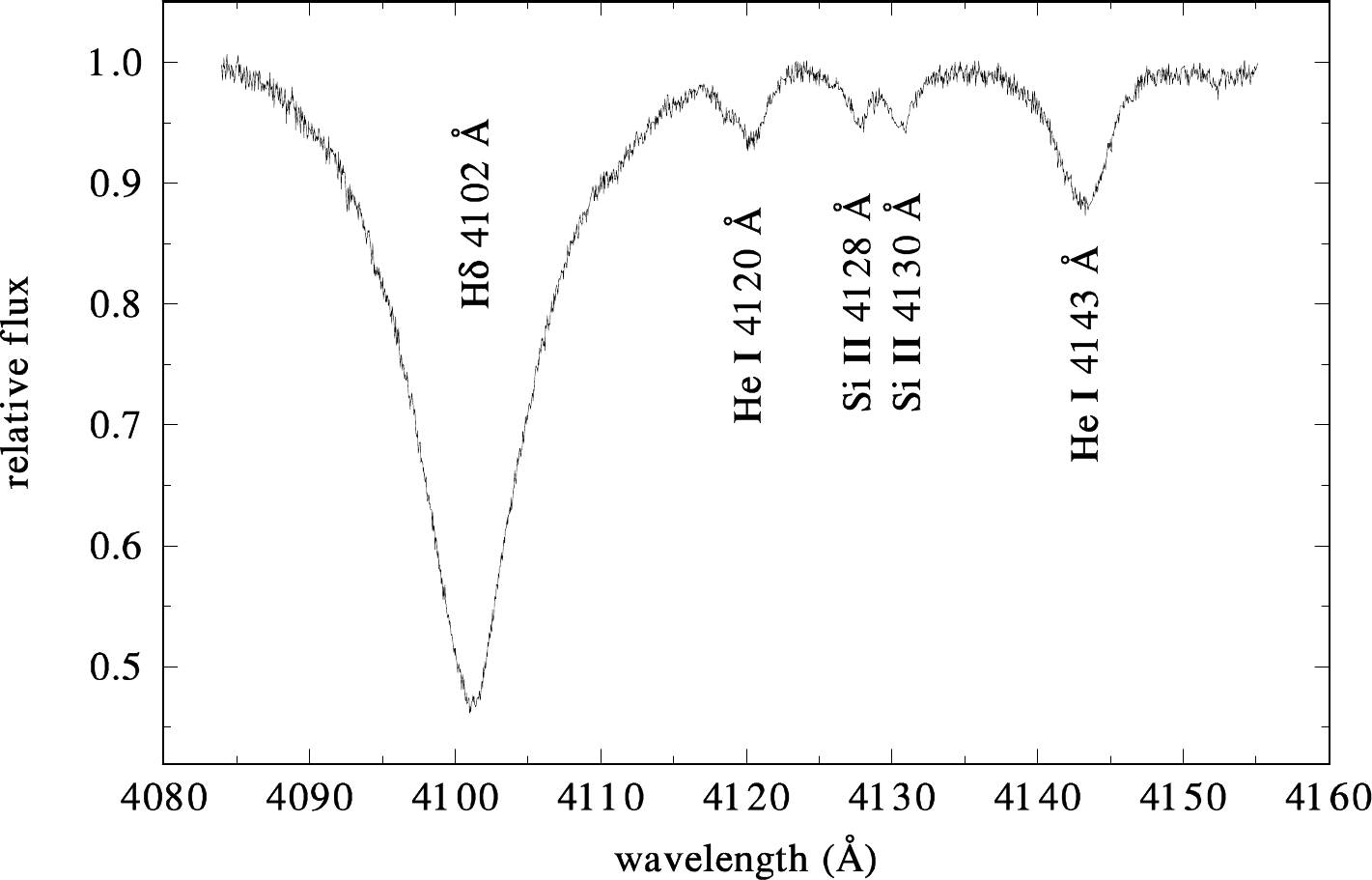}
\caption{Top: Ond\v{r}ejov spectrum taken on RJD~56928.3090 is
shown, and several stronger spectral lines are identified. All sharp lines in the
vicinity of \ha are telluric or weak interstellar lines.
Bottom: The Haute Provence Aurelie spectrum taken on
RJD~51011.5748, the stronger lines are again identified.}\label{spectra}
\end{figure}

While the nature of global magnetic fields in late-type stars with convective
envelopes seems to be -- at least in principle -- understood as a consequence of
the dynamo mechanism, the origin of organized magnetic fields detected in
some O and B stars is less obvious. Given the large progress in the
instrumentation allowing the detection of even weak magnetic fields, it is
understandable that a very systematic search for the presence of magnetic
fields among WR, O, B, and Be stars has recently be conducted by a large
team of collaborators. This project, known as MiMES (Magnetism in MassivE
Stars), was conducted between 2008 and 2013, and the first summary report
was published by \citet{wade2016}. These authors report that unlike the
magnetic fields of late-type stars, the magnetic fields of hot stars do not
show any clear correlations with basic stellar properties such as
their mass or rotation rate. We note that the
realisation has grown in recent years that a large portion of O and B stars are
binaries or even multiple systems, which must have significant
impact on the understanding of their properties
\citep[see, e.g.][]{chini2012,demink2013}. The role of
multiplicity in the magnetism of hot stars is worth consideration.
Recently, \citet{neiner2014} reported one preliminary result of the MiMES
survey: the discovery of a magnetic field of the triple system \ve.
Since the properties of this system were not well known, we were
motivated to this
study.

\vc (HD~1976, HR~91, BD+51$^\circ$62, HIP~1921, Boss~67) is a bright ($V=5$\m6)
B5\,IV star. Its radial-velocity (RV) variations were discovered
by \citet{adams12} and confirmed by \citet{plaskett31}. \citet{blaauw63}
derived RVs from a new series of McDonald spectra and published the
first orbital elements; see Table~\ref{oldsol}. In addition to their
spectra, they also used the DAO spectra. \citet{abt70} published six previous
Mt.~Wilson RVs, which partly overlap with those published by \citet{adams12}.
\citet{abt90} measured RVs of 20 new KPNO spectra and published another set
of elements, deriving a period of 25\fd44$\pm$0\fd03.
Their solution is based solely on the KPNO RVs, but the
authors claim that it also fits the previous Mt.~Wilson, DAO, and McDonald RVs.
\citet{mcswain2007} analysed their 15 new KPNO RVs along with earlier
published data and found two comparable
periods near 25\fd4 and 27\fd6. We note that these two periods are 1~yr
aliases of each other. \citet{mcswain2007} preferred the shorter period
and obtained another solution, this time based on most of previously
published RVs, which is also given in Table~\ref{oldsol}.

\begin{table}
\caption[]{Previously published orbital solutions for the \ve\ close binary.}
\label{oldsol}
\begin{center}
\begin{tabular}{rcccccl}
\hline\hline\noalign{\smallskip}
Element         &  1 & 2 & 3\\
\noalign{\smallskip}\hline\noalign{\smallskip}
$P_1$ (d)         &27.8&25.44(3)&25.4176(4)\\
$T_{\rm peri\,1}$(RJD)&not given&43840.3(3)&35783.5(1)\\
$e_1$             &0.2&0.14(9)&0.12(3)\\
$\omega_1$ ($^\circ$) &140&165(5)&172(2)\\
$K$ (\ks)       &30&23.4(1.9)&23.6(7)\\
$\gamma$ (\ks)  &$-$18.&$-$15.8(1.3)&$-$9.7(5)\\
rms (\ks)       &not given&4.5&4.81\\
No. of RVs      &16&20&15+32\\
\noalign{\smallskip}\hline\noalign{\smallskip}
\end{tabular}
\tablefoot{All epochs are in RJD=HJD-2400000.0;
the rms is the rms of one observation of unit weight.\\
Solution Numbers:
1. \citet{blaauw63}, 2. \citet{abt90}, 3. \citet{mcswain2007}
}
\end{center}
\end{table}

\begin{table}
\begin{center}
\caption[]{Journal of RV data sets}\label{jourv}
\begin{tabular}{ccrll}
\hline\hline\noalign{\smallskip}
Spg.&Time interval&No.   &Source \\
 No.&             &of    &       \\
    &(RJD)&RVs   &       \\
\noalign{\smallskip}\hline\noalign{\smallskip}
 1&19026.89--23004.70& 7&A\\
 2&24010.41--24891.13& 6&B\\
 3&35795.62--35833.55&16&C\\
 4&43411.86--44231.62&20&D\\
 5&53657.79--53696.79&15&E\\
\noalign{\smallskip}\hline\noalign{\smallskip}
 6&50967.59--51099.45&17&F\\
 7&52538.56--53026.28& 3&F\\
 8&56175.56--56214.42&13&F\\
 9&56746.65--57328.69&41&F\\
\noalign{\smallskip}\hline\noalign{\smallskip}
\end{tabular}
\tablefoot{ Column Spg. No.: the rows show\ \
1. Mount Wilson Solar Observatory 1.52~m reflector, three-prism spg;
2. Dominion Astrophysical Observatory 1.88~m reflector, prism spg.;
3. McDonald 2.08~m reflector, coud\'e grating spg. 34~\Ame;
4. Kitt Peak 1~m coud\'e feed telescope, grating spg. 16.9~\Ame;
5. Kitt Peak 2.1~m reflector, grating spg.;
6. Haute Provence 1.52 m reflector, Aurelie linear electronic spg.;
7. Haute Provence 1.52 m reflector, Elodie echelle spg.;
8. Bernard Lyot 2 m reflector, echelle spg.;
9. Ond\v{r}ejov 2.0~m reflector, coud\'e grating spg. 17.2~\Ame.\\
Column Source: the rows show\ \
A. \citet{adams12,abt70};
B. \citet{plaskett31};
C. \citet{blaauw63};
D. \citet{abt90};
E. \citet{mcswain2007};
F. this paper.}
\end{center}
\end{table}

  The first attempt to detect the spectral lines of the secondary was
reported by \citet{gomez82}. They studied 160~\AA\ long CCD spectra in
two wavelength regions: one containing \ion{He}{i}~5876~\AA\ and the
\ion{Na}{i} doublet at 5889 and 5895~\AA, and the other containing
the \ion{Si}{ii} doublet at 6347 and 6371~\AA, \ion{the Ne}{i}~6402~\AA\ line,
and several \ion{Fe}{ii}, \ion{Fe}{i}, and \ion{Ca}{i} lines, to be able
to restrict the possible spectral class of the secondary. They did not show the
actual line profiles and published only a few comments. They concluded that
the spectral type of the secondary must be earlier than F6, and suspected a  possible weak secondary component in the
\ion{He}{i}~5876~\AA\ line.

\vc is also the brighter member of the visual system ADS~328.
Doboco \& Andrade \citep[in][]{doboco2005} derived the orbit of the wide
pair with a period of 169.29~yr (61832~d), semimajor axis 0\farcs214,
eccentricity 0.163, inclination $64.\!\!^\circ8$, argument of periastron
$311.\!\!^\circ8,$ and periastron passage at 1955.06 (JD~2435130). The distant
component B is by some 0\m9 fainter than \ve.

The light variability of \vc in the range from 5\m54 to 5\m56 was discovered
by the Hipparcos team \citep{esa97}, and several investigators
reported the presence of two principal periods, 1\fd065 and
2\fd504, and alternatively a few additional periods as well
\citep[for instance][see Sect.~\ref{rapid}
for a more detailed account]{waeletal98,andrews2000,decat2007,dukes2009}.
They all classified the object as a slowly pulsating B-type star (SPB).
Before the light variability was discovered, \citet{glushneva1992}
published spectrophotometry of \vc and several other stars and suggested that
\vc could serve as a secondary spectrophotometric standard.

\citet{neiner2014} reported the discovery that the line spectra of \vc are
composed of a narrow component, for which they found clear signatures of
a magnetic field, and a broad component that is much more variable in RV.
They identified the narrow lines and the magnetic field with the primary
and the broad-line component with the secondary of the 25\fd4 binary,
concluding that the lines of the distant tertiary are not seen
in the spectra. They allowed, however, that the narrow-lined star
might also be the visual tertiary or a combination of the primary and
tertiary.

Our initial motivation for this study was to resolve the remaining
ambiguity in the value of the orbital period. We started to collect new
CCD spectra in the red spectral region. Their subsequent analysis led
us to a more complex investigation and hopefully to a better understanding
of this remarkable triple system, which we report here.


\section{Available observational data and their reductions}

\subsection{Spectroscopy}
Our observational material consists of 41 Ond\v{r}ejov CCD spectra
secured in 2014 -- 2015 (S/N 160--300; three underexposed spectra
with 27, 37, and 100), 17 OHP Aurelie spectra with an S/N of 130-230, the first one
of 36 only \citep{gillet}, which were obtained and studied by
\citet{mathiasetal01}, 3 archival OHP Elodie spectra with S/N 270--280 in the
red parts of the spectra
\citep{moultaka2004}, and 13 publicly available echelle spectra from
the Bernard Lyot telescope with an S/N greater than 200 in the red parts
of the spectra \citep{polar}.
The initial reduction of all Ond\v{r}ejov spectra (bias subtraction,
flat-fielding, creation of 1D spectra, and wavelength calibration) was
carried out in {\tt IRAF}. For the Bernard Lyot and OHP Elodie spectra,
extracted from public databases, we adopted the original reductions, verifying
that the zero-point of the wavelength scale was corrected via telluric
lines, as was the case for the Ond\v{r}ejov spectra. This was more
complicated with the Aurelie spectra, which are not available from public
archives. P. Mathias kindly provided us with the old DAT tapes, which
were reconstructed in Prague. Their new reduction was carried out with the
help of simple dedicated programs, the wavelength calibration being
carried out with the program \spefo \citep{sef0,spefo}, namely the latest
version 2.63 developed by J.~Krpata. Median values of each corresponding
sets of offsets and flat spectra were used.

Rectification and removal of residual cosmics and flaws for all sets of spectra
were carried out in \spefoe.

 We first extracted the red parts of the Elodie and Bernard Lyot spectra
($\sim$6360 -- 6740~\AA) to have the same spectral region as is covered
by the Ond\v{r}ejov spectra. One Ond\v{r}ejov spectrum is shown in
Fig.~\ref{spectra}. As was previously noted by
\citet{neiner2014}, the line profiles are not symmetric but contain
broad and narrow components, the broad component varying in RV over a wide
velocity range. The profiles might
also be affected by subfeatures that are related to rapid light and
line-profile variations and move across the line profiles
as a result of stellar rotation. For all Elodie and Bernard Lyot spectra,
we also extracted the blue, green, and yellow parts of the spectra over the
wavelength range from 4000 to 6360~\AA. We also extracted the Aurelie
spectra, which cover the wavelength range from 4095 to 4155~\AA.

Finally, we also collected RV measurements obtained by
several investigators and published in the literature. The journal
of all available RVs is listed in Table~\ref{jourv} and all individual RVs are
presented in Table~\ref{rvlit} for the data from the literature.
When they were not available in the original source, we converted
the observation dates into HJDs. For brevity, we use reduced Julian dates
\centerline{RJD = HJD $-$ 240000.0}
throughout. \citet{abt70} published
six RVs from Mount Wilson secured between December 1910 and November 1921.
They include RVs published by \citet{adams12}, with the exception of the
very first observation, which was secured on 1910 December 21. We therefore adopted
the first RV from \citet{adams12} and all remaining Mount Wilson RVs
(spg.~1) from \citet{abt70}.

\subsection{Standard RV measurements}
To resolve the problem of the true orbital period and to obtain
reliable orbital elements, we first tested several techniques of RV
measurements. For the most numerous set of the red spectra, we measured
the two strongest lines, \ha and \ion{He}{i}~6678~\AA,\ in \spefoe.
This program displays direct and flipped traces of the line
profiles superimposed on the computer screen that the user can slide
to achieve a precise overlapping of the parts of the profile whose RV
is to be measured. We separately measured the line cores and outer line
wings. We note that two of the Ond\v{r}ejov spectra (RJDs 56920 and 57228) are
underexposed and we were only able to measure the RVs of the wings of \hae.
These RV measurements are listed in Table~\ref{indrvs}.
For the Aurelie spectra, we also measured RVs of the
broad wings of the \ion{He}{i}~4144~\AA\ line, and these measurements are listed
in Table~\ref{rvaure} in Appendix~\ref{apa}.


While the uncertainty of the \spefo measuring procedure alone seems
small (of the order of a few \ks), as determined from three independent
measurements, there are three additional sources of uncertainties:
photon noise, systematics arising from the rectification, and
systematics from line blending. The latter is probably not critical for RVs of a particular spectral line, but they may dominate the uncertainty
budget for a mean RV from several spectral lines
(as discussed below). In any case, the total uncertainty
is higher than the formal rms errors and might reach up to
10~\kms in the least favourable cases.

For the Elodie and Bernard Lyot spectra, we also measured RVs of the following
stronger lines in \spefoe: \ion{He}{i}~4009, 4143, 4471, 4713, 4922, 5016, 5047,
and 5876~\AA, \ion{C}{ii}~4267~\AA, and \ion{Mg}{ii}~4481~\AA. We did not
use the \ion{He}{i}~4026~\AA, 4120~\AA, and  4387~\AA\ lines, since the first
is affected by an inter-order jump in all Bernard Lyot spectra, while the
profiles of the other two lines are affected by strong blends in their neighbourhood.
Finally, for the Aurelie spectra we measured RVs of H$\delta$ and
\ion{He}{i}~4143~\AA.


\subsection{Photoelectric observations and their homogenisation}
There are three principal sets of photoelectric observations suitable
for period analyses:
\begin{enumerate}
\item The Hipparcos \hp\ observations \citep{esa97},
\item the \uvby\ observations from the Four-College APT secured by one of us;
a subset of these observations has previously been analysed by \citet{dukes2009},
and
\item the Geneva 7-C observations obtained and studied by \citet{decatetal04}
and \citet{decat2007}.
\end{enumerate}
One of us, R.J.~Dukes, was able to recover the APT \uvby\ observations,
or more precisely, a more numerous data set than was used in his published studies.
It was deemed useful to repeat the period analysis with a complete and
homogenised set of the data. The yellow-band observations (\hp\ and Geneva $V$
magnitudes converted into Johnson $V$, and Str\"omgren $y$) are the most numerous
and represent the best set for the analysis. More details on the data
processing and their homogenisation can be found in Appendix~\ref{apc},
where the journal of these observations is also listed in Table~\ref{jouubv}.
After the removal of data from non-photometric nights, we are left with
2060 individual yellow-band observations spanning an interval of nearly 9800~d.


\section{Towards true orbital periods of the system}
\begin{figure}[H]
\includegraphics[width=9.0cm]{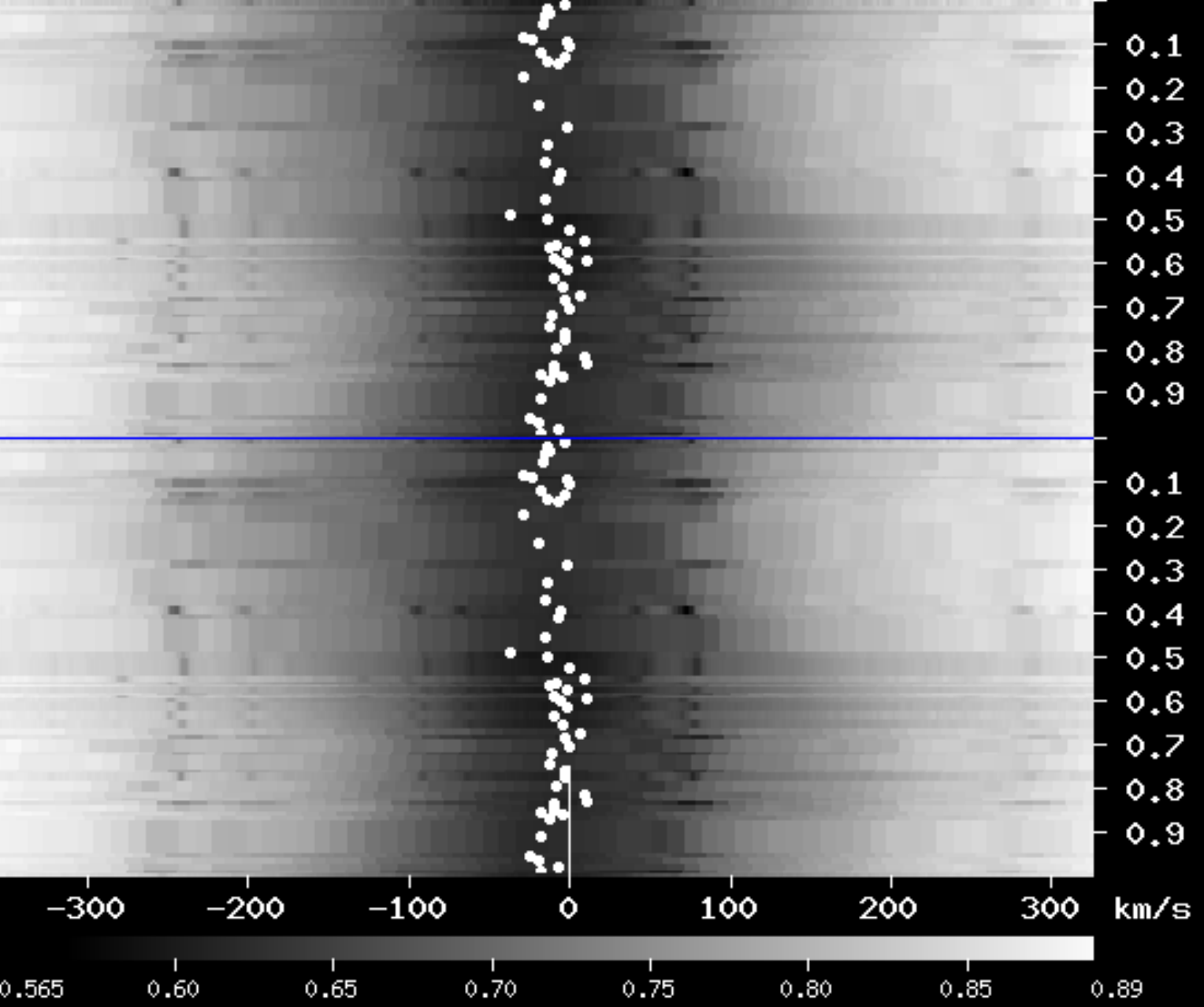}
\includegraphics[width=9.0cm]{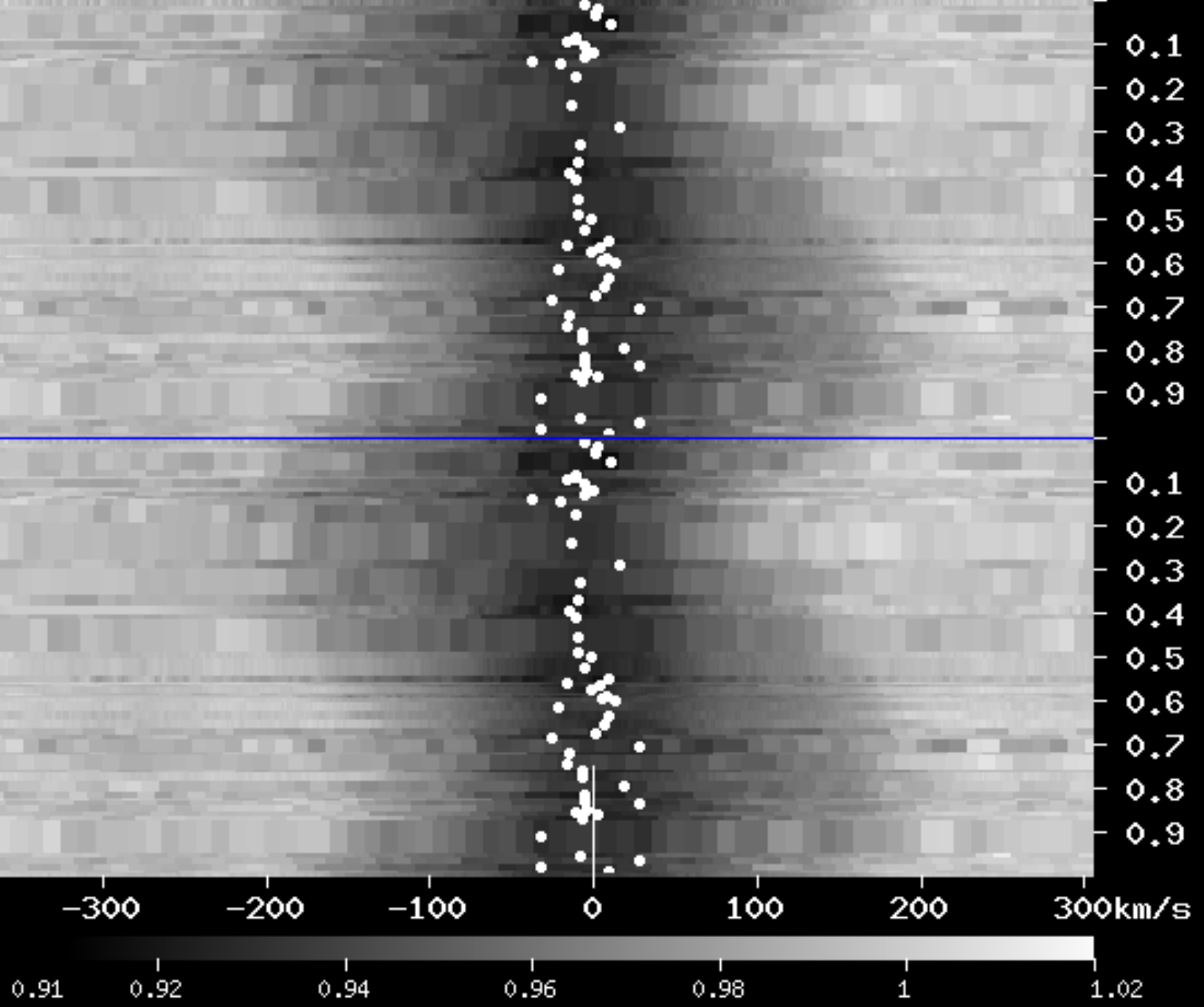}
\caption{Dynamical spectra plotted vs. orbital phase of the 25\fd416
period with phase zero at superior conjunction. Lines of \ha (top panel) and
\ion{He}{i}~6678~\AA\ (bottom panel) are shown. The white dots denote the
RV position of the deepest point of each line profile to show that
the position of the line cores does not vary with the 25\fd416 period.}
\label{dynsp}
\end{figure}

Figure~\ref{dynsp} shows dynamical spectra for the \ha and \he line
versus phase of the 25\fd416 period. The orbital motion in
the line wings of both lines is clearly visible. In contrast, the line cores
do not exhibit clear RV variations with the same amplitude.
We also note that there is no indication
of an antiphase RV variation expected for the secondary in the 25\fd416 orbit.
This means that the bulk of the line cores {\sl is not} associated with
{\sl the primary,} as suggested by \citet{neiner2014}, but with {\sl the tertiary}
of the visual orbit.

As explained above, we measured RVs separately for the broad and narrow
parts of the \ha and \he lines. We immediately noted,
in accord with \citet{neiner2014}, that the range of RV variations is
much smaller for the narrow parts of the lines,
and we verified that the RVs of the wide parts of the lines vary
with the known ${\sim}25.4$~d period.

The \spefo RVs of the line cores of the Balmer lines
seemingly follow the $25.4$~d RV curve of the wide parts of the lines,
in phase with the wide wings, but with a strongly reduced amplitude.
This implies that they do not belong to the secondary in the $25.4$~d orbit but
to the distant tertiary, with a nearly constant RV. Their apparent RV changes
are caused only by the strong line blending of the lines of the primary and
tertiary.

\subsection{Period analysis of RVs}
\begin{figure}[H]
\includegraphics[width=8.0cm]{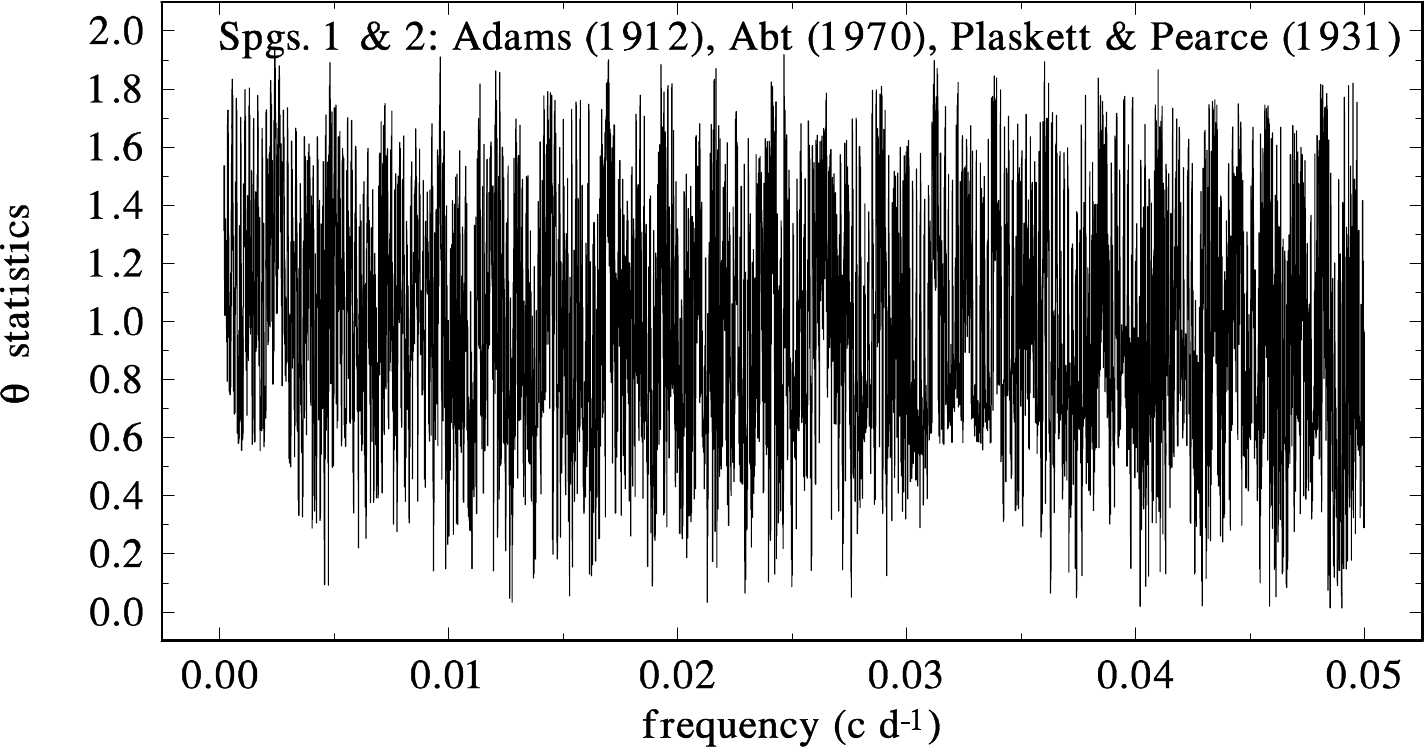}
\includegraphics[width=8.0cm]{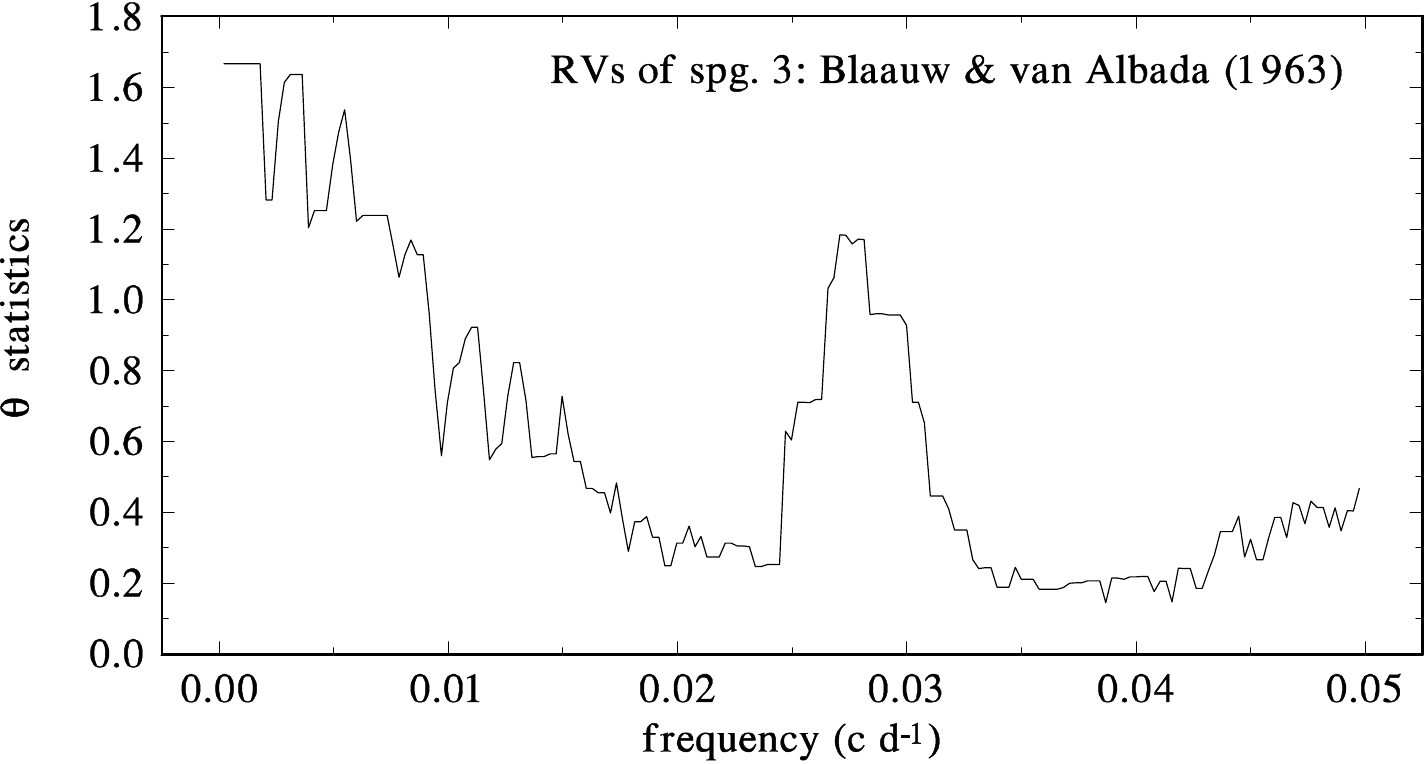}
\includegraphics[width=8.0cm]{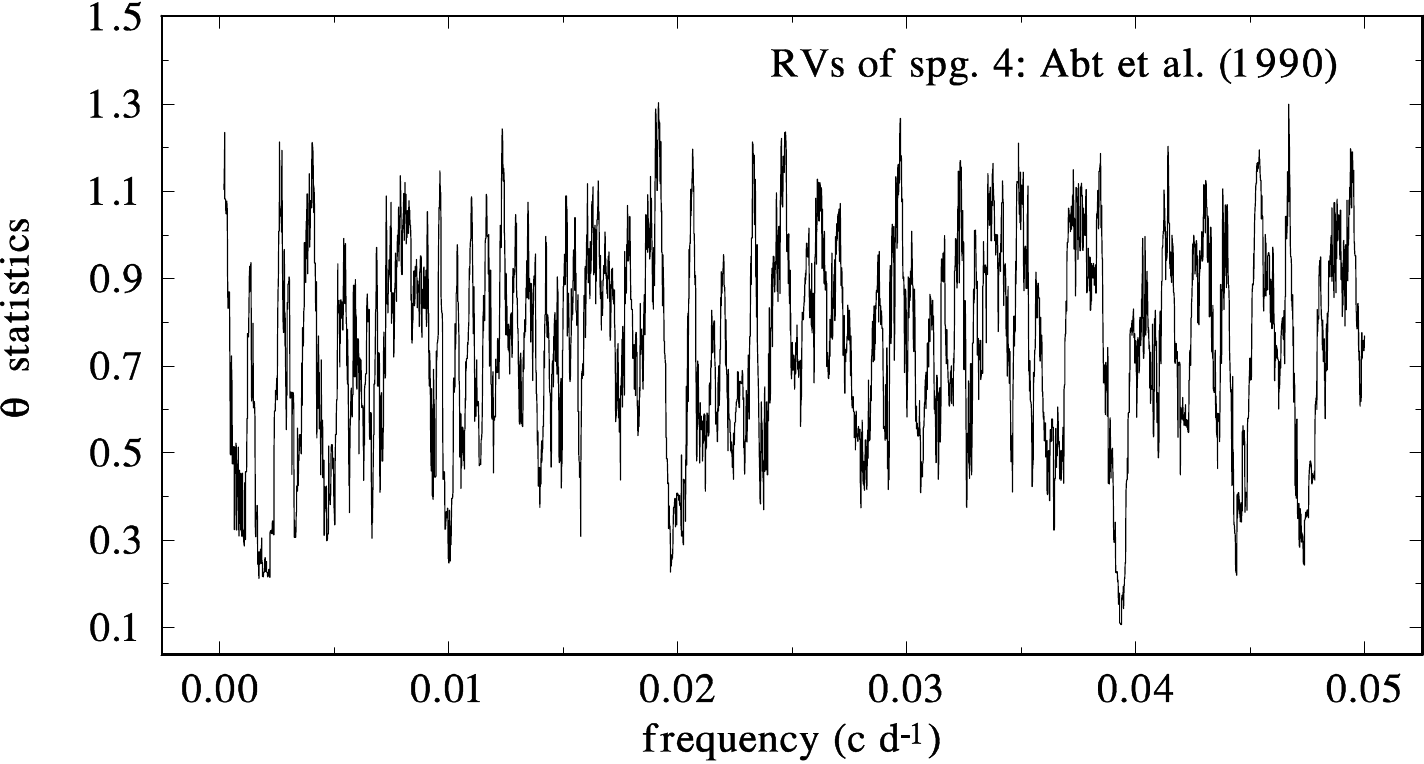}
\includegraphics[width=8.0cm]{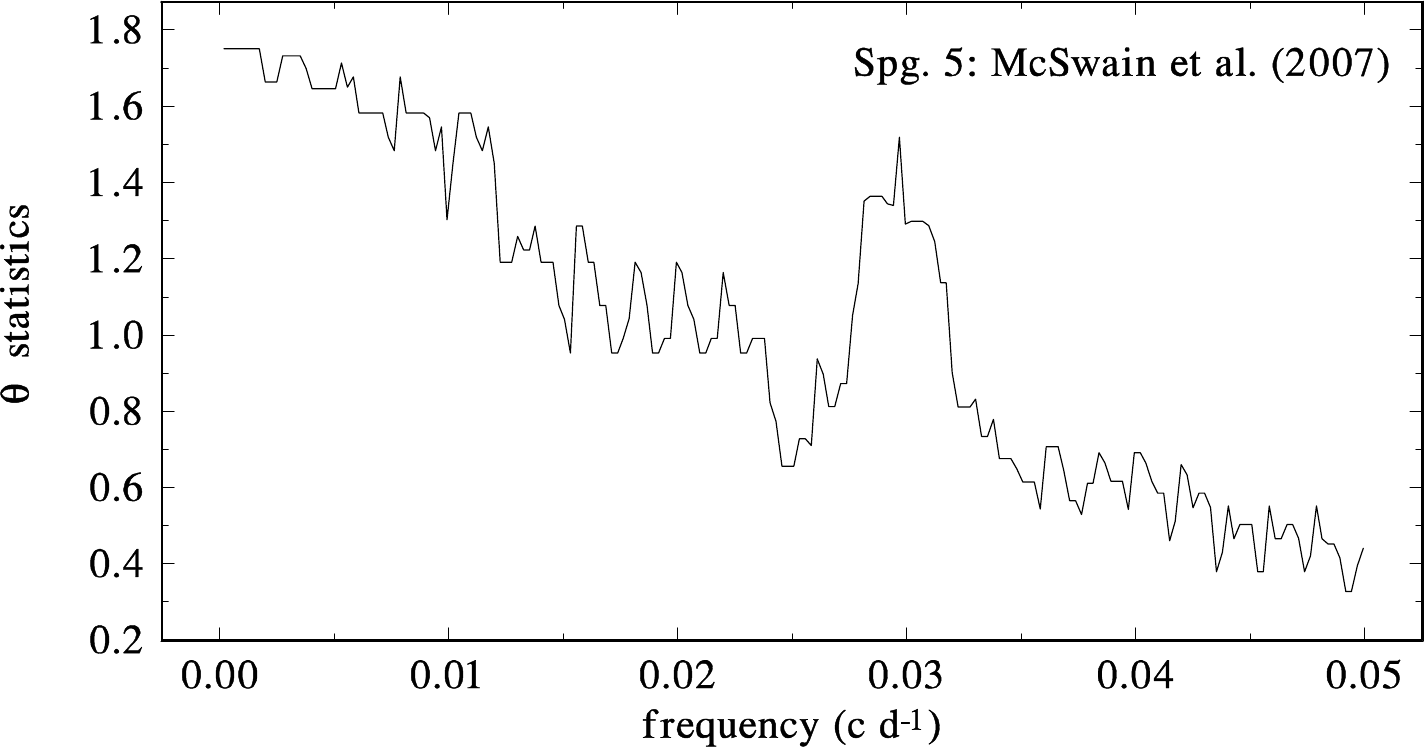}
\includegraphics[width=8.0cm]{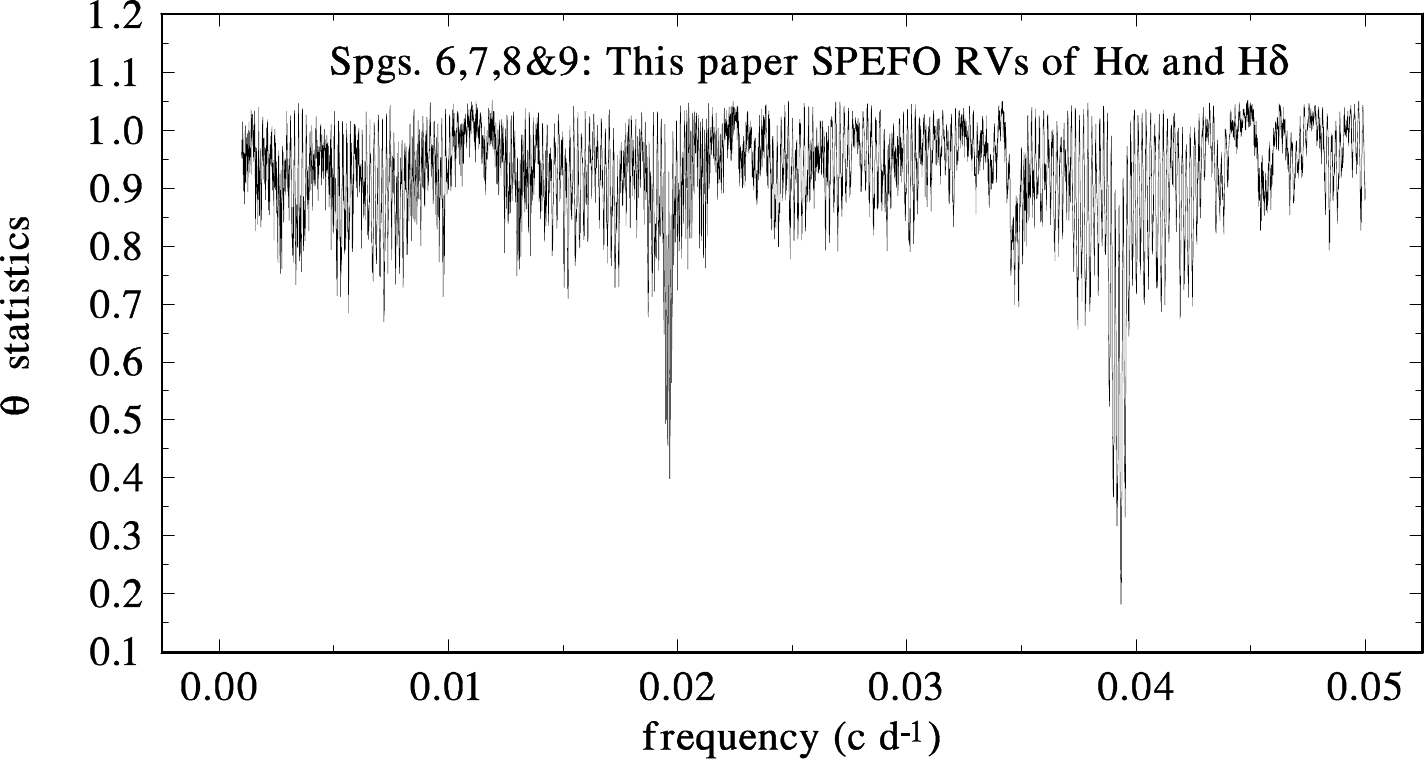}
\caption{\citet{stell78} $\theta$ statistics periodograms for several
data sets of RVs published by various authors, and the \spefo RVs
of the outer wings of the \ha and H$\delta$ lines.}\label{thetas}
\end{figure}

To resolve the question of the true value of the orbital
period of the closer pair, we calculated PDM periodograms \citep{stell78} for
individual data subsets. They are shown in Fig.~\ref{thetas}.  It is
immediately seen that neither the \cite{blaauw63} nor the \citet{mcswain2007} RVs
alone are able to restrict the value of the orbital period.
\citet{blaauw63} were obviously aware of the limitations of their
data set since they estimated the uncertainty of the 27\fd8 period
to be as much as one~day. The time interval covered by their own observations
is shorter than two orbital periods.
We also note that the first data set from spectrograph~1, although
it spans a long time interval, is probably of limited accuracy and cannot
be used to determine a~unique period.

Only observations by \citet{abt70} and our new RVs are numerous enough
to identify the true orbital period. Their periodograms are mutually
similar. The dominant period of 25\fd416 ($f=0.0393$~c\,d$^{-1}$)
and its first harmonics are clearly visible. A~smaller minimum at 0.0360~c\,d$^{-1}$ corresponds
to a 1~yr alias of the 25\fd416 period.

\begin{table}
\caption[]{Exploratory \fotel solutions for RVs from
the literature and from our red spectra.}
\label{newsol}
\begin{center}
\begin{tabular}{rcccccl}
\hline\hline\noalign{\smallskip}
Element         & Spgs. 2 -- 8&Gauss \ion{He}{i}&\spefo \ion{He}{i}\\
\noalign{\smallskip}\hline\noalign{\smallskip}
$P$ (d)             &25.41569(42)  &25.41569 fix. &25.41569 fix.\\
$T_{\rm periastr.}$ &43838.5(1.0)  &54385.9(1.2)  &54385.16(84)\\
$T_{\rm super.c.}$  &43836.1       &54383.7       &54383.92\\
$T_{\rm RV max.}$   &43829.9       &54377.6       &54378.33\\
$e$                 &0.104(23)     &0.163(40)     &0.162(28)\\
$\omega$ ($^\circ$) &131(15)       &132(17)       &115(12)\\
$K_1$ (\ks)         &24.0(1.2)     &37.1(4.4)     &41.6(1.5)\\
$\gamma_{2}$ (\ks)  &$-$29.3(4.8)  & --           & --\\
$\gamma_{3}$ (\ks)  &$-$15.9(1.7)  & --           & --\\
$\gamma_{4}$ (\ks)  &$-$16.5(1.2)  & --           & --\\
$\gamma_{5}$ (\ks)  &$-$8.4(1.0)   & --           & --\\
$\gamma_{6}$ (\ks)  &$-$16.4(1.1)  &$-10.2(1.8)$  &$-15.0(1.7)$\\
$\gamma_{7}$ (\ks)  &$-$8.4(1.3)   &$-17.0(3.4)$  &$-10.8(4.7)$\\
$\gamma_{8}$ (\ks)  &$-$12.1(1.3)  &$-8.9(1.9)$   & $-9.1(2.1)$\\
$\gamma_{9}$ (\ks)  &$-$12.35(0.75)&$-12.3(1.6)$  &$-13.7(1.3)$\\
rms (\ks)           &4.45          &8.27          &7.41\\
No. of RVs          &131           &72            &72\\
\noalign{\smallskip}\hline\noalign{\smallskip}
\end{tabular}
\end{center}
\tablefoot{All epochs are in RJD;
rms is the rms of one observation of unit weight.
The error of epoch is shorter for a reference epoch close to
the centre of the interval covered by data. To facilitate comparison of the
first solution with the remaining two, we also provide corresponding
recalculated epochs: RJD~54386.0, RJD~54383.6, and RJD~54377.4\,.
}

\end{table}

\subsection{Linear ephemeris for the inner-orbit period}
To obtain as precise a value of the orbital period $P_1$ of the inner
binary as possible, we derived some trial orbital solutions with the
program \fotel \citep{fotel1,fotel2}.
We first calculated an orbital solution based on our new \spefo RVs.
The semiamplitude for the \ion{He}{i}~6678~\AA\ RVs
is higher than that obtained from the \ha RVs. It is a well-known effect
for early-type stars that the lines with appreciably Stark-broadened
wings like the Balmer lines give lower RV amplitudes and are not
suitable for the determination of true orbital elements
\citep[e.g.][]{ander75,ander83}. This clearly represents another
complication on the way to obtaining realistic binary properties.

 Since our first task is, however, to derive the correct value of the orbital
period, we adopted the \spefo \ha RVs and H$\delta$ RVs for the Aurelie
spectra, which give a semiamplitude closer to those found
by previous investigators; see Table~\ref{oldsol}. We combined
them with all RVs from the literature and ran a solution in which we
allowed the calculation of individual systemic velocities ($\gamma$s) for
individual data sets. Using the rms errors of individual sets,
we derived their weights inversely proportional to the squares of
the corresponding rms errors. Investigating the phase plots, we found that
the RVs of spectrograph~1 (Mt.~Wilson) probably refer to a combination of
the broad and narrow parts of the line profiles and cannot contribute
meaningfully to constrain the orbital period.

\begin{figure}
\includegraphics[width=9.0cm]{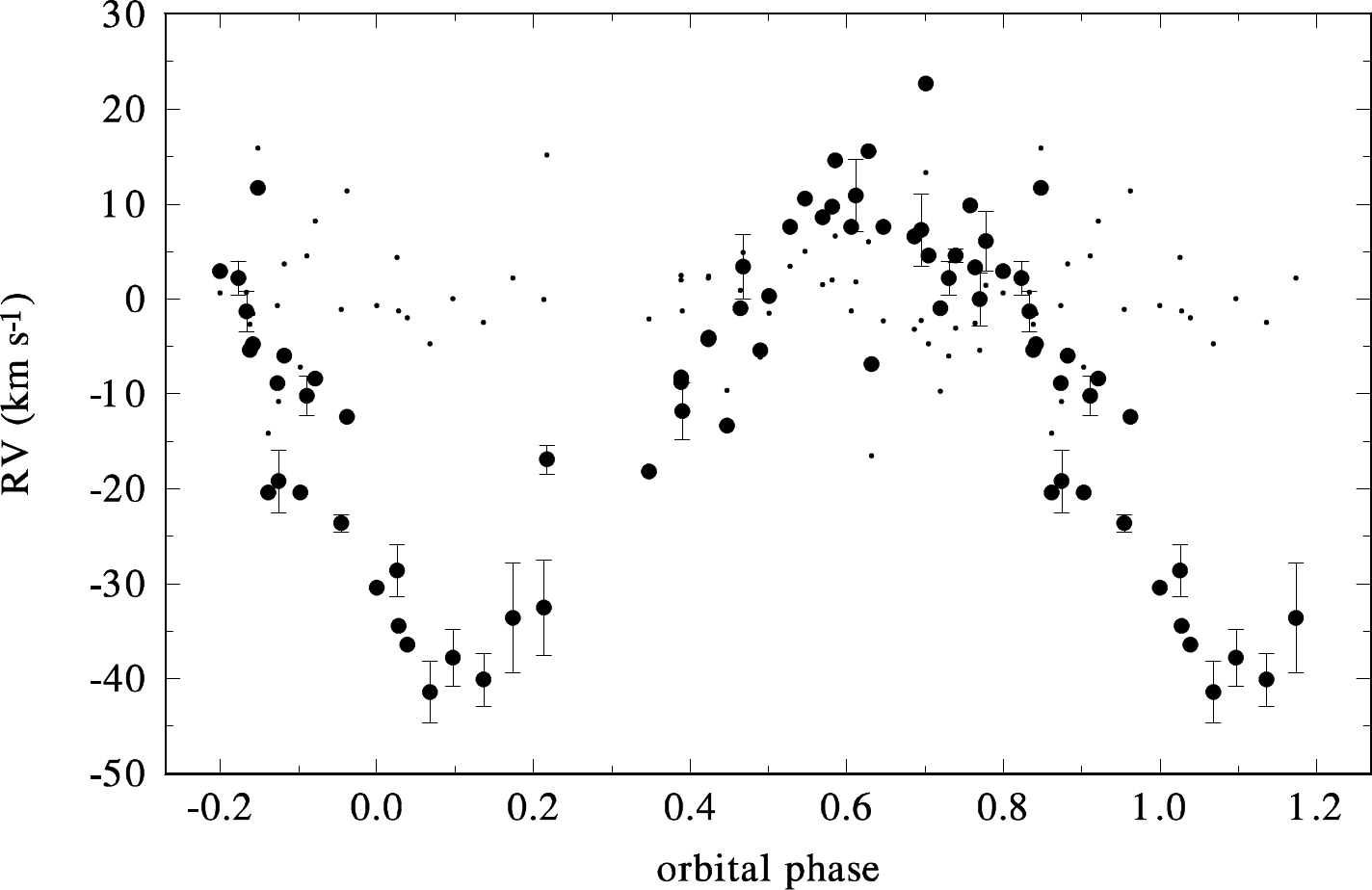}
\includegraphics[width=9.0cm]{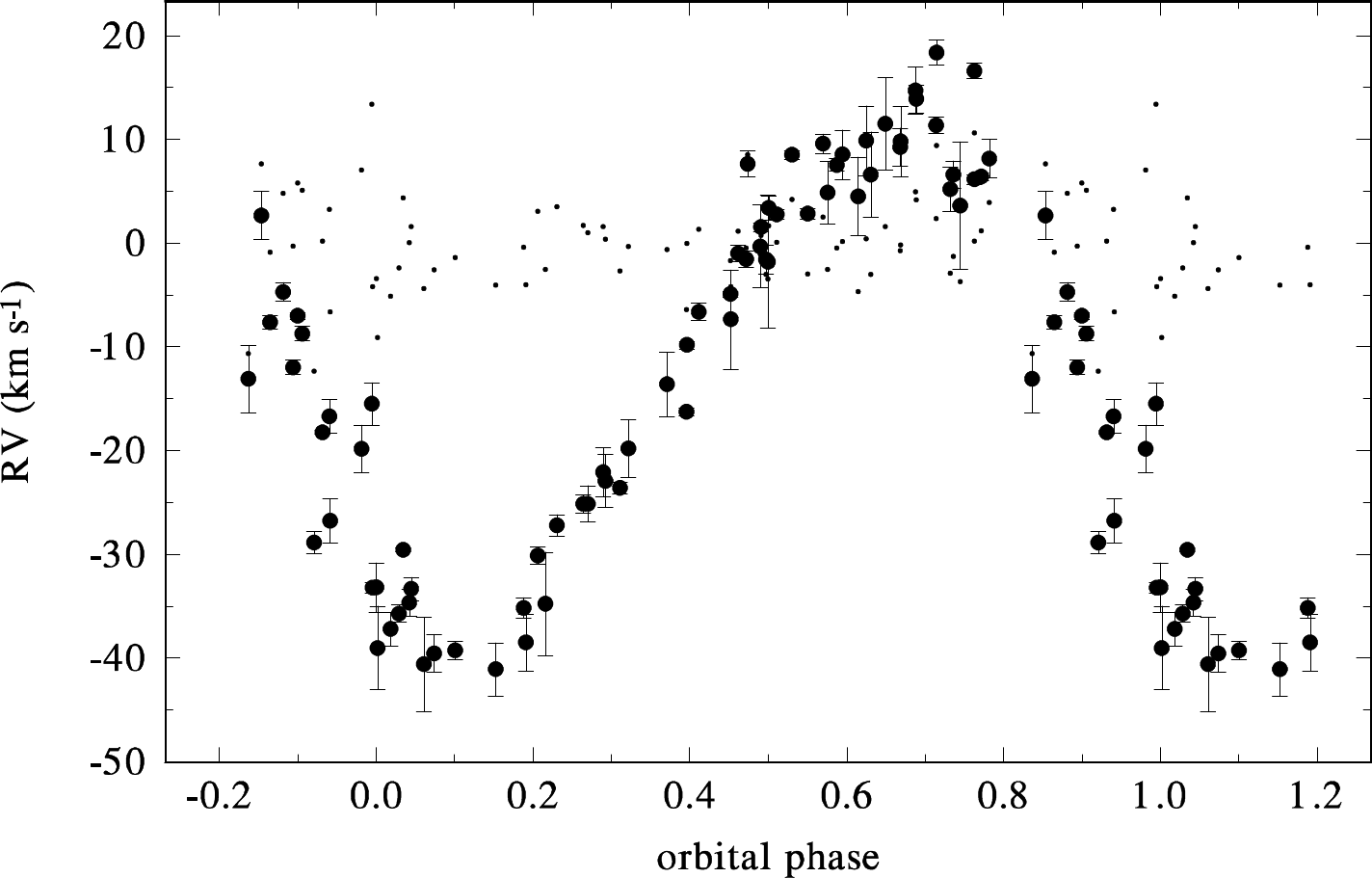}
\caption{%
Top: A phase plot of all file 2 to file 5 RVs (i.e. data
from the literature).
Bottom: A phase plot of all our new file 6 to file 9 RVs.
\spefo \ha and H$\delta$ RVs were used for the new spg. 6 to 9 spectra.
Small dots denote the \oc\ residuals from the orbital solution. All RVs
in the two lower panels were corrected for the difference in $\gamma$
velocities, adopting $\gamma$ of the most numerous file~9 as the reference.
In all plots, the orbital period of 25\fd41569 and the reference epoch of
the periastron passage RJD = 43838.65 from the first solution of
Table~\ref{newsol} were used. The rms errors of individual RVs are shown
whenever available.}\label{rv12}
\end{figure}

We then calculated a joint solution for all weighted RVs
from  spgs.~2 -- 8
(see the first solution in Table~\ref{newsol}), and we adopted the
orbital period from this solution
\begin{equation}
T_{\rm super.conj.}={\rm HJD}~2454384.65(44)+25\fd41569(42)\,.\label{efe}
\end{equation}
Phase plots for this joint solution are shown
in Fig.~\ref{rv12}, separately for the RVs from the literature and for
our new \ha and H$\delta$ RVs.

Keeping the orbital period fixed, we then derived two other solutions,
this time based on our new \he RVs from spgs.~6 -- 9 and
using alternatively the RV set based on the Gaussian fits and
on the \spefo RVs. The results are also listed in Table~\ref{newsol}.
The elements show that it is not easy
to choose the correct value of $K_1$ and the mass function.
The semiamplitude and eccentricity differ for the individual solutions,
depending on the line(s) measured and also on the measuring technique.
A part of the problem is the fact that almost all spectral lines are
to some extent affected by neighbouring blends.

We also tried to disentangle the lines of the system components using the \korel
program \citep{korel1,korel2,korel3}. However, since the spectra at our
disposal cover only one-tenth of the visual orbit for the \ion{He}{i}~4143~\AA\
line and even less for the other spectral lines, the semiamplitudes of
the bodies in the outer orbit could not be meaningfully converged
and we have no firm clue how to fix them. We only verified
that no trace of the secondary could be found. In our experience,
this implies that the secondary must be for more than three~magnitudes fainter than the
combined light of the primary and tertiary.

\subsection{Speckle-interferometry and the visual orbit}
Adding one speckle-interferometric observation from 2007
\citep{mason2009} to the existing set of 37 observations of \ve,
we were able to derive a new visual orbit of the third body, but our
preliminary solution confirmed the orbit published by Docobo and Andrade
\citep[see][]{doboco2005}. For reference, it is summarised in Table~\ref{vis}.
The reported uncertainties are the nominal ones, corresponding
to a single local minimum of the respective $\chi^2$. However,
the values of $P_2$ and $e_2$ seem to be strongly correlated. Moreover,
they strongly depend on the five previous astrometric observations from
the beginning of the 20th century. Their real uncertainty
can probably be greater than 0.02\,arcsec, which we assumed for them.

\begin{table}
\caption[]{Improved visual orbit of the \ve\ tetriary.}
\label{vis}
\begin{center}
\begin{tabular}{rcccccl}
\hline\hline\noalign{\smallskip}
Element         & Value with rms error\\
\noalign{\smallskip}\hline\noalign{\smallskip}
$P_2$ (d/yr)         &$62004\pm1202/169.76\pm3.29$\\
$T_{\rm peri\,2}$(RJD)&$34613\pm820$\\
$e_2$             &$0.176\pm0.023$\\
$\omega_2$ ($^\circ$) &$308.2\pm9.4^{**}$\\
$\Omega_2$ ($^\circ$) &$29\pm12$\\
$i_2$ ($^\circ$)      &$64\pm16$\\
$a_2$ (mas)           &$211\pm97$\\
$m_1+m_2+m_3$ (\ms) &9.34 (range 5.50--17.8)$^*$\\
\noalign{\smallskip}\hline\noalign{\smallskip}
\end{tabular}
\tablefoot{$^*)$ Assuming a revised Hipparcos parallax of
0\farcs00326$\pm$0\farcs00063 after \citet{leeuw2007a,leeuw2007b}.
$^{**})$ The argument of periastron refers to
the visual tertiary. For the close binary it would be $128.\!\!^\circ2$, of
course.}
\end{center}
\end{table}


\section{Nature of the rapid photometric variations}\label{rapid}
\subsection{Overview and analysis of previous results}

\begin{table}
\caption[]{Frequencies and periods of light variations of \vc found
by various investigators.}\label{freq}
\begin{center}
\begin{tabular}{ccccl}
\hline\hline\noalign{\smallskip}
Frequency         &  Period & Source & Identification\\
 (c\,d$^{-1}$)     &   (d)   \\
\noalign{\smallskip}\hline\noalign{\smallskip}
     0.94  & 1.06 & 1 & $f_1$\\
\noalign{\smallskip}\hline\noalign{\smallskip}
     0.939 & 1.06  &2& $f_1$\\
     0.399 & 2.51  &2& $f_2$\\
\noalign{\smallskip}\hline\noalign{\smallskip}
0.93914(3) &1.06480&3& $f_1$\\
0.39934(4) &2.50413&3& $f_2$\\
\noalign{\smallskip}\hline\noalign{\smallskip}
0.39923(4) &2.50413&4& $f_2$\\
0.93895(4) &1.06502&4& $f_1$\\
0.79906(9) &1.25147&4& $2f_2$\\
0.96630(9) &1.03488&4&\\
\noalign{\smallskip}\hline\noalign{\smallskip}
0.39946(4) &2.50338 &5& $f_2$\\
0.93895(4) &1.06502 &5& $f_1$\\
1.20346(6) &0.830937&5&$3f_2$?\\
\noalign{\smallskip}\hline\noalign{\smallskip}
0.9390     &1.065   &6& $f_1$\\
0.3994     &2.504   &6& $f_2$\\
0.9309     &1.074   &6&      \\
0.7988     &1.252   &6&$2f_2$\\
\noalign{\smallskip}\hline\noalign{\smallskip}
\end{tabular}
\tablefoot{Column Source: the rows show
1. \citet{waeletal98}, \hp\ photometry;
2. \citet{andrews2000}, Four College APT and \hp\ photometry;
3. \citet{mathiasetal01}, \hp\ photometry;
4. \citet{decatetal04}, Geneva 7C and \hp\ photometry;
5. \citet{decat2007}, Geneva 7C photometry;
6. \citet{dukes2009}, Four College APT, Geneva 7C and \hp\ photometry.
}
\end{center}
\end{table}

Table~\ref{freq} provides an overview of various frequencies of light
variations of \vc reported by several investigators. There seems to be
no doubt about the two principal frequencies, detected by
all investigators, 0.939 and 0.399~\cd. Notably, \citet{mathiasetal01}
were unable to find any spectroscopic signatures of these two frequencies
in their Aurelie spectra covering the wavelength interval from 4085
to 4155~\AA. The authors state, however, that all their spectra have S/Ns
lower than 140. Two groups, \citet{decatetal04} and \citet{dukes2009},
reported another frequency, 0.799~\cd. \citet{decatetal04} denoted it
as a possible alias. We note that it is an exact harmonics
of the 0.399~\cd\ frequency. In our opinion, this indicates
that the light curve with the 0.399~\cd\ frequency deviates from
a~sinusoidal shape. Therefore, the 0.799~\cd\ frequency is not
an independent frequency. Another weak frequency of 1.20346~\cd\
reported by \citet{decat2007} (which in our opinion could be the second
harmonics of 0.399~\cd) could not be confirmed by \citet{dukes2009},
who analysed their own Str\"omgen \uvby\ photometry from the Four College
Automatic Photoelectric telescope (APT) along with the \hp\ and
Geneva~7-C photometries. They found a frequency of 0.9309~\cd,
however, which is quite close to $f_1=0.939$~\cd. We note that 0.9390 and 0.9309~\cd\
mutually differ for one-third of the frequency of a sidereal year, so we
suspect that the 0.9309~cd$^{-1}$ frequency is an alias.
The above facts led us to suspect that the light changes of \vc are
modulated only by two independent periods.

\subsection{New period analysis of available photometry}
For the purpose of the period analysis, we subtracted seasonal mean values
from all data to remove possible slight secular variations of either
instrumental systems of the telescopes or true small secular changes
(see Appendix~\ref{apc} for details). We used the programs {\tt PERIOD04}
\citep{lenz2005} and \fotel for the modelling and fits. We found that
the longer, non-sinusoidal period of 2.5~d has larger amplitudes for
the second and fifth harmonics as well (periods 1\fd25194, and 0\fd50077), which
were then included into fits. Inspecting cases with more observations
within short time-intervals like 0.03 d, we conclude that in all three data
sets there are cases when the rms error of single observations exceeds
0\m007. The rms error of the fit is only slowly decreasing when more frequencies are added, and it remains on the 0\m007 level.

We arrive at the following ephemerides for the 1\fd065, and 2\fd504 periods:
\begin{eqnarray}
T_{\rm max.light} &=& {\rm RJD}~52540.2866(61) + 1\fd0649524(40)\,,\label{efej}\\
T_{\rm max.light} &=& {\rm RJD}~52242.585(32) + 2\fd503867(19)\,.\label{efed}
\end{eqnarray}

\begin{figure}
\includegraphics[width=9.0cm]{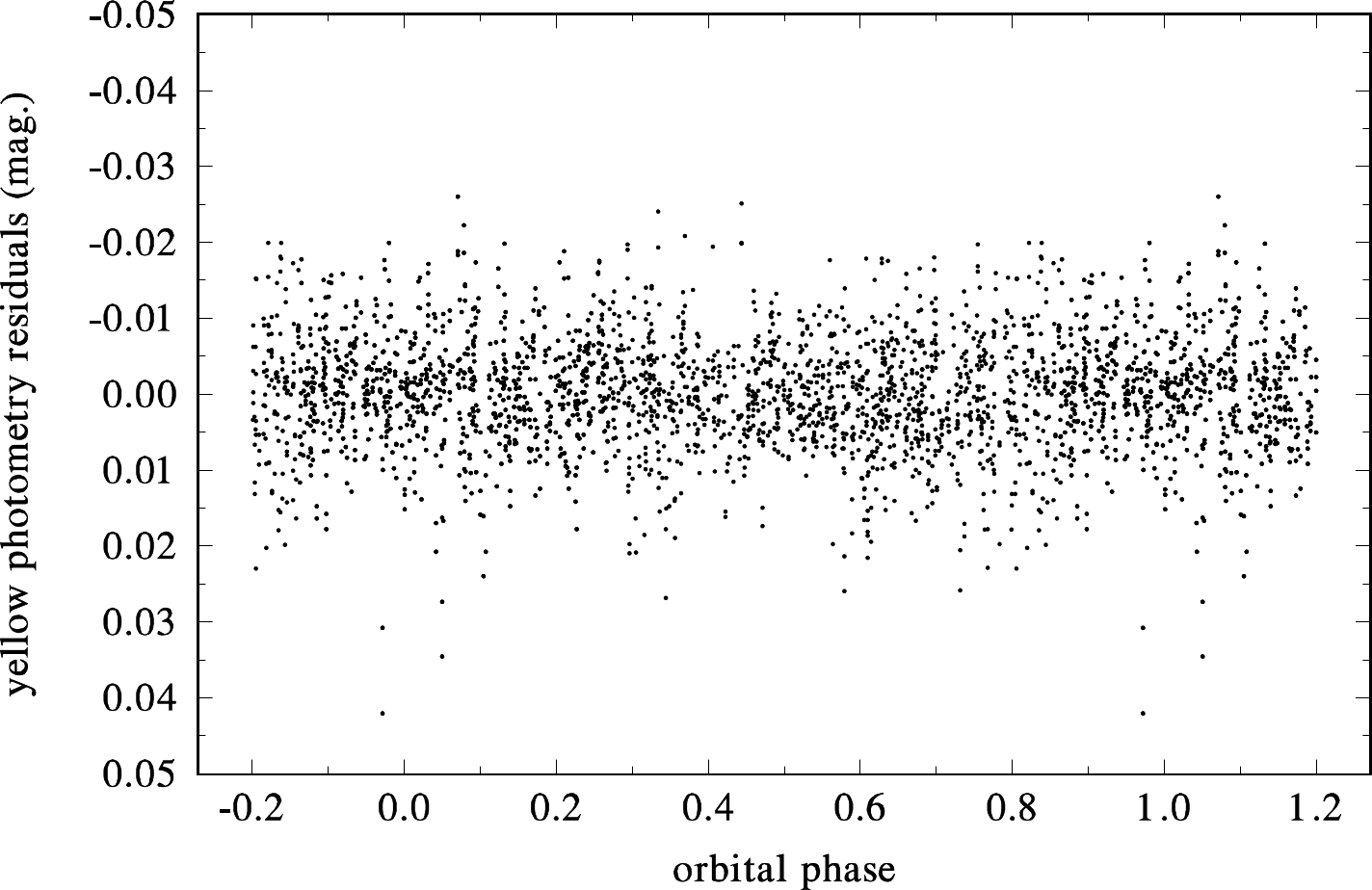}
\caption{All yellow-band photometric observations prewhitened for both the 1\fd065
period, and the 2\fd5039 period and its harmonics, plotted vs.
orbital phase from ephemeris~(\ref{efe}).}\label{lcorbit}
\end{figure}

Binary eclipses can also be safely excluded from the available
photometry (see Fig.~\ref{lcorbit}). Our simulations of the light curve for
the 25\fd4 orbit with the program \phoebe~1 \citep{prsa2005} for plausible
values of the stellar radii (see below) then show there is a rather strict
{\em upper\/} limit for the orbital inclination, $i_1 \le 85^\circ$.

\subsection{Periodic variations of the magnetic field}
\begin{figure}
\includegraphics[width=9.0cm]{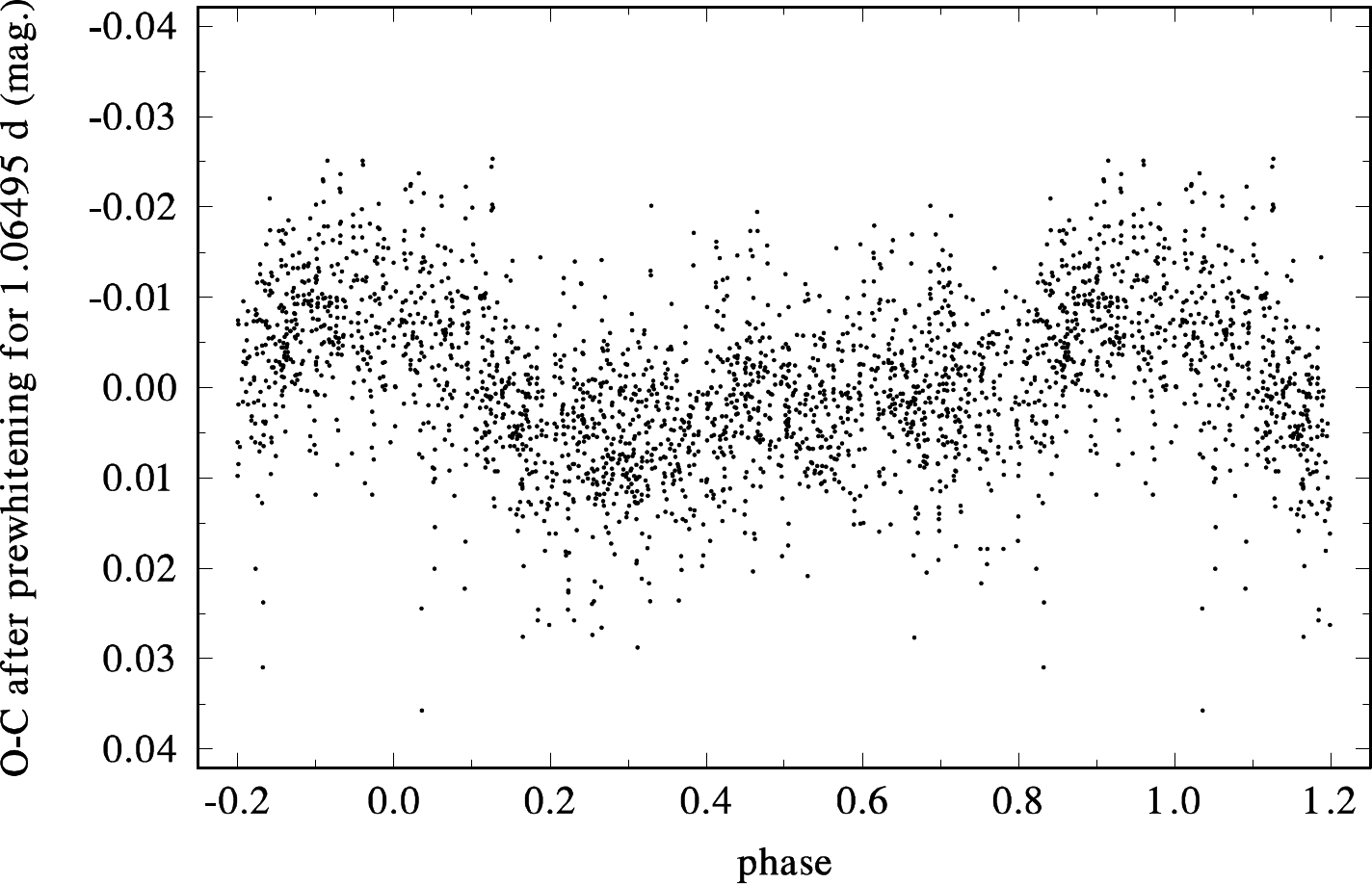}
\includegraphics[width=9.0cm]{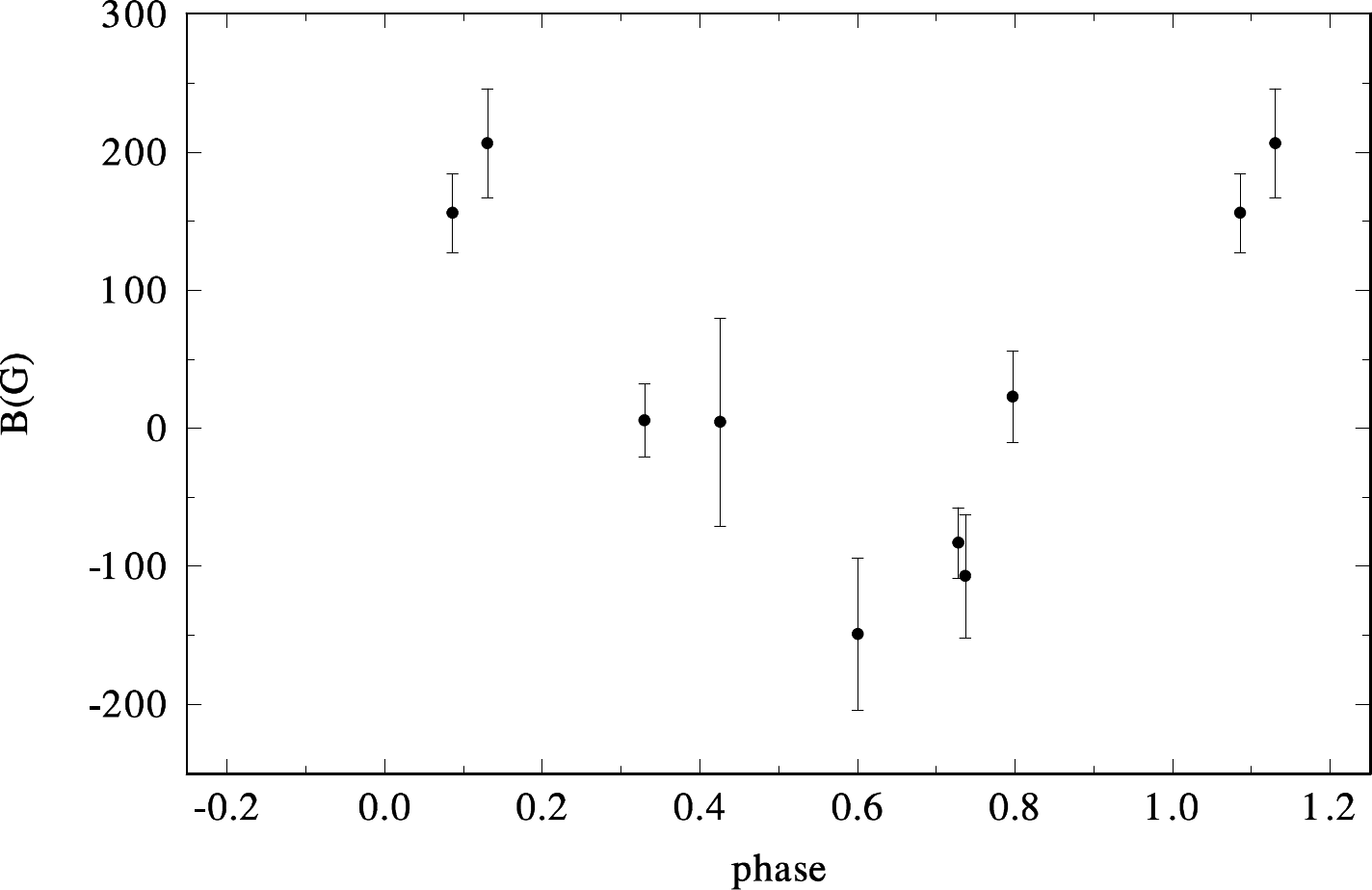}
\includegraphics[width=9.0cm]{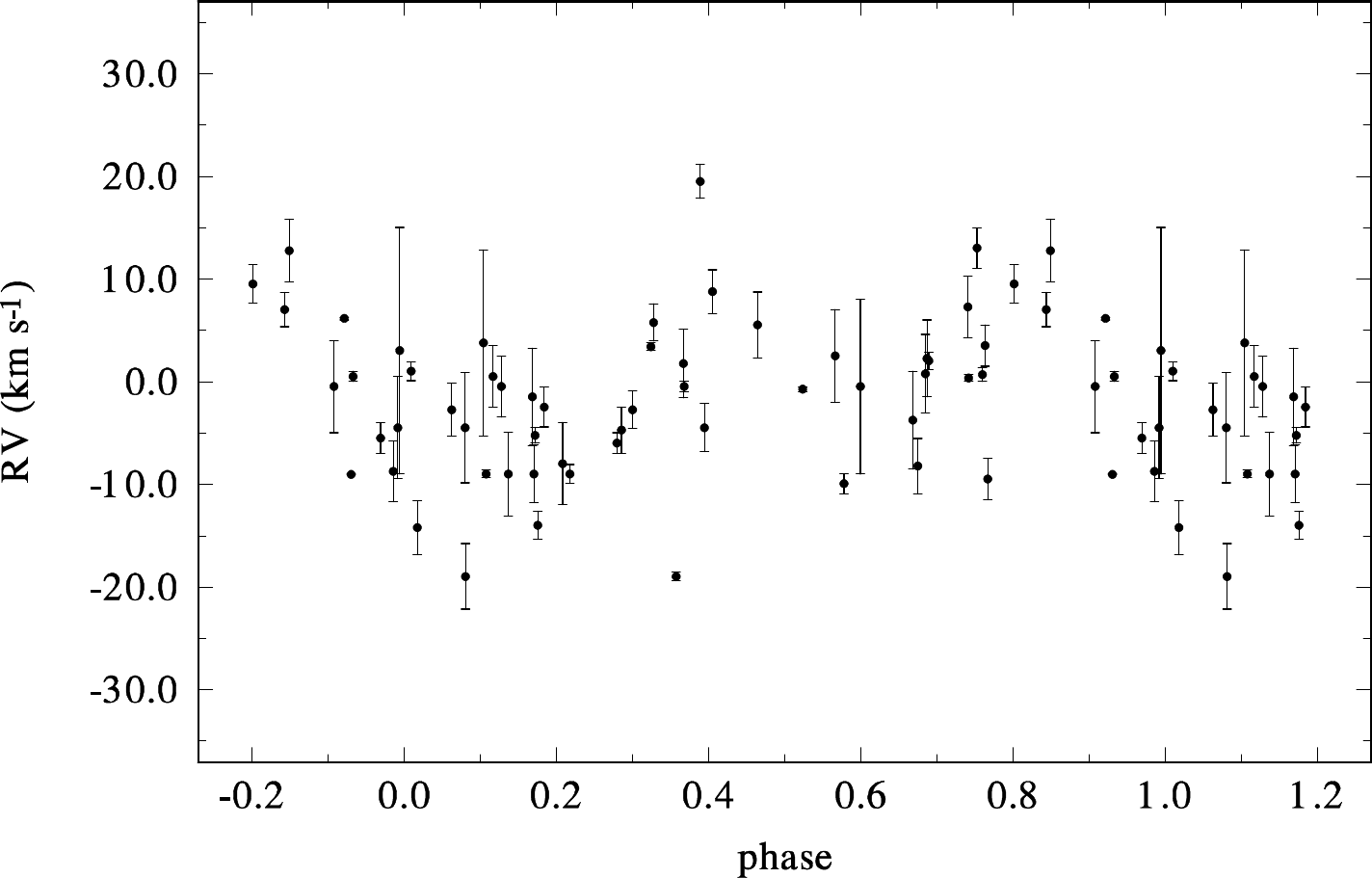}
\caption{Phase plots for the 2\fd503867 period calculated using
ephemeris~\ref{efed}.
Top: A phase plot of all yellow-band photometric observations
prewhitened for the 1\fd06495 period.
Middle: A phase plot of the magnetic field observations
from measurements of \citet{neiner2014} from spectra not affected by technical
problems and for the mean values from all lines. The rms errors of
the measurements are shown by the error bars.
Bottom: A phase plot of the Gaussian-fit \ion{He}{i}~6678~\AA\ RVs.
}
\label{mgfield}
\end{figure}

We tried to determine whether the observed variations of the magnetic field found
by \cite{neiner2014} might not be related to the known photometric changes.
We quickly found that the magnetic field indeed varies with the longer of the
photometric periods, 2\fd50387; see Fig.~\ref{mgfield}.

A sinusoidal variation of the magnetic field
intensity is commonly interpreted as a dipole field inclined to the rotational axis of
the star and varying with the stellar rotational period. \citet{neiner2014}
associated the magnetic field with the narrow-line component, which
they considered to be the primary, but which is -- as we have shown here --
the distant tertiary component moving in the long orbit with the 25\fd4
binary.

In the bottom panel of Fig.~\ref{mgfield} we also plot the Gaussian-fit
RVs of the narrow component of \ion{He}{i}~6678~\AA\ (i.e. RVs of
the magnetic-field tertiary) line vs. phase
of the 2\fd504 period. The scatter is rather large, but that
the RV variations clearly show some similarity to the light changes. This could be
due to the corotating structures related to the magnetic field of the tertiary.
We mention this to alert future observers that it might be rewarding
to obtain systematic whole-night series of spectra at high resolution to
verify the reality of the phenomenon and possibly to apply techniques such
as Doppler imaging if the phenomenon is confirmed to be real.

\begin{figure}
\includegraphics[width=9.0cm]{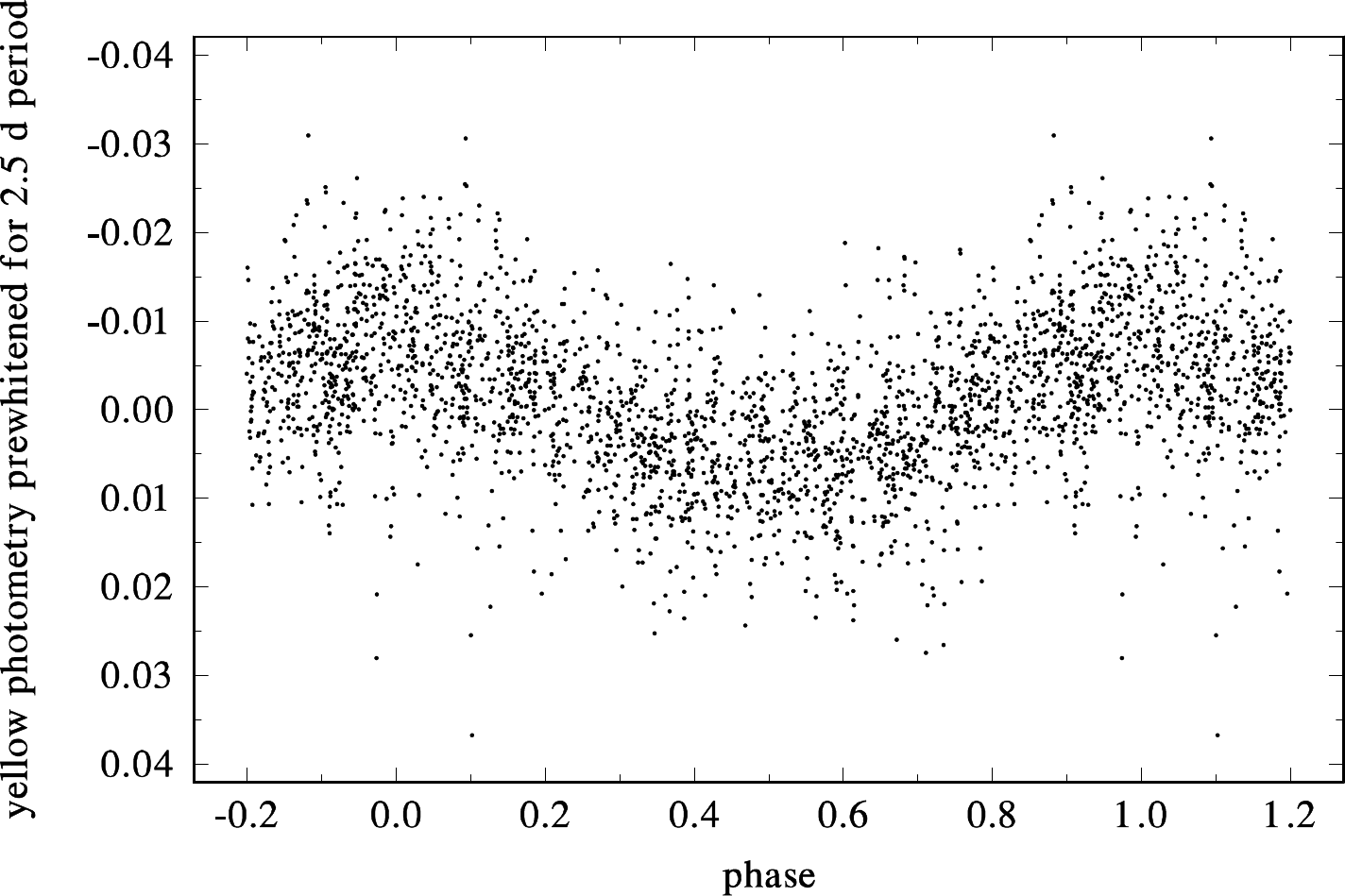}
\includegraphics[width=9.0cm]{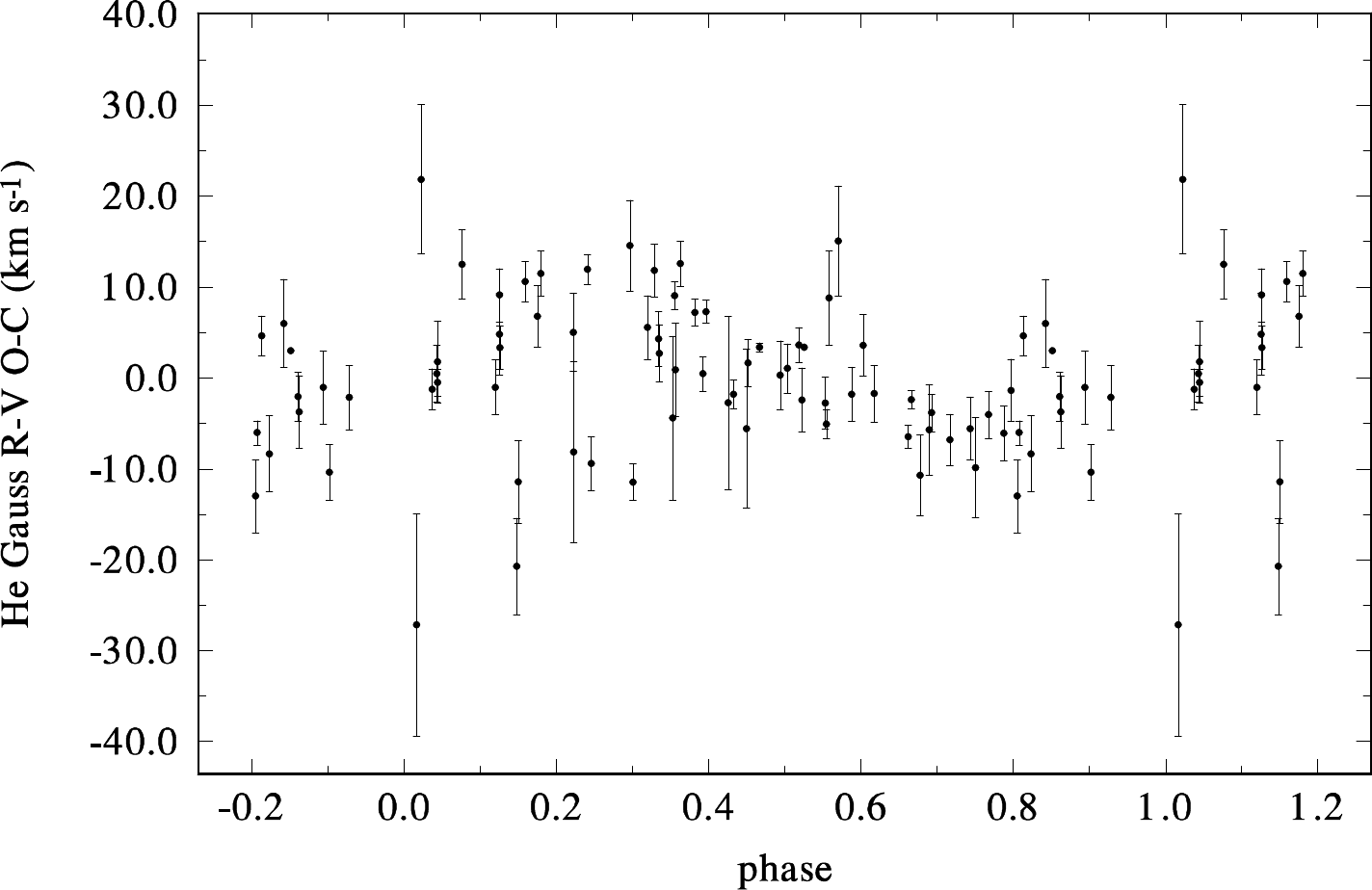}
\includegraphics[width=9.0cm]{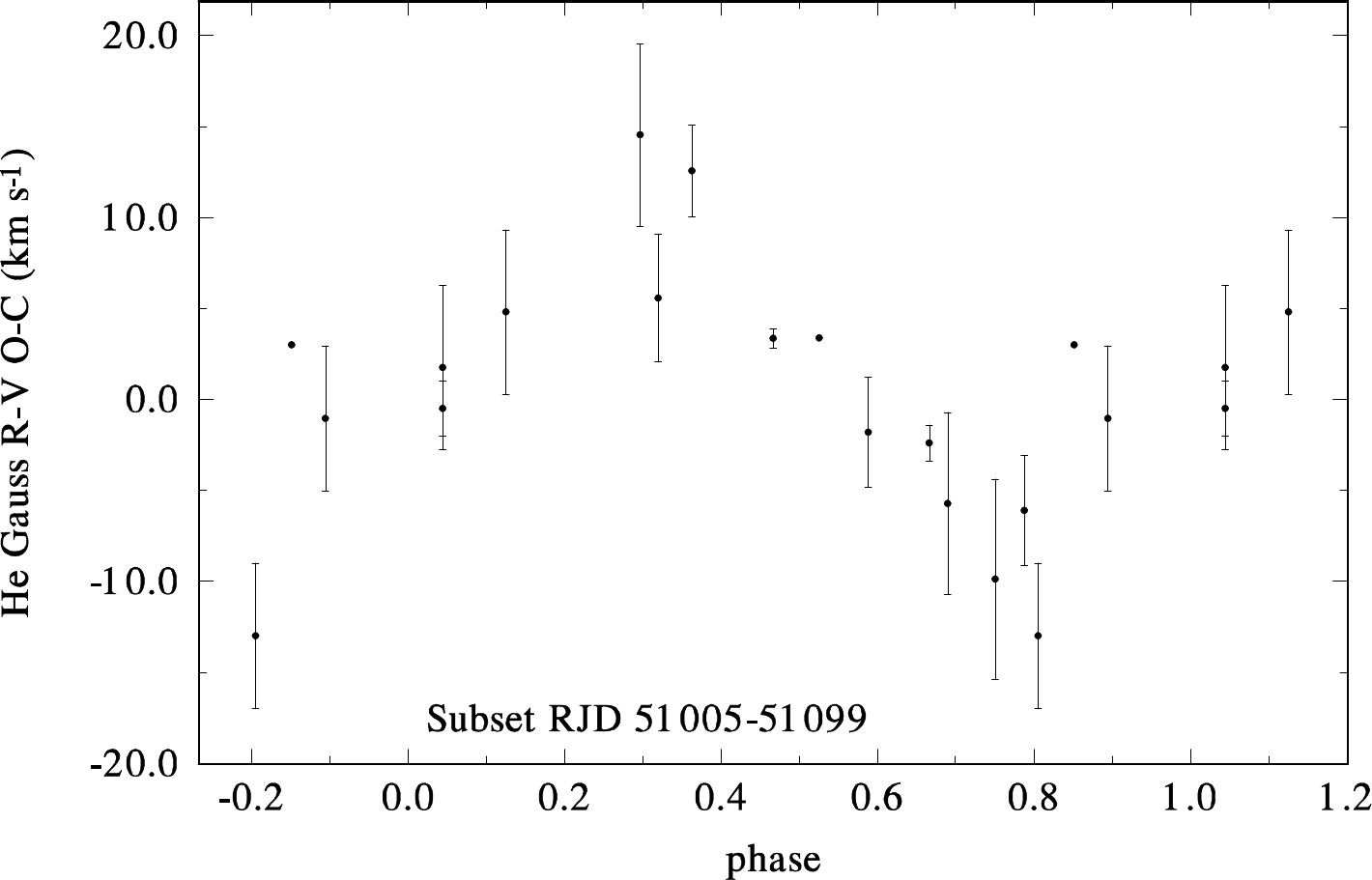}
\caption{Phase plots for the 1\fd06495 period calculated using
ephemeris~\ref{efej}.
Top: All yellow-band photometry \oc\ residuals prewhitened for
the 2.5 d period and its harmonics are shown.
Middle: A plot of all the residuals from the orbital solution
for the Gaussian-fit \he RVs.
Bottom: A plot of the residuals from the orbital solution
for the Gaussian fit \he RVs for the subset of Aurelie spectra.
}\label{oc}
\end{figure}

\subsection{Rapid line-profile changes of the primary?}
If some rapid line-profile changes are also present in the spectra
of the primary, they could manifest themselves as additional
RV disturbances, overlapped over the orbital RV changes.
While the technique of the settings on the outer line
wings in \spefo might tend to mask such changes (while affecting the true RV
amplitude), such disturbances should be best detected in the Gaussian fits,
which use two fixed Gaussian line profiles, mutually moved only in RV. We
therefore inspected the \oc\ RV residuals from the 25\fd4 orbit for
the Gaussian \he RVs based on repeated measurements.
In Fig.~\ref{oc} we compare the plot of these RV residuals with the plot
of yellow-band photometry prewhitened for the 2\fd50387 period.
There is some indication of sinusoidal residual RV changes with
the photometric period of 1\fd065, especially for RVs with lower rms
errors. The modulation is nicely seen in the subset of Aurelie high-S/N
and high-resolution spectra. While it is clear that the ultimate
confirmation may only come from new whole-night series of high-resolution
spectra, we do believe there is a reason to identify the 1\fd065 period
with the rotational period of the primary. The modulation of measured
RVs could be caused by some structures on the surface of the primary, moving
across the stellar disk as the star rotates. We note that there is growing
evidence for rotational modulation for a number of B stars from the photometry
obtained by the Kepler satellite and by ground-based surveys
\citep[see, e.g.][]{mcn2012,nie2013,kou2014,balona2015,balona2016}.


\section{Initial estimates of the basic physical elements of the system}

\subsection{Radiative properties of the primary and tetriary}\label{synt}

To determine the radiative properties of the two dominant components,
we used the Python program \pyte,
which interpolates in a pre-calculated grid of synthetic spectra. Using
a set of observed spectra, it tries to find the optimal fit between
the observed and interpolated model spectra with the help
of a~simplex minimization technique. It returns the radiative properties of
the system components such as \teff, \vsin or $\log~g$, but also the
relative luminosities of the stars and
RVs of individual spectra.\footnote{The program \pyt is available with
a~tutorial at \\ https://github.com/chrysante87/pyterpol/wiki\,.}
The function of the program is described in detail in \citet{jn2016}.

\begin{table}
\caption[]{Radiative properties of the \ve\ primary and tertiary derived from a comparison
of selected wavelength segments of the observed and interpolated synthetic
spectra of the primary and tertiary.}
\label{synpar}
\begin{center}
\begin{tabular}{rcccccl}
\hline\hline\noalign{\smallskip}
Element         & Primary&Tertiary  \\
\noalign{\smallskip}\hline\noalign{\smallskip}
\tef (K)        &15850(400)&14920(490)\\
\lgg [cgs]      &3.68(12)  &3.80(15)\\
$L_{4000-4280}$&0.709(54) &0.295(54) \\
$L_{4308-4490}$&0.700(48) &0.302(48) \\
$L_{4700-5025}$&0.701(65) &0.305(64) \\
$L_{6670-6690}$&0.704(60) &0.295(60) \\
\vsin (\ks)     &179(6)    &72(4)     \\
\noalign{\smallskip}\hline\noalign{\smallskip}
\end{tabular}
\end{center}
\end{table}

In our particular application, two grids of synthetic spectra,
\citet{bstars} for \teff$>$15000~K, and \citet{pala2010} for \tef$<$15000~K,
were used to estimate basic properties of the primary and tertiary
for all 16~Aurelie, 3~ Elodie, and 13~Bernard Lyot high-S/N spectra.
The following spectral regions containing numerous spectral lines, but avoiding
the inter-order transitions and regions with stronger telluric lines,
were modelled simultaneously:

\smallskip
\centerline{4000--4025~\AA, 4097--4155~\AA, 4260--4280~\AA,}
\centerline{4308--4405~\AA, 4450--4490~\AA, 4700--4725~\AA,}
\centerline{4817--4935~\AA, 5008--5025~\AA, and 6670--6690~\AA.}

Relative component luminosities were fitted separately in four
spectral bands: 4000--4280~\AA, 4308--4490~\AA, 4700-5025~\AA, and
6670--6690~\AA.
Uncertainties of radiative properties were obtained through
Markov chain Monte Carlo (MCMC) simulation implemented
within {\tt emcee}\footnote{The library is
available through GitHub~\url{https://github.com/dfm/emcee.git}
and its thorough description is
at~\url{http://dan.iel.fm/emcee/current/}.}
Python library by~\citet{fore2013}.
They are summarised in Table~\ref{synpar}.
We note that in contrast to the N-body model, which we use below
for the final estimate of the most probable properties of the system,
\pyt derives the RV from individual spectra without any assumption
about orbital motion. It is therefore reassuring that when we allowed
for a free convergence of the orbital period in a trial solution based
on the \pyt RVs, we arrived at a value of 25\fd4156(13), in excellent agreement
with the value of ephemeris (\ref{efe}) based on all available RVs.

\subsection{Additional properties of the primary}\label{sec:primary}
All available MKK spectral classifications of \vc summarized
in the SIMBAD bibliography, including a recent one by \citet{tamaz2006}
based on high-resolution spectra, agree on the spectral type B5\,IV.
Dereddening of the mean all-sky standard \ubv\ magnitudes from Hvar
(see Table~\ref{jouubv}) gives $V_0= $5\m43, (\bv)$_0 = -$0\m170,
(\ub)$_0 = -$0\m635, and E(\bv) = 0\m055, which corresponds to a
B4-5IV star after the calibration by \citet{golay74}.

We can then use the observed Hipparcos parallax $p$ of \vc
to further constrain the primary. According
to the improved reduction \citep{leeuw2007a,leeuw2007b}, $p=0\farcs00326(63)$.
Assuming the above mentioned $V_0=5$\m43 for the whole system, the observed
magnitude difference between the primary and tertiary and neglecting the light
contribution from the unseen secondary, we arrive at $V_0^1=5$\m82 for the
primary. Using the above-mentioned range of the possible values of
the parallax, 0\farcs00263 to 0\farcs00394, we obtain
$M_{\rm V}^1=-2$\m08 and $-1$\m20, respectively. The bolometric corrections
for the considered range of the effective temperature of the primary
are from $-1$\m433 to $-1$\m557 so that the extreme allowed values
of the bolometric magnitude of the primary $M_{\rm bol}^1$ are
$-2$\m63 to $-3$\m64.

If 1\fd065 is the rotational period of the primary, then
for $v\,\sin i_1 = 179\,{\rm km}\,{\rm s}^{-1}$, derived from
the \pyt solution for the primary, $R_1 \geq 2\pi\,v\sin i/P = 3.77$\,\rs.
For instance, if $i_1=70^\circ$ , we obtain $R_1 = 4.01$\,\rs. For the expected
range of the primary mass, this would imply \lgg = 3.98 to 4.04 [cgs] for
the primary, which is slightly offset at the 2.5$\sigma$ level
with respect to the range deduced from the fit by synthetic spectra
(Table~\ref{synpar}). We note, however, that the \lgg values are very
sensitive to an accurate placement of the continuum, which is especially
difficult for Balmer lines from the echelle spectra.
On the other hand, if the true rotational period is a photometric
double-wave curve with a period twice longer, that is, 2\fd130,
we would obtain \lgg = 3.38 to 3.44 [cgs]; again offset from the nominal range.
This would also correspond to significantly larger radius $R_1 = 8.02$\,\rs\
and an even lower upper limit of~$i$. Given all the uncertainties involved,
however, both possibilities need to be kept in mind.


\subsection{Tertiary component}\label{sec:tetriary}

The more recent estimates of the magnitude difference between
the close binary and tertiary from astrometry \citep{esa97,mason2009} agree on
$m_{\rm 3\,V}-m_{\rm (1+2)\,V}\sim0\m9$. The relative luminosities of the
primary and tertiary estimated with \pyt (see Table~\ref{synpar}) imply
$m_{\rm 3\,V}-m_{\rm (1+2)\,V}=0\m92$, in remarkable agreement with the
astrometric estimates. This once more confirms our identification of the
narrow-line component seen in the spectra with the tertiary.

Using the $v\,\sin i_3 = 72\,{\rm km}\,{\rm s}^{-1}$ estimated from the \pyt solution for
the tertiary and adopting the 2\fd504 period as its period of rotation, we can
similarly estimate the radius of the tertiary $R_3\geq3.56$\,\rs.
Assuming that the inclination of the rotational axis of the tertiary
is identical with the orbital inclination of the outer orbit $64^\circ$,
we estimate \lgg from 3.90 to 4.01 for it. This agrees quite well
with the observed range of \lgg from the line-profile modelling.


\begin{figure}
\centering
\includegraphics[width=9cm]{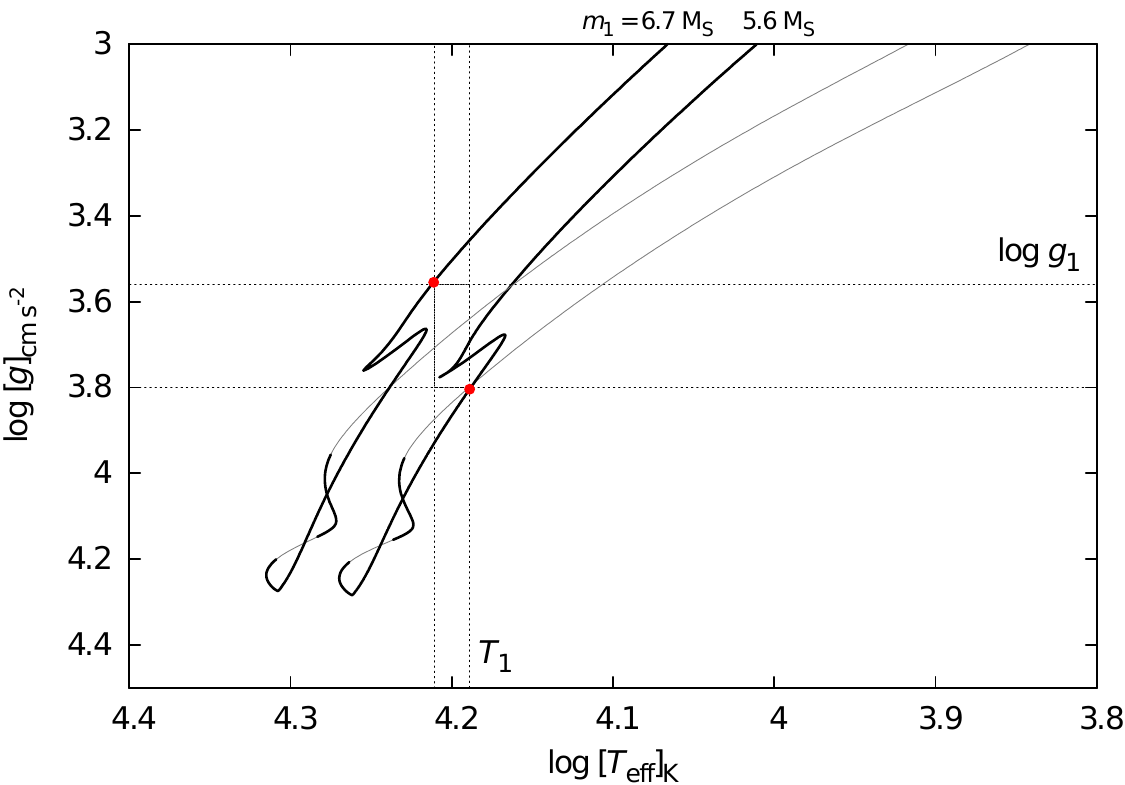}
\includegraphics[width=9cm]{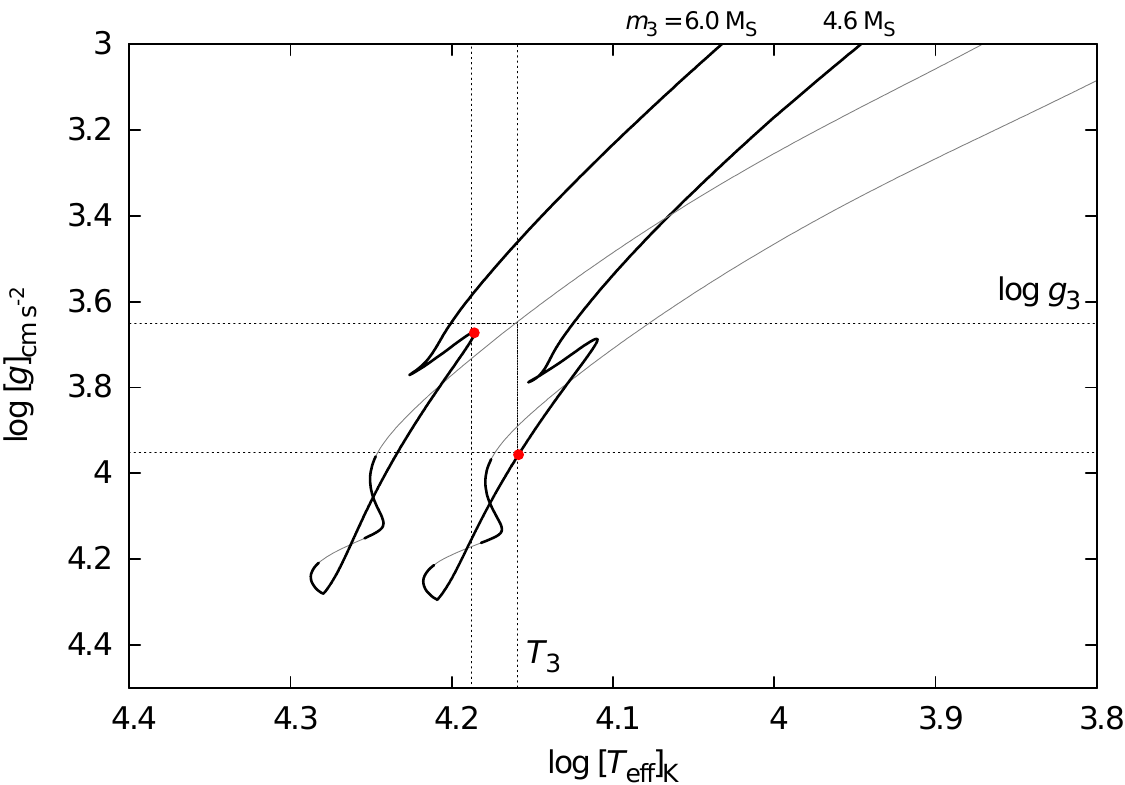}
\caption{Evolutionary tracks in
effective temperature $T_{\rm eff}$ vs
surface gravity $\log g$ plots,
including pre-MS (thin line),
MS, and SGB phases (thick line).
The horizontal and vertical dotted lines correspond to the
allowed $1\sigma$ ranges of the respective parameters, according
to Table~\ref{synpar}. The top panel shows the primary component of \ve
for the two values of its mass $m_1 = 5.6\,M_\odot$ and $6.7\,M_\odot$.
The bottom panel shows the tertiary
for $m_3 = 4.6\,M_\odot$ and $6.0\,M_\odot$.}
\label{logg13}
\end{figure}

\subsection{Component masses}
In order to preliminarily estimate the component masses,
we calculated evolutionary tracks using the Mesastar program \citep{pax2015}
and compared them with the observed range of the values of $T_{\rm eff}$ and
$\log g$, deduced from the line-profile fits, in Table~\ref{synpar}.
In all cases we assumed a helium abundance $Y = 0.274$,
metallicity $Z = 0.0195$ (i.e. very close to the standard values of
\cite{caughlan1988}), and mixing length parameter $\alpha = 2.1$.
We also accounted for element diffusion, Reimers red giant branch (RGB) wind with $\eta = 0.6$,
Blocker AGB wind with $\eta = 0.1$, even though neither is very relevant on
the zero-age main sequence (ZAMS). Semiconvection and convective overshooting were both switched off.
The maximum output time step was $\Delta t = 10^5\,{\rm yr}$ to resolve the terminal main sequence (TAMS).
Mesastar program, rev. 8845 \citep{pax2015},
was used for these computations.

From the comparison presented in Fig.~\ref{logg13}, we can estimate the mass
of the primary to be from $m_1 = 5.6$ to $6.7\,M_\odot$ ($1\sigma$ range).
We verified that the evolutionary tracks in the considered
region close to the TAMS are not
overly sensitive to the value of metallicity. For $Z = 0.04$,
the corresponding mass range changes by as much as $0.1\,M_\odot$;
lower metallicities are not likely for relatively young stars.
The same is true for the mixing-length parameter $\alpha$
or the semiconvection parameter $\alpha_{\rm sc}$,
since both stars are in a radiative equilibrium.
For instance, $\alpha = 1.6$ or $\alpha_{\rm sc} = 0.01$
lead to essentially the same results.
For the overshooting parameters $f_{\rm ov} = 0.014$ and $f_0 = 0.004$
(i.e. within the range discussed in \cite{herwig2000}),
the evolution around the TAMS is notably different, nevertheless,
the range of masses would be shifted only slightly upwards,
by less than $0.1\,M_\odot$,

A similar analysis for the
tertiary component led to the mass $m_3 = 4.6$ to $6.0\,M_\odot$.
Within the uncertainties, the primary and tertiary stars can even be equally massive
and have very similar effective temperatures. The similarity of the primary and tertiary
is also supported by the fact that their relative luminosities over the whole
optical range seem to be the same; see Table~\ref{synpar}.


\section{N-body model of the V746 Cas system}
\subsection{Formulation}
To analyse all available observational data in the most consistent way,
we attempted to use the N-body model of~\cite{broz2017}, which was recently
developed and successfully applied to the $\xi$~Tauri quadruple system \citep{jn2016}.
Here, we used a significantly extended version of it.
While a detailed technical description is given in the latter paper,
we repeat here a subset of equations relevant for our problem:
\begin{equation}
\ddot{\bf r}_{{\rm b}i} = -\sum_{j\ne i}^{N_{\rm bod}} {Gm_j\over r_{ji}^3}\,{\bf r}_{ji} + {\bf f}_{\rm tidal} + {\bf f}_{\rm oblat} + {\bf f}_{\rm ppn}\,,
\end{equation}
\begin{equation}
I_\lambda' = \sum_{j=1}^{N_{\rm bod}} {L_j\over L_{\rm tot}} \,I_{\rm syn}\!\left[\lambda\left(1-{v_{z{\rm b}j+\gamma}\over c}\right), T_{{\rm eff}\,j}, \log g_j, v_{{\rm rot}\,j}, {\cal Z}_j\right],
\end{equation}
\begin{equation}
F_V' = \sum_{j=1}^{N_{\rm bod}} \left({R_j\over d}\right)^{\!2} \int_0^\infty F_{\rm syn}\!\left[\lambda, T_{{\rm eff}\,j}, \log g_j, v_{{\rm rot}\,j}, {\cal Z}_j\right] f_V(\lambda) {\rm d}\lambda\,,
\end{equation}
\begin{equation}
m_V' = -2.5\log_{10} {F_V'\over F_{V{\rm calib}} \int_0^\infty f_V(\lambda){\rm d}\lambda}\,,
\end{equation}
where the notation is as follows.
The index~$i$ always corresponds to observational data,
$j$~to individual bodies,
$N_{\rm bod} = 3$ is the number of bodies,
${\bf r}_{\rm b}$ barycentric coordinates,
$m$~component mass,
${\vec f}_{\rm tidal}$, ${\vec f}_{\rm oblat}$ and ${\vec f}_{\rm ppn}$
contributions from tidal, oblateness, and
parametrized post-Newtonian (PPN) accelerations,
which are included for the sake of completeness,
even though they are negligible in the V746~Cas system;
$I_\lambda$, $I_{\rm syn}$ normalized synthetic spectrum (intensity) of the whole system and component,
with appropriate Doppler shifts,
$L, L_{\rm tot}$ component luminosity and the total luminosity,
$T_{\rm eff}$ effective temperature,
$\log g$ surface gravity [in cgs],
$v_{\rm rot}$ projected rotational velocity,
${\cal Z}$ metallicity,
$F_{\rm syn}$ absolute monochromatic flux (in ${\rm erg}\,{\rm s}^{-1}\,{\rm cm}^{-2}\,{\rm cm}^{-1}$ units)
for any of the standard UBVRIJHK or non-standard bands,
$F_{V{\rm calib}}$ calibration flux, and
$f_V(\lambda)$ filter transmission function.
Both normalized and absolute synthetic spectra
were interpolated on-the-fly by \pyte, as described
in Nemravov\'a et al. (2016). For the absolute
spectra, the grids BSTAR \citep{bstars}
and PHOENIX \citep{husser2013} were used.

Synthetic data and observation are compared by means of a combined $\chi^2$ metric
\begin{equation}
\chi^2 = w_{\rm sky}\chi^2_{\rm sky} + w_{\rm syn}\chi^2_{\rm syn} + w_{\rm sed}\chi^2_{\rm sed} + \chi^2_{\rm mass}\,,\label{eq:chi2}
\end{equation}
with individual contributions defined as
\begin{equation}
\chi^2_{\rm sky} = \sum_{j=1}^{N_{\rm bod}} \sum_{i=1}^{N_{{\rm sky}\,j}} \left\{ {(\Delta x_{ji})^2 \over \sigma_{{\rm sky\,major}ji}^2} + {(\Delta y_{ji})^2 \over \sigma_{{\rm sky\,minor }ji}^2}\right\}\,, \label{chisky}
\end{equation}
\begin{equation}
(\Delta x_{ji}, \Delta y_{ji}) = {\bf R}\left(-\phi_{\rm ellipse}- {\pi\over2}\right) \times
\begin{pmatrix}
x_{{\rm p}\,ji}' - x_{{\rm p}\,ji} \\
y_{{\rm p}\,ji}' - y_{{\rm p}\,ji}
\end{pmatrix}
\,,
\end{equation}
\begin{equation}
\chi^2_{\rm syn} = \sum_{i=1}^{N_{\rm syn}} {\left(I_{\lambda\,i}' - I_{\lambda\,i}\right)^2 \over \sigma_{{\rm syn}\,i}^2}\,, \label{chisyn}
\end{equation}
\begin{equation}
\chi^2_{\rm sed} = \sum_{i=1}^{N_{\rm sed}} {\left(m_{Vi}' - m_{Vi}\right)^2 \over \sigma_{{\rm sed}\,i}^2}\,, \label{chised}
\end{equation}
\begin{equation}
\chi^2_{\rm mass} = \sum_{j=1}^{N_{\rm bod}} \left({2m_j-m_j^{\rm min}-m_j^{\rm max} \over m_j^{\rm max}-m_j^{\rm min}}\right)^{\!\!100}\!\!,\label{eq:mass_limit}
\end{equation}
where
$x_{\rm p}, y_{\rm p}$ denote 1+2 photocentric sky-plane coordinates,
$\sigma_{\rm sky\,major,\,minor}$ uncertainty of the astrometric position,
$\phi_{\rm ellipse}$ position angle of the respective ellipse,
${\bf R}(\dots)$ the corresponding $2\times 2$ rotation matrix,
$\sigma_{\rm syn}$ uncertainty of the normalized intensity,
$\sigma_{\rm sed}$ uncertainty of the spectral-energy distribution, and
$m_j^{\rm min}, m_j^{\rm max}$ minimum and maximum allowed masses.
The last term is an artificial function with
a sufficiently steep and smooth behaviour;
the high exponent prevents the simplex from drifting away.
Optionally, we can use weights~$w$ to enforce a convergence
for selected less-numerous data sets (e.g. with $w_{\rm sky} = 10$).

\subsection{Application to \ve}
The number of free parameters in our model is $L_{\rm free} = 26$;
namely the masses~$m_j$,
orbital elements of the two orbits $P_j$, $e_j$, $i_j$, $\Omega_j$, $\omega_j$, $M_j$,
systemic velocity~$\gamma$,
distance~$d$,
gravitational acceleration~$\log g_j$,
effective temperatures~$T_{{\rm eff}\,j}$, and
projected rotational velocities~$v_{{\rm rot}\,j}$.
Radii are thus dependent quantities computed as $R_j = \sqrt{Gm_j/g_j}$.
As usual, the parameter space is very extended
and we can expect a~number of local minima.
As reasonable starting points, we used the results
of previous observation-specific models.
We then employed a simplex algorithm for the minimisation,
with various starting points and several restarts.
We also verified the results by simulated annealing, even though neither of
the algorithms can guarantee that the true global minimum was found.

We used a complete observational data set:
(i)~37~speckle-interferometric and astrometric observations of the third body,
including some older and rather uncertain ones;
(ii)~72~spectra from the Elodie, Aurelie, Lyot, and Ond\v rejov datasets; this would represent 184\,666 individual data points,
but for our analysis we used 21\,506 mean points
in order to have a faster computation with a comparable resolution of all spectra;
(iii)~93~spectral-energy distribution measurements
from Hvar, Geneva and \cite{glushneva1992}.
For the last data set we had to perform dereddening,
assuming $A_V = 3.1\,E(B-V)$ relation,
$E(B-V) = 0.055$,
and the wavelength dependence from \cite{scheffler1987}, Tab.~4.1.

On the other hand, we did {\em not\/} use radial velocities,
as these are derived quantities possibly affected by some systematics,
but rather the rectified spectra directly.
We incorporated the same spectral regions, unaffected by any stronger
telluric lines as in Sect.~\ref{synt}.
Our approach is even better than using Pyterpol alone
(which fits a single spectrum at a time),
because the orbital motion is simulated by the N-body model,
which links the spectra observed at different times.
As a consequence, we are more sensitive to uncertainties
and systematics of the rectification procedure,
which would otherwise be hidden.

We encountered two serious problems, which required some modifications of
the program.
First, the secondary is practically unconstrained, except for the radial motion
of the primary. Its convergence is thus impossible and we used
 the  \cite{mr88} relations instead, that is to say,
we used only its effective temperature $T_{\rm eff\,2}$ as a free parameter
and computed the remaining ones, as
$\log m_2(X)$,
$\log R_2(X)$,
$\log g_2(X)$,
where $X \equiv \log T_{\rm eff\,2}$,
to constrain the secondary as a main-sequence star.
If the secondary becomes too hot in the course of convergence,
it will become visible in the synthetic spectra, of course.

Another problem arises from missing direct estimates of the radii $R_1$ and $R_3$
(there are no eclipses, no spectro-interferometry).
We experienced a tendency to converge towards unrealistic hot subdwarfs
and therefore had to use a mass constraint inferred from Mesastar modelling,
in particular $m_1^{\rm min} = 4.6\,M_\odot$,
which corresponds to a $3\sigma$ lower limit.

 Finally, a technical note: We prefer to inspect first the $\chi_2$
values, which are not reduced, that is, not divided by the respective
number of degrees of freedom $\nu$. This gives a better chance to search
for possible reasons of systematic errors (or model deficiencies).

We also prefer to use the nominal uncertainties
in our modelling (not directly the "realistic" ones) because otherwise
we would not be sensitive to any systematics at all; they would be completely
hidden in large $\sigma$s and we might be falsely satisfied with the model.

\subsection{Nominal model}
\begin{figure*}
\centering
\includegraphics[width=18cm]{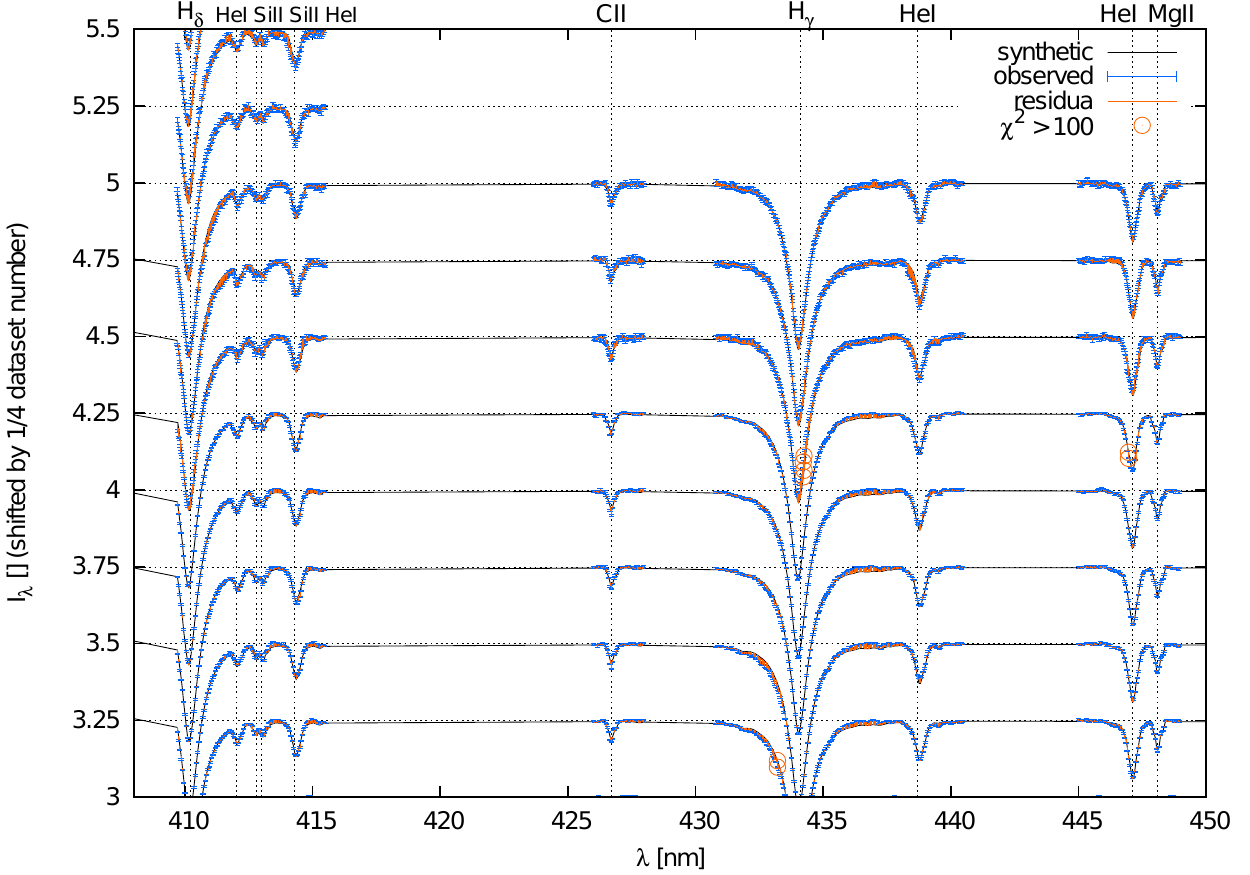}
\caption{Subset of synthetic spectra $I_\lambda'$ of V746~Cas (black line)
for the nominal model with $\chi^2 = 157\,756$
compared to the normalized observed ones (blue error bars).
Only a region between 408 and 450\,nm is shown, with a number
of indicative lines: H$_\delta$, HeI, SiII, SiII, HeI, CII, H$_\gamma$, HeI, HeI, and MgII.
The synthetic spectra were interpolated to the observed wavelengths.
The differences are denoted by red lines, or even red circles
if the respective contribution to $\chi^2_{\rm syn}$ is larger than~$100$.
Note it is difficult to improve the solution because our N-body model
does not fit the spectra one-by-one or individual lines,
but rather fits all the spectra at once. For example, the problematic
synthetic spectrum of H$_\gamma$ in the middle of the figure
cannot be simply shifted to the left. These systematics may
at least partly be caused by the rectification procedure.}
\label{fig:chi2_SYN_ZOOM}
\end{figure*}

The nominal model is presented in Table~\ref{tab:nbody} and
Figure~\ref{fig:chi2_SYN_ZOOM}.
Its total $\chi^2$ value seems too high, $\chi^2 = 157\,756$,
that is, much higher than the corresponding number of degrees of freedom
$\nu \equiv N_{\rm data}-L_{\rm free} = 21\,647$,
the reduced $\chi^2_{\rm R} \equiv \chi^2/\nu = 7.3$\,
and the resulting probability is thus essentially zero.
However, we explain the mismatch as follows.
While the $\chi^2_{\rm sky}$ contribution seems perfectly reasonable
and it is indeed not difficult to reach the value as low as $N_{\rm sky}$,
it may become larger in the course of fitting
because the other data sets are much more numerous
and are likely affected by systematics.

\begin{figure}
\centering
\includegraphics[width=9cm]{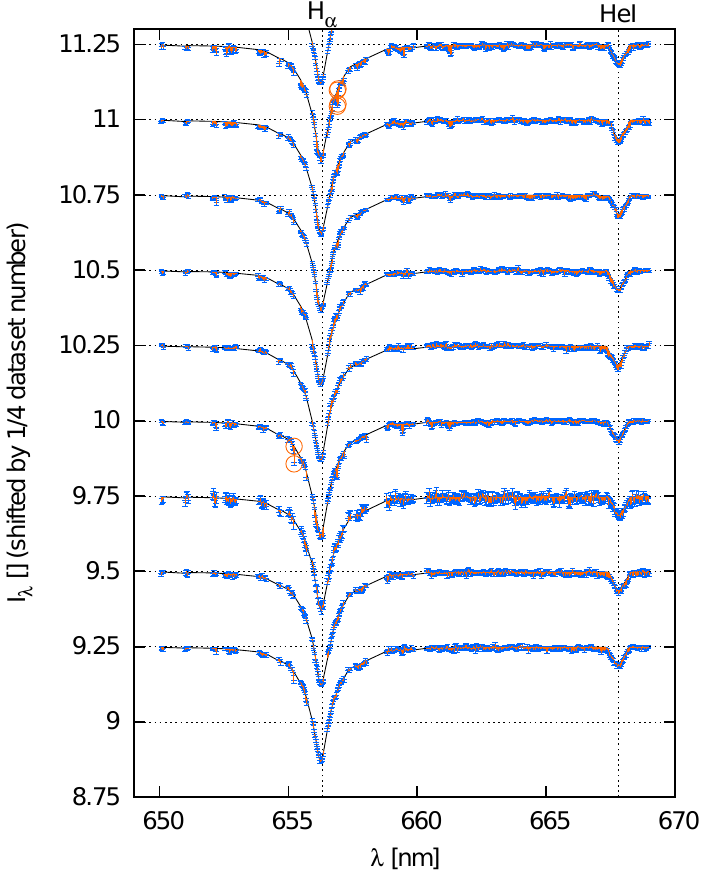}
\caption{Subset of synthetic spectra $I_\lambda'$ of V746~Cas (black line)
for the preferred model with $\chi^2 = 97\,733$ compared to the normalized
observed ones (blue error bars). A~region between 650 and 670\,nm is shown,
with $H_\alpha$ and \ion{He}{i} lines. Some parts of the observed spectra
were discarded due to the presence of many telluric lines.}
\label{fig:chi2_SYN_ZOOM2}
\end{figure}

\begin{figure}
\centering
\includegraphics[width=9cm]{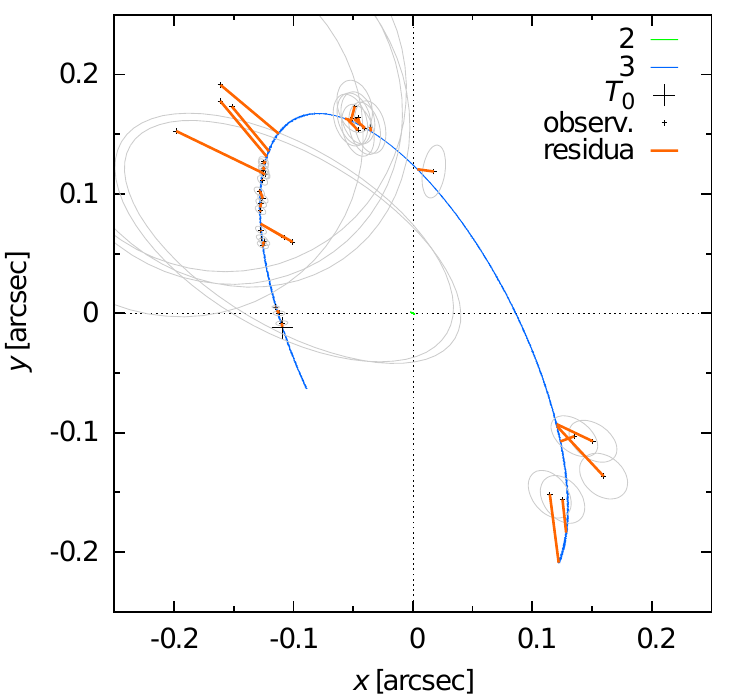}
\caption{One of the allowed solutions for the orbit of V746~Cas tertiary,
shown in photocentric coordinates $x_{\rm p}$, $y_{\rm p}$ (blue curve)
and compared to speckle-interferometry and astrometric measurements
(black crosses with gray uncertainty ellipses). The large cross denotes
the position at the epoch $T_0 = 2454384.65$ of osculation.
There are clearly several uncertain measurements (cf. red residua)
that do not contribute much to~$\chi^2_{\rm sky}$.
Note that normalized spectra (RVs)
and spectral-energy distribution were fitted at the same time
and there is no easy way to improve the solution further
without increasing $\chi^2_{\rm syn}$ or $\chi^2_{\rm sed}$.}
\label{fig:chi2_SKY_ARCSEC}
\end{figure}

\begin{figure}
\centering
\includegraphics[width=9cm]{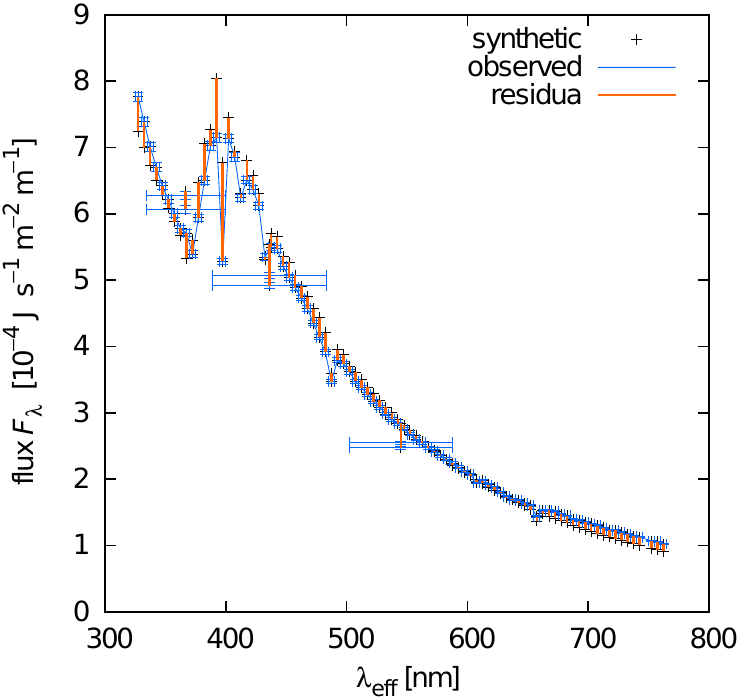}
\caption{Spectral-energy distribution of V746 Cas expressed as the absolute flux~$F_\lambda'$
(in ${\rm J}\,{\rm s}^{-1}\,{\rm m}^{-2}\,{\rm m}^{-1}$ units; black crosses)
and compared with wide-band UBV measurements from Hvar and Geneva,
and narrow-band SED measurements by \cite{glushneva1992}.
All the data sets were dereddened using $E(B-V)=0$\m060
and standard procedures. While the model seems consistent with
$U$ data from Hvar and SED in the region 550 to 650\,nm, there are
still some systematics in $B$, $V,$ and both short- and long-wavelength
parts of the SED. It is not possible to improve $\chi^2_{\rm sed}$,
e.g. by changing temperatures of the components, because it would
inevitably increase~$\chi^2_{\rm syn}$.}
\label{fig:chi2_SED_FLUX}
\end{figure}

The systematics in $\chi^2_{\rm syn}$ are probably caused
by the rectification procedure, even though it was performed
as carefully as possible. For example, we sometimes see differences
for a given spectrum that is surrounded by two or more spectra
that are fitted well enough. The same is true for individual lines in a single spectrum,
for instance, we have a good fit of the \ion{}{ii} and \ion{Mg}{ii} lines,
while H$_\gamma$ (with its extended wings) and \ion{He}{i} are somewhat offset.
In principle, we cannot exclude the presence of rapid line-profile variations,
but a self-consistent model would be needed for them, otherwise such
unconstrained variations could explain any departures from the model
(including any systematics).

The value of $\chi^2_{\rm sed}$ seems also higher than expected.
This is probably caused by systematics in calibrations of the
absolute fluxes (even the Hvar and Geneva photometry from Table C1
differ by more than $3\sigma$). Alternatively, a significant
systematic uncertainty can be hidden in the dereddening procedure (see below).
If we accept the arguments above and the high value of the nominal $\chi^2$,
then $1\sigma$ probability level would correspond to an increase
of up to $158\,659$, and $3\sigma$ to $162\,201$.


Parameter uncertainties can be obtained by a bootstrap or $\chi^2$ mapping.
In this case, we used the latter method, with the notion that
the respective uncertainties correspond to a {\em local\/} minimum only.
Several correlations are still present in the model, however.
In particular, a~lower inclination $i_1$ usually requires
a higher $m_{\rm tot}$. A~similar correlation exists between
the distance~$d$ and $m_{\rm tot}$.
Moreover, there is a non-zero possibility of a long orbit of the tertiary,
with $P_2 \simeq 211\,000\,{\rm d}$, a high eccentricity
$e_2 \simeq 0.65,$ and small~$d$, with only marginally worse
$\chi^2_{\rm sky} = 124$, but the total $\chi^2$ is then relatively large.

\begin{table*}
\caption{Osculating orbital elements at the epoch $T_0 = 2454384.65$
and radiative parameters derived for V746 Cas using the nominal N-body model,
where
$m_j$ denote the component masses,
$P_j$ osculating periods,
$e_j$ eccentricities,
$i_j$ inclinations,
$\Omega_j$ longitudes of ascending nodes,
$\omega_j$ arguments of pericentres, and
$M_j$ mean anomalies.
Note $P_1$ is indeed different from Eq.~(\ref{efe}), as explained in the main text.
The resulting $\chi^2 = 157\,756$,
the total number of measurements $N_{\rm data} = 21\,673$.
The individual contributions are
$\chi^2_{\rm sky} = 75$,
$\chi^2_{\rm syn} = 152\,600$, and
$\chi^2_{\rm sed} = 5\,079$.
Uncertainties were determined by a~$\chi^2$ mapping,
so they correspond to a local minimum.}
\label{tab:nbody}
\centering
\begin{tabular}{lllllll}
\hline
\hline
\vrule width 0pt height 10pt
Parameter & Value & & & & & Unit \\
\hline
\vrule width 0pt height 10pt
$m_1$                   & $4.71_{-0.16}^{+0.17}         $ &
$m_2$                   & $1.31^{\rm h}                 $ &
$m_3$                   & $2.65_{-0.25}^{+0.25}         $ & $M_\odot$   \\[4pt]
$P_1$                   & $25.3068_{-0.0054}^{+0.0051}  $ &
$P_2$                   & $56823_{-9744}^{+11636}       $ & & & day \\[4pt]
$e_1$                   & $0.258_{-0.130}^{+0.120}      $ &
$e_2$                   & $0.030_{-0.030}^{+0.068}      $ & & & \\[4pt]
$i_1$                   & $85_{-19}^{+0}                $ &
$i_2$                   & $57_{-10}^{+9}                $ & & & deg \\[4pt]
$\Omega_1$              & $31^{\rm u}                   $ &
$\Omega_2$              & $33_{-6}^{+7}                 $ & & & deg \\[4pt]
$\omega_1$              & $92_{-6}^{+6}                 $ &
$\omega_2$              & $301_{-6}^{+6}                $ & & & deg \\[4pt]
$M_1$                   & $243_{-5}^{+5}                $ &
$M_2$                   & $131_{-7}^{+7}                $ & & & deg \\[4pt]
$T_{{\rm eff}1}$        & $15534_{-69}^{+69}            $ &
$T_{{\rm eff}2}$        & $6385_{-657}^{+565}           $ &
$T_{{\rm eff}3}$        & $14750_{-66}^{+66}            $ & K \\[4pt]
$\log g_1$              & $3.59_{-0.01}^{+0.01}         $ &
$\log g_2$              & $4.27^{\rm h}                 $ &
$\log g_3$              & $3.80_{-0.03}^{+0.03}         $ & [cgs] \\[4pt]
$v_{{\rm rot}1}$        & $178_{-6}^{+6}                $ &
$v_{{\rm rot}2}$        & $165^{\rm f}                  $ &
$v_{{\rm rot}3}$        & $63_{-8}^{+9}                 $ & ${\rm km}\,{\rm s}^{-1}$ \\[4pt]
$\gamma$                & $-9.7_{-1.8}^{+1.7}          $ & & & & & ${\rm km}\,{\rm s}^{-1}$ \\[4pt]
$d$                     & $330.3_{-4.6}^{+4.6}          $ & & & & & pc \\[4pt]
\hline
\vrule width 0pt height 11pt
$R_1$                   & $5.74_{-0.16}^{+0.18}         $ &
$R_2$                   & $1.40^{\rm h}                 $ &
$R_3$                   & $3.37_{-0.26}^{+0.29}         $ & $R_\odot$ \\[2pt]
\hline
\end{tabular}
\tablefoot{$^{h}$ derived from $T_2$ assuming the \cite{mr88} relations;
$^{u}$ unconstrained parameter;
$^{f}$ fixed parameter.}
\end{table*}

\begin{table*}
\caption{Parameters derived for V746 Cas with the N-body model
but without spectral data in the vicinity of the Balmer lines
H$_\beta$, H$_\gamma$, and H$_\delta$, which are prone to several
instrumental problems and rectification systematics.
On the other hand, H$_\alpha$ from the Ond\v{r}ejov linear spg.~9
was included. The notation is the same as in Table~\ref{tab:nbody}.
The resulting $\chi^2 = 97\,733$,
and the total number of measurements $N_{\rm data} = 30\,614$.
The individual contributions are
$\chi^2_{\rm sky} = 36$,
$\chi^2_{\rm syn} = 93\,146$, and
$\chi^2_{\rm sed} = 4\,551$.
We consider this to be the preferred model.}
\label{tab:nbody2}
\centering
\begin{tabular}{lllllll}
\hline
\hline
\vrule width 0pt height 10pt
Parameter & Value & & & & & Unit \\
\hline
\vrule width 0pt height 10pt
$m_1$                   & $6.45_{-0.17}^{+0.17}         $ &
$m_2$                   & $1.83^{\rm h}                 $ &
$m_3$                   & $6.10_{-0.26}^{+0.27}         $ & $M_\odot$   \\[4pt]
$P_1$                   & $25.3245_{-0.0045}^{+0.0043}  $ &
$P_2$                   & $66206_{-7796}^{+8425}        $ & & & day \\[4pt]
$e_1$                   & $0.171_{-0.102}^{+0.100}      $ &
$e_2$                   & $0.223_{-0.030}^{+0.031}      $ & & & \\[4pt]
$i_1$                   & $85_{-27}^{+0}                $ &
$i_2$                   & $65_{-4}^{+4}                 $ & & & deg \\[4pt]
$\Omega_1$              & $68^{\rm u}                   $ &
$\Omega_2$              & $27_{-4}^{+4}                 $ & & & deg \\[4pt]
$\omega_1$              & $202_{-5}^{+5}                $ &
$\omega_2$              & $310_{-3}^{+3}                $ & & & deg \\[4pt]
$M_1$                   & $146_{-5}^{+6}                $ &
$M_2$                   & $109_{-4}^{+4}                $ & & & deg \\[4pt]
$T_{{\rm eff}1}$        & $16526_{-82}^{+100}           $ &
$T_{{\rm eff}2}$        & $8071_{-352}^{+266}           $ &
$T_{{\rm eff}3}$        & $13620_{-144}^{+154}          $ & K \\[4pt]
$\log g_1$              & $3.81_{-0.01}^{+0.01}         $ &
$\log g_2$              & $4.18^{\rm h}                 $ &
$\log g_3$              & $3.92_{-0.02}^{+0.02}         $ & [cgs] \\[4pt]
$v_{{\rm rot}1}$        & $170_{-4}^{+4}                $ &
$v_{{\rm rot}2}$        & $165^{\rm f}                  $ &
$v_{{\rm rot}3}$        & $73_{-5}^{+6}                 $ & ${\rm km}\,{\rm s}^{-1}$ \\[4pt]
$\gamma$                & $-10.8_{-1.5}^{+1.5}          $ & & & & & ${\rm km}\,{\rm s}^{-1}$ \\[4pt]
$d$                     & $345.6_{-3.1}^{+2.9}          $ & & & & & pc \\[4pt]
\hline
\vrule width 0pt height 11pt
$R_1$                   & $5.24_{-0.10}^{+0.14}         $ &
$R_2$                   & $1.83^{\rm h}                 $ &
$R_3$                   & $4.48_{-0.20}^{+0.20}         $ & $R_\odot$ \\[2pt]
\hline
\end{tabular}
\end{table*}

We emphasize that the results {\em can\/} differ from previous
observation-specific models. In Table~\ref{tab:nbody},
we use osculating elements corresponding to the epoch~$T_0 = 2454384.65,$
which differ from fixed elements. In particular, $P_1$ is different from Eq.~(\ref{efe})
because the very definition of osculation is ``without any perturbation'',
that is to say without the third body; the period perceived by an observer
is close to that in Eq.~(\ref{efe}). We note, however, that for
the period $P_2$ of the long orbit, the difference between the osculating
and the sidereal value is negligible.
Moreover, we realized that at least several Aurelie spectra
(RJD between 51005.6 and 51010.6) were affected by the rectification
systematics (overcorrection), since the spectra do not cover
the blue H$_\delta$ wing completely, and this made the $K_1$ estimates
too large. The true $K_1$ value seems closer to $30\,{\rm km}\,{\rm s}^{-1}$
according to the N-body model. In a similar way, these systematics can
increase (or decrease) the value of~$e_1$.

\paragraph{Mirror solutions.}
Obviously, there are several mirror solutions that cannot be
easily resolved with the current limited data set (no eclipses,
no spectro-interferometry). In particular, we have
$i_1' = 180^\circ-i_1 = 95^\circ$, which gives the same $\chi^2 = 157\,757$.
Unfortunately, with the current data set, $\Omega_1$ is unconstrained,
but we can expect a mirror solution $\Omega_1' = \Omega_1+180^\circ$ anyway.
Finally, the third body can have $i_2' = -i_2 = -57^\circ$,
resulting in $\chi^2 = 157\,815$, which is again statistically the same.

\paragraph{Higher reddening.}
We also tried to assume the reddening $E(B-V) = 0.101$,
that is,~at a~typical $1\sigma$ uncertainty. This would allow for
higher luminosities of the primary and tertiary.
Indeed, the value of $\chi^2_{\rm sed} = 3\,797$ is somewhat lower,
but the total value $\chi^2 = 156\,777$ is neither significantly
higher nor lower. We therefore conclude that increasing the reddening
alone is not a definitive solution. There might be additional
systematics related to the wavelength dependence of extinction~$A_\lambda$,
in other words, a presence of additional interstellar matter
with different~$\kappa_\lambda$ in the direction towards \ve.

\paragraph{The complex interplay among the various types of observational data.}
Finally, we explain that the N-body model is strongly constrained by
the SED data: high masses $m_1$, $m_3$, with $\log g_1$, $\log g_3$ fixed
by the line spectra, lead to large radii $R_1$, $R_3$,
which would result in too bright stars, or a~large distance~$d$.
At the same time, we fit both the speckle-interferometric data
(astrometry of the third body) and the spectra (also known as RVs),
where large~$d$ inevitably requires high $m_{\rm tot}$, and this
would contradict the former set of constraints.
It may seem that making hot stars cooler is an option,
but significantly lower $T_{{\rm eff}\,j}$
are incompatible with the observed (rectified) spectra.
These sometimes unexpected but inevitable relationships are
the main reasons why some of the parameters may be different from
observation-specific models.

On the other hand, it is obvious that the result in terms of surprisingly
low stellar masses is not satisfactory. Upon closer inspection, we concluded
that the problem lies in a strong sensitivity of the result to the exact
values of the surface gravities, which might be affected mainly by
the instrumental problems of the Balmer lines from the echelle spectra.
For this reason, we derived another model that we discuss below.

\subsection{Model with the H$_\alpha$ line, but without other Balmer lines}

Further examinations of the rectification procedure confirmed
that the corrections of a slowly varying atmospheric extinction,
instrument response, and sharp inter-order jumps in the echelle spectra at
the same time are difficult to deal with. Consequently, we decided to also compute a~model without the H$_\beta$, H$_\gamma$, and
H$_\delta$ Balmer lines,
which are especially prone to such effects. On the other hand, using
no hydrogen lines at all would mean to loose the most sensitive indicator
of the gravity acceleration. Therefore, we decided to use H$_\alpha$ from
the Ond\v{r}ejov (spg.~9) linear spectra, for which the rectification is
straightforward. However, it was necessary to remove a~number of regions
affected by the water vapour telluric lines.
Specifically, we included the following spectral regions:
4125--4155,
4260--4280,
4380--4405,
4450--4490,
4700--4725,
4912--4935,
5008--5025,
6500.6--6501.6,
6509.9--6511.4,
6520.5--6522.3,
6525.3--6530.0,
6538.2--6541.7,
6546.2--6547.2,
6549.7--6552.3,
6554.7--6557.1,
6559.2--6563.8,
6565.1--6571.7,
6573.2--6574.3,
6576.0--6580.5,
6588.0--6598.7, and
6603.5--6690~\AA.

The resulting model is presented in Table~\ref{tab:nbody2}
and Figures~\ref{fig:chi2_SYN_ZOOM2} to~\ref{fig:chi2_SED_FLUX}.
After several simplex runs we reached $\chi^2$ as low as $97\,733$,
which should be compared with the larger number of degrees of freedom,
$\nu = 30\,614$.  The differences with respect to the previous Table~\ref{tab:nbody}
seem to be acceptable and within uncertainties, except for
$m_1$, $m_3$, $e_2$, $\log g_1$, $\log g_3$, $R_1$ , and $R_3$.
The substantially higher and more realistic masses are allowed
for by higher $\log g$ values, enforced by the H$_\alpha$ line.
The differences in radii simply correspond to relatively higher $T_1$ and
lower $T_3$, in order to fit the same SED. The mass of the tertiary
$m_3$ is now in better agreement with the high $T_{{\rm eff}\,3}$,
and $R_3$ fulfils the condition from Sect.~\ref{sec:tetriary},
which makes our model more self-consistent.
The $e_2$ value reflects more freedom given
to the outer orbit due to higher $m_{\rm tot}$. Last but not least,
the reduced $\chi^2_{\rm R}$ decreased from 7.3 down to 3.2,
which indicates that we succeeded to remove a substantial part
of the systematic uncertainties. We thus consider this model
to be the preferred one.


\section{Conclusions}
Using several approaches to the analysis of the rich set of
spectral, photometric, spectro-photometric, and astrometric observations
available to us, we attempted to provide a new interpretation of the
interesting massive triple system V746~Cas with a tertiary, which possesses
a measurable magnetic field. The existing principal geometric limitations,
such as that there are no eclipses and the flux of the secondary component
was not detected, prevented us from deriving unique physical properties
for all the components of the system. However, by combining all types of
observations and bounding them mutually with the help of the N-body model,
we were able to present a reasonably self-consistent picture of the system.

Our main findings are summarised below.
\begin{enumerate}
\item The rapidly rotating B4-B5 primary moves in the orbit with an
unseen secondary. The fact that the secondary could not be detected in
the optical spectra, even when using the spectra disentangling and
the observed mass function imply that it probably is an A or F star.
A~direct detection of its spectrum by standard observing techniques is
probably impossible.

\item The bipolar magnetic field, discovered by \citet{neiner2014}, varies
with the photometric period of 2\fd50387. It is associated with the
tertiary, which is in a wide orbit with the 25\fd4 binary. The tertiary is
a $\sim$B5-6IV star, which contributes 30\% of the light in
the optical region.

\item The photometric period of 2\fd504 can almost certainly be
identified with the rotational period of the tertiary. We note that for
the radius and the projected rotational velocity of the tertiary
from Table~\ref{tab:nbody2}, this assumption leads to an inclination of
the rotational axis identical (within the error limits) with the orbital
inclination of the wide orbit.

\item The photometric period 1\fd065 is tentatively identified
with the rotational period of the primary, but this identification
is much less certain and needs to be proved or disproved
by future high S/N whole-night series of spectra. (At the moment,
the radius and the projected rotational velocity of the primary
from Table~\ref{tab:nbody2} would require an inclination of
the rotational axis as low as $\sim44^\circ$.) The ultimate proof
of line-profile variations with the 1\fd065 period and a reliable and
more accurate determination of $\log~g$ of the primary with the help
of several Balmer lines are both needed.

\item Very accurate systematic photometric observations relative to
some truly constant comparison star and new spectral observations consisting
of whole-night series, which would constrain putative rapid line-profile
variations, are also needed to progress in the understanding of this system.

\item The classification of \vc as an SPB variable should be critically
re-examined.
\end{enumerate}


\begin{acknowledgements}
We profited from the use of 13 echelle spectra from the Bernard Lyot
telescope, made publicly available via the Polar Base web service and
Geneva~7-C photometry made available via HELAS service, to which we were
kindly directed by C.~Aerts. P.~Mathias kindly provided us with old
archival DAT tapes with the original Aurelie spectra of ten B stars and some
advice about their content and structure. The spectra were reconstructed
from the tapes with the help of R.~Vesel\'y. H.~Bo\v{z}i\'c,
K.~Ho\v{n}kov\'a, and D.~Vr\v{s}njak kindly obtained some of
the calibrated \ubv\ observations of \vc and its comparison stars for us.
We acknowledge the use of the public versions of programs \fotel and \korele,
written by P.~Hadrava and \phoebe~1.0 written by A. Pr\v{s}a.
The research of PH, MB, PM, and JN was supported by the grant P209/10/0715
of the Czech Science Foundation. JN and PH were also supported by
the grants GA15-02112S of the Czech Science Foundation and No.~250015 of
the Grant Agency of the Charles University in Prague. Research of DK was
supported by a~grant GA17-00871S of the Czech Science Foundation.
Our thanks are due to M.~Wolf, who obtained one Ond\v{r}ejov spectrum
used here and who provided a few useful comments to this paper.
A~persistent but constructive criticism by an anonymous referee
helped us to re-think the whole study, present our arguments and analyses
more clearly and convincingly, and to improve the layout of the text
and figures as well.
The use of the NASA/ADS bibliographical service and SIMBAD electronic database
are gratefully acknowledged.
\end{acknowledgements}


\bibliographystyle{aa}
\bibliography{citace}



\begin{appendix}
\section{Details of the spectral data reduction and measurements}\label{apa}

All RVs collected from the astronomical literature are provided in
Table~\ref{rvlit}. Whenever necessary, we derived heliocentric Julian dates
(HJDs) for them. The RVs derived by us are provided in Tables~\ref{indrvs}
and \ref{rvaure}.

\begin{table*}
\centering
\caption[]{Individual RVs of \vc from the astronomical literature.
Observing instruments in column Spg. are identified by the same numbers
as in Table~\ref{jourv}. All RVs are in \ks.}
\label{rvlit}
\begin{tabular}{crccrcrrrr}
\hline\hline\noalign{\smallskip}
RJD&RV&Spg.&RJD&RV&Spg.\\
\noalign{\smallskip}\hline\noalign{\smallskip}
19026.8885&$    5.0$& 1  &43461.679 &$ -37.8 $& 4 \\
19027.6501&$  -35.0$& 1  &43462.790 &$ -21.1 $& 4 \\
19058.7623&$    0.4$& 1  &43508.748 &$ -32.8 $& 4 \\
19382.7240&$    1.9$& 1  &43726.972 &$   6.7 $& 4 \\
19409.8989&$   -1.8$& 1  &43808.852 &$  -5.5 $& 4 \\
22915.9555&$   -3.7$& 1  &43809.896 &$ -23.4 $& 4 \\
23004.7039&$   -8.0$& 1  &43810.821 &$ -14.4 $& 4 \\
24010.4054&$   -5.3$& 2  &43830.758 &$   3.1 $& 4 \\
24362.4915&$    5.7$& 2  &43831.648 &$  -2.0 $& 4 \\
24439.1971&$  -18.0$& 2  &43832.647 &$  -4.2 $& 4 \\
24767.3810&$  -23.9$& 2  &44068.973 &$ -45.6 $& 4 \\
24777.4394&$  -51.5$& 2  &44116.922 &$ -27.8 $& 4 \\
24891.1328&$  -16.7$& 2  &44187.697 &$   0.4 $& 4 \\
35795.6197&$    7.0$& 3  &44188.666 &$   1.9 $& 4 \\
35796.6027&$   11.0$& 3  &44189.825 &$  -2.0 $& 4 \\
35797.6796&$   12.0$& 3  &44229.655 &$ -16.0 $& 4 \\
35799.6366&$    1.0$& 3  &44231.618 &$  -0.8 $& 4 \\
35803.6214&$  -24.0$& 3  &53657.787 &$ -14.2 $& 5 \\
35804.6644&$  -24.0$& 3  &53658.833 &$  -4.8 $& 5 \\
35819.5897&$   -9.0$& 3  &53658.835 &$  -4.3 $& 5 \\
35820.5586&$    4.0$& 3  &53659.720 &$  -0.3 $& 5 \\
35821.6155&$    5.0$& 3  &53659.721 &$  -0.1 $& 5 \\
35822.5385&$    4.0$& 3  &53660.754 &$   3.0 $& 5 \\
35823.5854&$    4.0$& 3  &53663.735 &$  13.7 $& 5 \\
35824.5884&$    3.0$& 3  &53685.735 &$  -9.4 $& 5 \\
35830.5580&$  -12.0$& 3  &53693.635 &$  13.8 $& 5 \\
35831.5949&$  -16.0$& 3  &53693.790 &$   7.3 $& 5 \\
35832.5589&$  -34.0$& 3  &53694.701 &$   6.9 $& 5 \\
35833.5528&$  -40.0$& 3  &53695.674 &$  -1.4 $& 5 \\
43411.860 &$ -36.7 $& 4  &53695.769 &$  -0.8 $& 5 \\
43459.731 &$ -42.0 $& 4  &53696.565 &$  -4.9 $& 5 \\
43460.730 &$ -44.3 $& 4  &53696.789 &$  -2.0 $& 5 \\
\noalign{\smallskip}\hline\noalign{\smallskip}
\end{tabular}
\end{table*}

\begin{table*}
\centering
\caption[]{Individual \spefo RVs of \vc for \ha and \he RVs
and Gaussian RVs for the \he line.
RVs of the line cores and line outer wings
are tabulated separately. All RVs are in \ks.
The uncertainties reported in this table correspond to the \spefo measurement
procedure alone. There are additional sources related to the
photon noise,
systematics from the rectification, and
systematics from line blending.
Given the differences between RVs inferred for the two lines,
the total uncertainty may reach up to $10\,{\rm km}\,{\rm s}^{-1}$
because it is not possible to separate contributions from the
primary and tertiary components.}
\label{indrvs}
\begin{tabular}{crrrrrrrrrr}
\hline\hline\noalign{\smallskip}
RJD &
\multicolumn{2}{c}{\ha 6563~\AA\ \spefoe} &
\multicolumn{2}{c}{\he \spefoe} &
\multicolumn{2}{c}{\he Gauss}&Spg.\\
&
\multicolumn{1}{c}{core} &
\multicolumn{1}{c}{wings} &
\multicolumn{1}{c}{core} &
\multicolumn{1}{c}{wings} &
\multicolumn{1}{c}{core} &
\multicolumn{1}{c}{wings} &
No.\\
\noalign{\smallskip}\hline\noalign{\smallskip}
52538.5637&$-13.14\pm2.52$&$-19.65\pm0.54$&$-15.04\pm0.34$&$-17.02\pm3.07$&$ -8.0 \pm 4.0$&$-19.5 \pm 1.5 $& 7\\
53025.2936&$-10.82\pm0.46$&$  2.96\pm0.77$&$ -8.05\pm0.45$&$ -7.01\pm2.17$&$ -0.5 \pm 8.5$&$ -3.5 \pm 9.5 $& 7\\
53026.2813&$ -9.05\pm0.26$&$  7.35\pm1.17$&$ -6.99\pm0.90$&$  9.70\pm2.50$&$  3.0 \pm12.0$&$  0.0 \pm 9.0 $& 7\\
56175.5586&$-13.57\pm0.58$&$ -6.36\pm0.84$&$ -1.28\pm0.09$&$ -8.41\pm1.19$&$  0.67\pm0.67$&$ -3.17\pm 2.82$& 8\\
56176.5910&$-14.05\pm0.11$&$ -4.63\pm0.28$&$ -2.64\pm0.29$&$  8.76\pm2.46$&$ -5.24\pm0.75$&$  3.50\pm 3.50$& 8\\
56177.5783&$-11.79\pm0.27$&$  1.81\pm0.39$&$ -5.04\pm0.87$&$ 11.77\pm1.14$&$  2.48\pm4.52$&$  5.73\pm 8.73$& 8\\
56178.5868&$ -8.67\pm0.73$&$  8.80\pm0.43$&$  3.22\pm0.33$&$ 24.27\pm0.18$&$ -5.51\pm1.49$&$ 23.29\pm 1.29$& 8\\
56179.5853&$-11.32\pm0.06$&$  9.86\pm0.93$&$  3.87\pm0.32$&$ 28.02\pm0.57$&$ -0.48\pm0.48$&$ 24.01\pm 3.01$& 8\\
56182.6157&$-10.95\pm0.42$&$ 14.17\pm1.35$&$ -3.48\pm0.21$&$ 33.54\pm0.24$&$ -9.96\pm0.96$&$ 35.48\pm 2.48$& 8\\
56188.5120&$-12.62\pm0.40$&$-28.60\pm1.05$&$ -5.19\pm0.13$&$-25.18\pm1.09$&$  0.50\pm0.50$&$-23.15\pm 2.85$& 8\\
56190.5364&$-18.99\pm0.15$&$-32.92\pm2.36$&$ -6.94\pm0.07$&$-46.13\pm1.60$&$  0.33\pm0.33$&$-40.87\pm 3.13$& 8\\
56202.5101&$-12.70\pm0.52$&$ -1.29\pm0.79$&$  2.72\pm0.48$&$  0.10\pm0.62$&$ -0.75\pm0.24$&$  6.69\pm 2.69$& 8\\
56203.5069&$ -7.94\pm0.10$&$  3.02\pm0.46$&$  4.54\pm1.13$&$  7.90\pm2.05$&$  6.13\pm0.13$&$ 12.43\pm 3.43$& 8\\
56204.5150&$-10.05\pm0.23$&$  3.10\pm0.52$&$  5.51\pm0.51$&$ 12.84\pm0.82$&$  3.37\pm0.37$&$ 12.47\pm 3.47$& 8\\
56213.5446&$-13.00\pm0.18$&$ -8.46\pm0.68$&$ -4.81\pm0.14$&$  0.97\pm0.39$&$ -9.04\pm0.04$&$ -6.69\pm 4.31$& 8\\
56214.4187&$-16.90\pm0.24$&$-16.45\pm1.62$&$ -7.63\pm0.37$&$-18.19\pm0.70$&$ -5.99\pm1.00$&$-21.84\pm 3.15$& 8\\
56746.6501&$ -7.21\pm0.64$&$ -4.74\pm0.91$&$ -2.85\pm0.46$&$ -6.82\pm0.79$&$  7.00\pm1.68$&$ -3.25\pm 2.17$& 9\\
56764.5933&$ -4.98\pm0.63$&$  7.51\pm0.61$&$  6.80\pm0.35$&$ 20.45\pm1.87$&$  1.00\pm0.91$&$ 11.25\pm 1.25$& 9\\
56765.5438&$  7.03\pm0.18$&$  9.85\pm3.35$&$ 15.41\pm0.75$&$ 16.45\pm1.64$&$ 20.50\pm1.65$&$ 14.75\pm 1.60$& 9\\
56772.5422&$-14.23\pm0.35$&$ -7.02\pm0.30$&$ -4.75\pm0.62$&$ -8.20\pm0.79$&$ -2.50\pm1.93$&$-10.00\pm 2.34$& 9\\
56782.5181&$-10.29\pm0.77$&$-22.95\pm2.54$&$ -2.33\pm1.83$&$-40.34\pm0.79$&$ -1.50\pm4.73$&$-25.25\pm 3.79$& 9\\
56810.5577&$ -8.21\pm1.68$&$-16.29\pm0.35$&$  1.35\pm1.04$&$-14.02\pm0.17$&$  1.75\pm3.35$&$-14.75\pm 4.19$& 9\\
56816.5317&$ -1.52\pm1.93$&$  6.57\pm4.13$&$ 15.81\pm0.52$&$ 14.13\pm0.71$&$ 13.00\pm1.95$&$ 18.25\pm 1.54$& 9\\
56817.4931&$  3.45\pm0.47$&$  9.78\pm3.35$&$ -5.16\pm0.30$&$ 35.79\pm1.30$&$-10.00\pm4.06$&$ 23.50\pm 3.17$& 9\\
56819.4233&$ -0.80\pm2.20$&$  3.59\pm6.15$&$ 16.24\pm4.58$&$ 28.33\pm1.20$&$ -0.50\pm4.50$&$ -3.00\pm 5.30$& 9\\
56822.4782&$-14.71\pm0.30$&$ -7.68\pm0.63$&$ -4.34\pm0.35$&$ -3.65\pm2.58$&$ -0.50\pm2.95$&$-30.50\pm12.22$& 9\\
56826.3772&$-30.03\pm0.47$&$-37.24\pm1.61$&$ -6.01\pm0.17$&$-49.38\pm1.67$&$  0.75\pm3.81$&$-57.00\pm 4.43$& 9\\
56827.4581&$-21.99\pm0.53$&$-40.63\pm4.52$&$-11.58\pm1.04$&$-73.09\pm0.96$&$  0.50\pm3.01$&$-56.00\pm 2.04$& 9\\
56851.3742&$-23.96\pm1.22$&$-39.08\pm4.04$&$-10.51\pm0.52$&$-45.58\pm0.62$&$ -3.75\pm4.76$&$-54.25\pm 4.55$& 9\\
56852.4071&$-28.71\pm0.18$&$-34.69\pm1.33$&$-21.05\pm0.52$&$-55.09\pm0.17$&$-20.00\pm3.18$&$-51.25\pm 3.01$& 9\\
56852.4657&$-22.29\pm3.96$&$-33.37\pm1.10$&$-31.67\pm0.30$&$-58.98\pm1.48$&$  3.75\pm9.05$&$-43.75\pm 3.37$& 9\\
56856.5629&$-25.75\pm0.47$&$-30.15\pm0.81$&$ -3.30\pm0.46$&$-52.72\pm0.79$&$  7.25\pm2.98$&$-20.00\pm 8.21$& 9\\
56861.4063&$-20.57\pm0.47$&$ -9.85\pm0.35$&$  1.69\pm0.35$&$ -8.86\pm0.46$&$ -8.25\pm2.68$&$  8.75\pm 6.06$& 9\\
56866.4436&$-20.86\pm0.30$&$  8.51\pm2.37$&$-13.05\pm0.35$&$ 15.47\pm1.21$&$  2.25\pm3.75$&$  6.75\pm 2.01$& 9\\
56889.3848&$-17.08\pm0.18$&$ -1.60\pm1.40$&$  4.38\pm0.37$&$  8.01\pm0.30$&$ 12.75\pm3.03$&$ 14.75\pm 4.87$& 9\\
56910.3387&$-15.09\pm3.96$&$-19.84\pm2.79$&$ -9.13\pm0.91$&$-26.24\pm1.21$&$-10.00\pm0.91$&$-16.25\pm 1.93$& 9\\
56920.3238&$ 16.22\pm1.84$&$ 18.33\pm1.23$&     --        &     --        &  --           &     --         & 9\\
56928.3090&$-17.42\pm0.13$&$-35.75\pm0.83$&$-13.61\pm0.33$&$-54.56\pm0.96$&$ -4.50\pm2.32$&$-47.75\pm 1.88$& 9\\
56949.2711&$-10.92\pm0.30$&$  2.62\pm2.31$&$ -2.68\pm0.35$&$ 16.15\pm1.48$&$-10.50\pm2.02$&$ 12.00\pm 3.82$& 9\\
56950.2946&$-15.20\pm0.18$&$-12.03\pm0.70$&$-13.94\pm0.69$&$-13.77\pm3.89$&$-14.00\pm1.35$&$-12.75\pm 2.25$& 9\\
56978.2935&$-14.62\pm0.35$&$-33.26\pm0.53$&$-24.65\pm0.79$&$-34.15\pm1.13$&$-19.00\pm0.40$&$-29.50\pm 2.90$& 9\\
57073.2605&$ -5.56\pm0.47$&$  5.16\pm2.13$&$ -6.07\pm0.17$&$ 21.75\pm1.58$&$ -4.75\pm2.25$&$ 19.75\pm 2.68$& 9\\
57073.3662&$ -2.40\pm0.52$&$  6.57\pm1.33$&$ 10.02\pm1.05$&$ 20.56\pm0.60$&$  5.75\pm1.79$&$ 22.00\pm 3.36$& 9\\
57074.2706&$ -4.87\pm0.53$&$  6.38\pm0.35$&$  5.43\pm0.60$&$ 20.80\pm0.35$&$  2.00\pm0.81$&$ 16.50\pm 2.59$& 9\\
57110.2598&$ -9.36\pm2.54$&$-35.20\pm0.98$&$-11.44\pm5.81$&$-47.73\pm1.65$&$ -2.75\pm2.56$&$-54.25\pm 2.98$& 9\\
57123.6370&$ -4.86\pm0.30$&$ 11.32\pm0.77$&$  5.53\pm0.75$&$ 10.71\pm1.73$&$  8.75\pm2.13$&$ 13.75\pm 1.31$& 9\\
57137.6098&$ -4.08\pm1.23$&$-25.18\pm0.88$&$-14.59\pm0.62$&$-23.40\pm1.25$&$ -8.75\pm2.95$&$-33.25\pm 3.54$& 9\\
57154.5792&$-12.83\pm0.47$&$-18.28\pm0.18$&$ -4.78\pm0.86$&$-24.82\pm0.79$&$  3.50\pm2.02$&$-26.75\pm 3.98$& 9\\
57228.4145&$ 10.53\pm0.63$&$-13.12\pm3.26$&$  3.76\pm2.17$&      --       &  --           &    --          & 9\\
57256.4904&$-19.43\pm0.18$&$-26.81\pm2.16$&$ -5.85\pm0.75$&$-30.56\pm0.46$&$  5.50\pm3.22$&$-17.25\pm 5.17$& 9\\
57260.5354&$-22.76\pm1.06$&$-39.29\pm0.88$&$-20.31\pm1.99$&$-51.07\pm1.87$&$ -4.50\pm5.36$&$-52.50\pm 5.13$& 9\\
57277.3647&$ -3.11\pm2.65$&$ 16.58\pm0.77$&$ 20.21\pm0.60$&$ 26.43\pm1.08$&$ 10.50\pm1.84$&$ 26.50\pm 2.21$& 9\\
57295.4332&$ -9.01\pm0.98$&$  7.60\pm1.24$&$ -7.86\pm0.87$&$ 21.81\pm1.34$&$-14.25\pm2.62$&$ 14.75\pm 2.92$& 9\\
57300.3763&$ -2.03\pm0.47$&$  9.22\pm1.85$&$  3.64\pm0.35$&$ 14.87\pm1.99$&$ -4.50\pm4.97$&$ 16.75\pm 2.59$& 9\\
57308.3352&$-13.20\pm1.76$&$-19.88\pm2.30$&$ -9.18\pm1.13$&$-23.01\pm4.22$&$-10.00\pm2.79$&$-25.75\pm 1.65$& 9\\
57328.2087&$ -8.13\pm3.26$&$  6.11\pm0.47$&$ -5.12\pm0.75$&$ 13.71\pm0.46$&$ -9.00\pm0.40$&$  5.50\pm 3.06$& 9\\
57328.6913&$ -4.37\pm0.18$&$  8.11\pm1.85$&$  2.34\pm1.21$&$ 29.29\pm1.21$&$ -2.75\pm1.84$&$ 22.50\pm 1.50$& 9\\
\noalign{\smallskip}\hline\noalign{\smallskip}
\end{tabular}
\end{table*}

\begin{table}
\centering
\caption[]{RVs of the broad wings of the \ion{He}{i}~4144~\AA\ line and
H$\delta$ of \vc measured in all Aurelie spectra with \spefo.
For the comment on uncertainties, see Table~\ref{indrvs}.
}\label{rvaure}
\begin{tabular}{crrrrrr}
\hline\hline\noalign{\smallskip}
RJD & RV$_{\spefoe}^{\ion{He}{i}\,4144}$ & RV$_{\spefoe}^{\ion{H}{$\delta$}}$\\
\noalign{\smallskip}\hline\noalign{\smallskip}
50967.5897&$ 14.96\pm2.82$&$  -5.86\pm6.39$ \\
51005.5778&$-39.13\pm0.69$&$ -19.57\pm2.05$ \\
51006.5970&$-49.73\pm1.20$&$ -33.64\pm0.19$ \\
51007.5940&$-48.03\pm0.57$&$ -43.62\pm1.82$ \\
51009.5946&$-52.66\pm2.46$&$ -45.15\pm2.56$ \\
51010.5765&$-60.59\pm0.57$&$ -42.57\pm2.72$ \\
51011.5748&$-52.10\pm1.75$&$ -31.28\pm1.02$ \\
51012.5773&$-42.08\pm2.36$&$ -29.20\pm1.76$ \\
51036.6214&$-46.21\pm2.26$&$ -38.85\pm4.98$ \\
51038.4965&$-35.73\pm1.60$&$ -26.14\pm2.35$ \\
51040.5680&$-24.47\pm4.67$&$ -17.69\pm3.08$ \\
51042.6335&$-13.14\pm0.55$&$ -11.42\pm4.79$ \\
51094.4222&$ -9.42\pm3.98$&$  -4.39\pm4.00$ \\
51096.5979&$ 15.09\pm1.94$&$   0.80\pm3.00$ \\
51097.5923&$ 28.24\pm1.28$&$   0.42\pm3.80$ \\
51098.4748&$ 25.06\pm2.97$&$   7.44\pm4.45$ \\
51099.4536&$ 26.19\pm1.81$&$  10.63\pm2.28$ \\
\noalign{\smallskip}\hline\noalign{\smallskip}
\end{tabular}
\end{table}

\section{Fitting the spectra with interpolated synthetic
spectra}\label{apb}

\begin{figure}
\includegraphics[width=7.8cm]{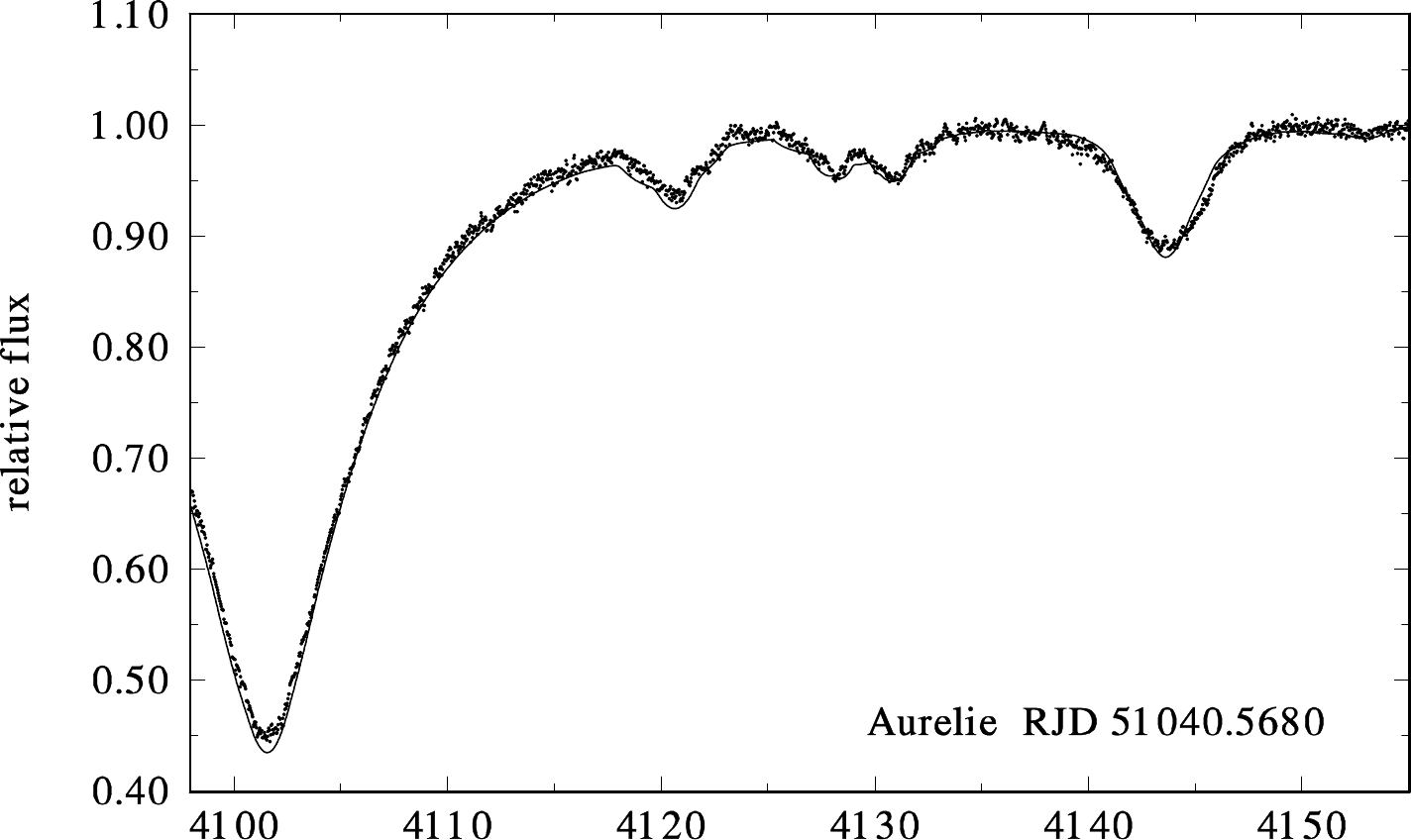}
\includegraphics[width=7.8cm]{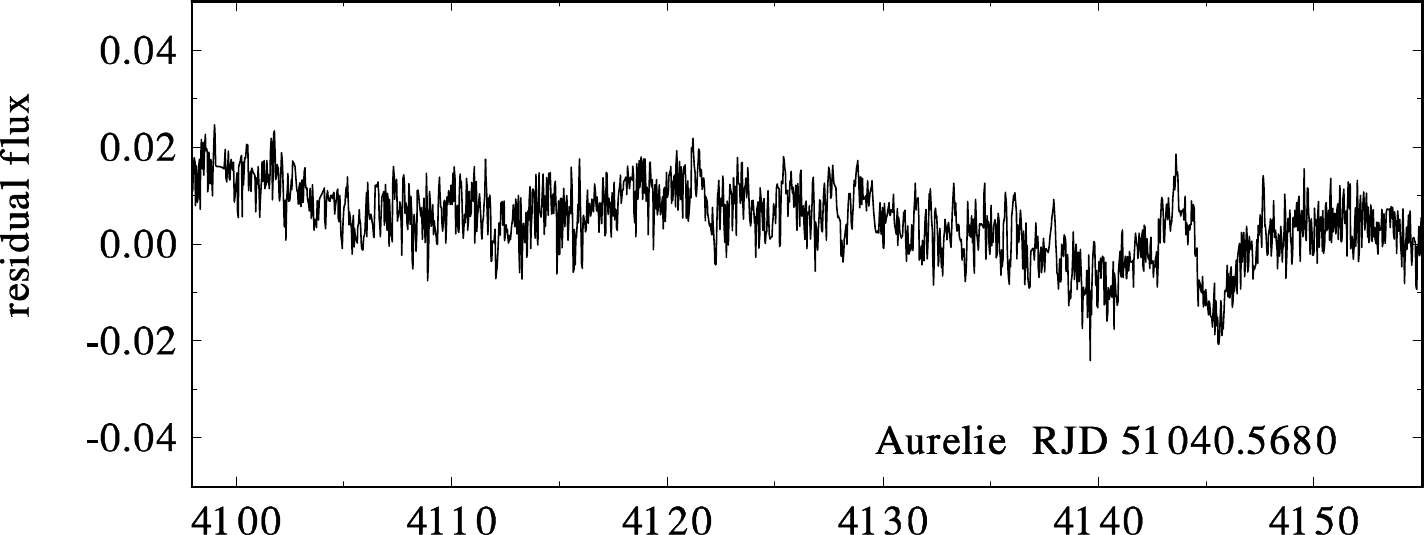}
\includegraphics[width=7.8cm]{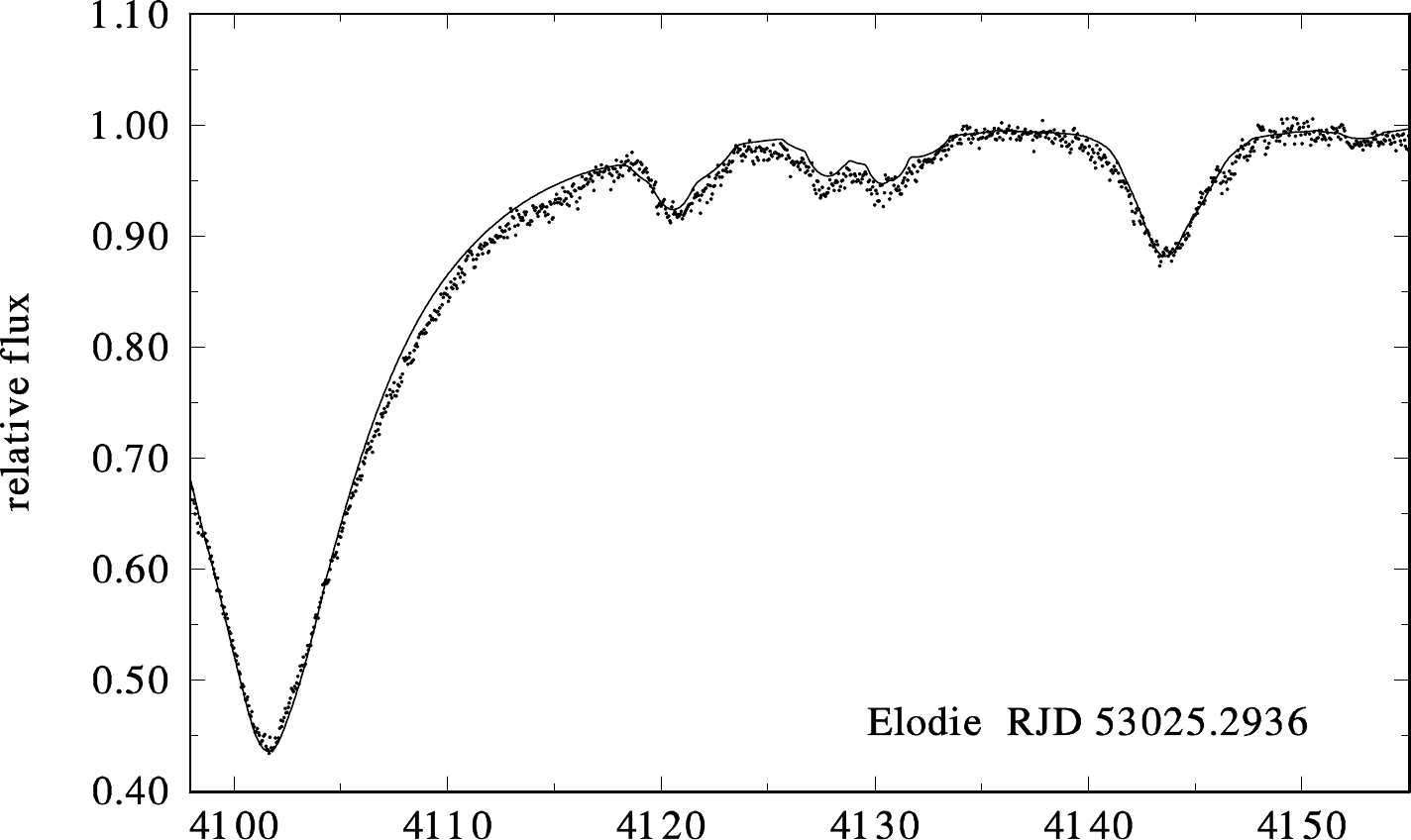}
\includegraphics[width=7.8cm]{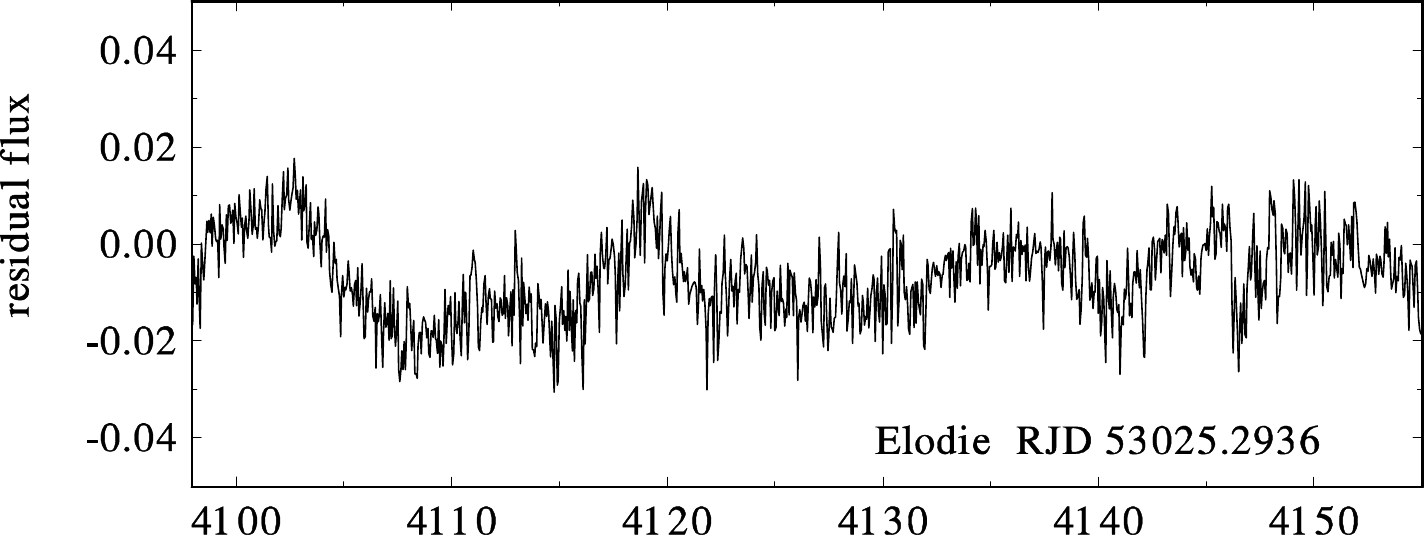}
\includegraphics[width=7.8cm]{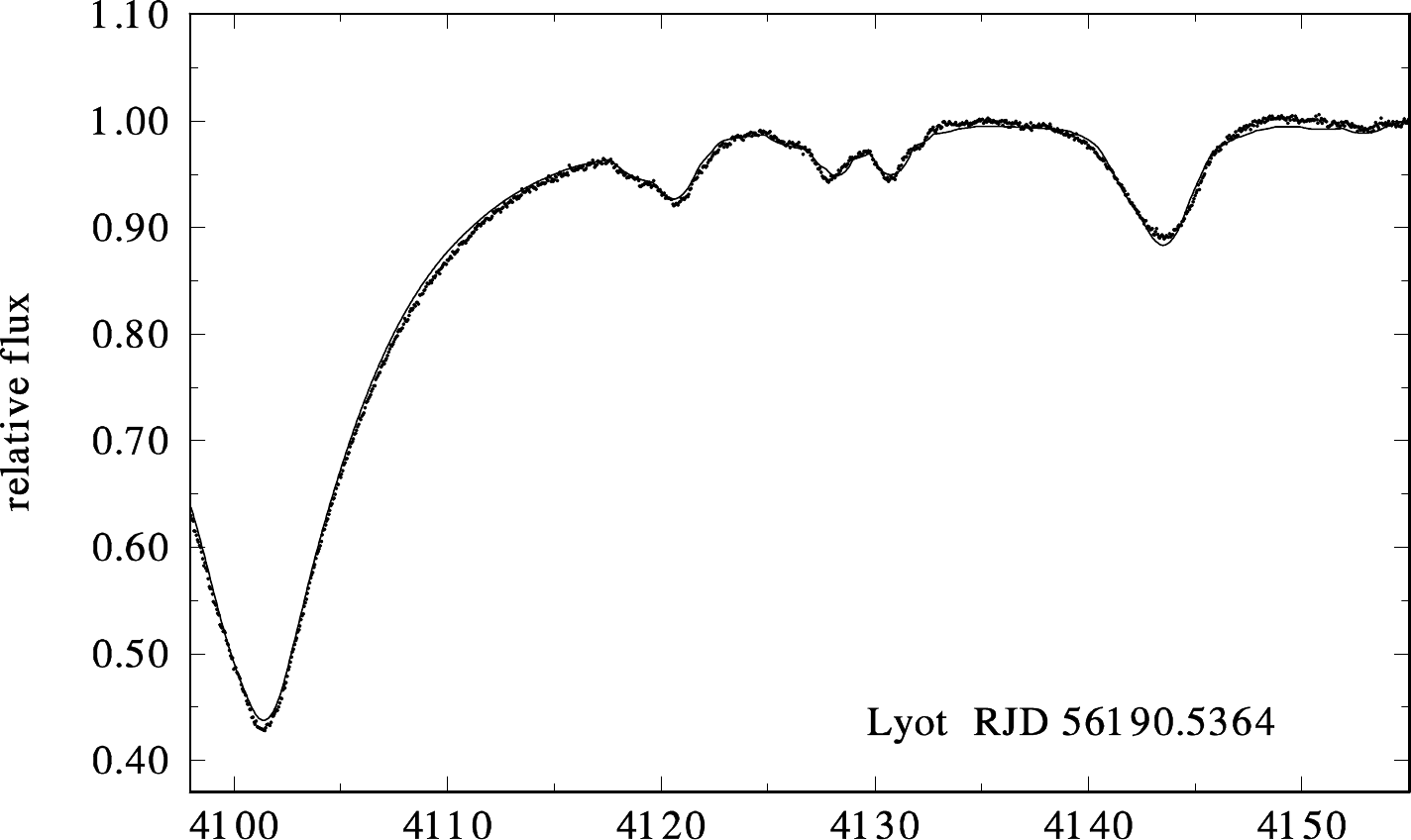}
\includegraphics[width=7.8cm]{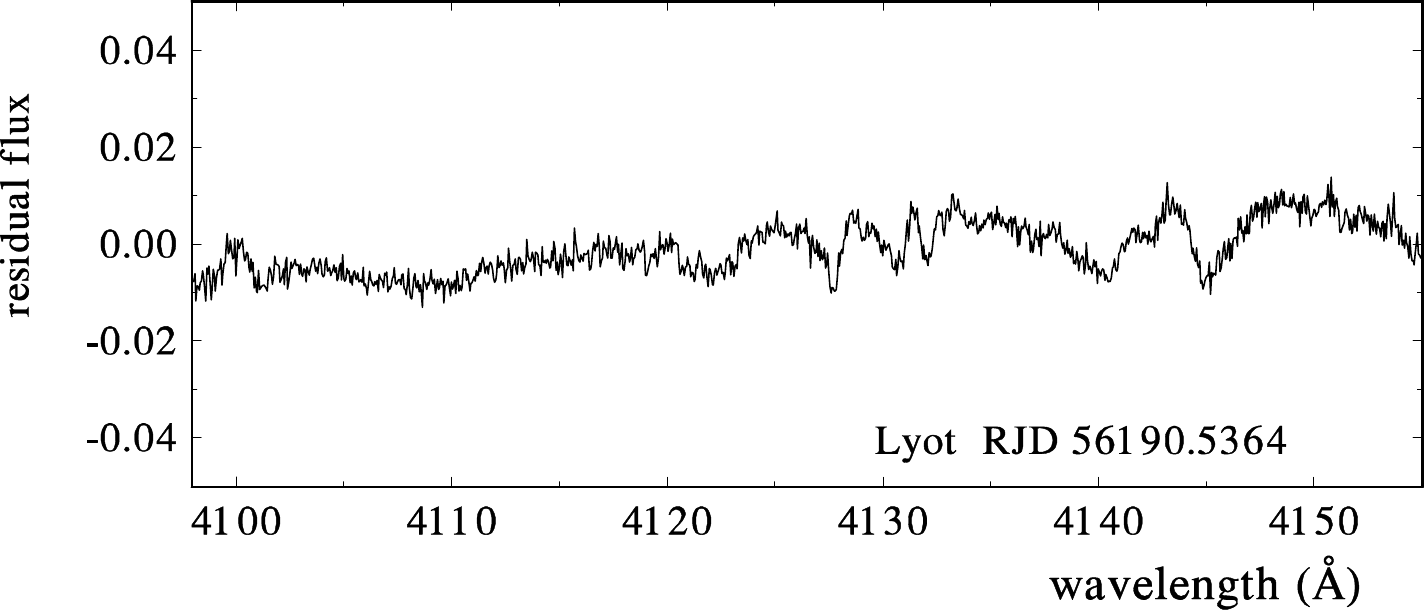}
\caption{Fit of several observed spectra (dots) by interpolated synthetic spectra (lines)
for the region that is also covered by the Aurelie spectra is shown. The residuals
from the fits are also shown below each spectrum (note the enlarged scale
used for the flux units).}\label{synta}
\end{figure}

\begin{figure}
\includegraphics[width=7.8cm]{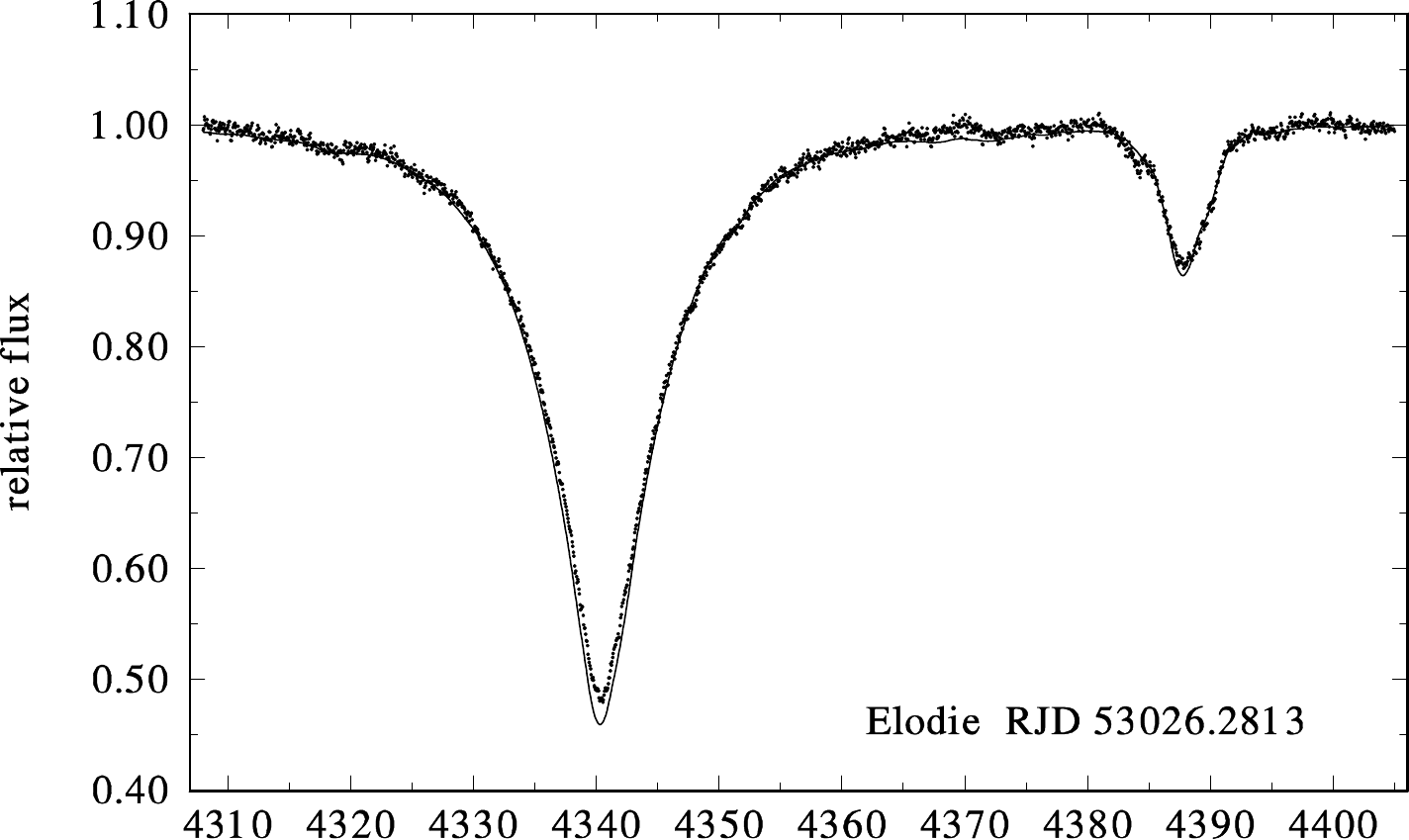}
\includegraphics[width=7.8cm]{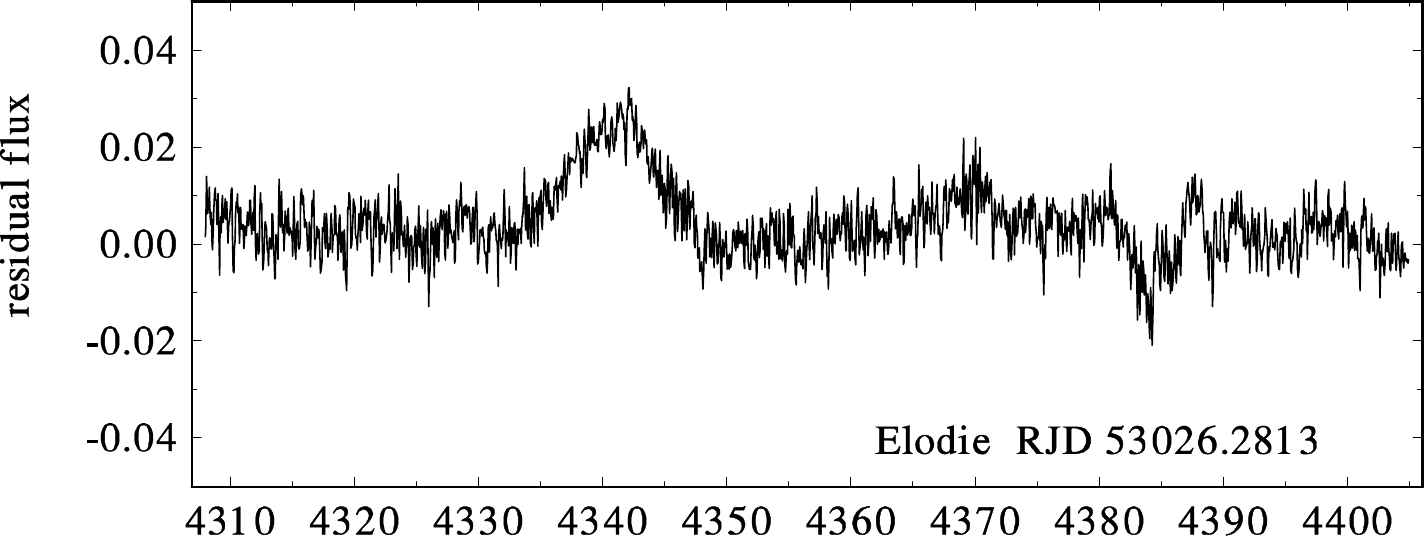}
\includegraphics[width=7.8cm]{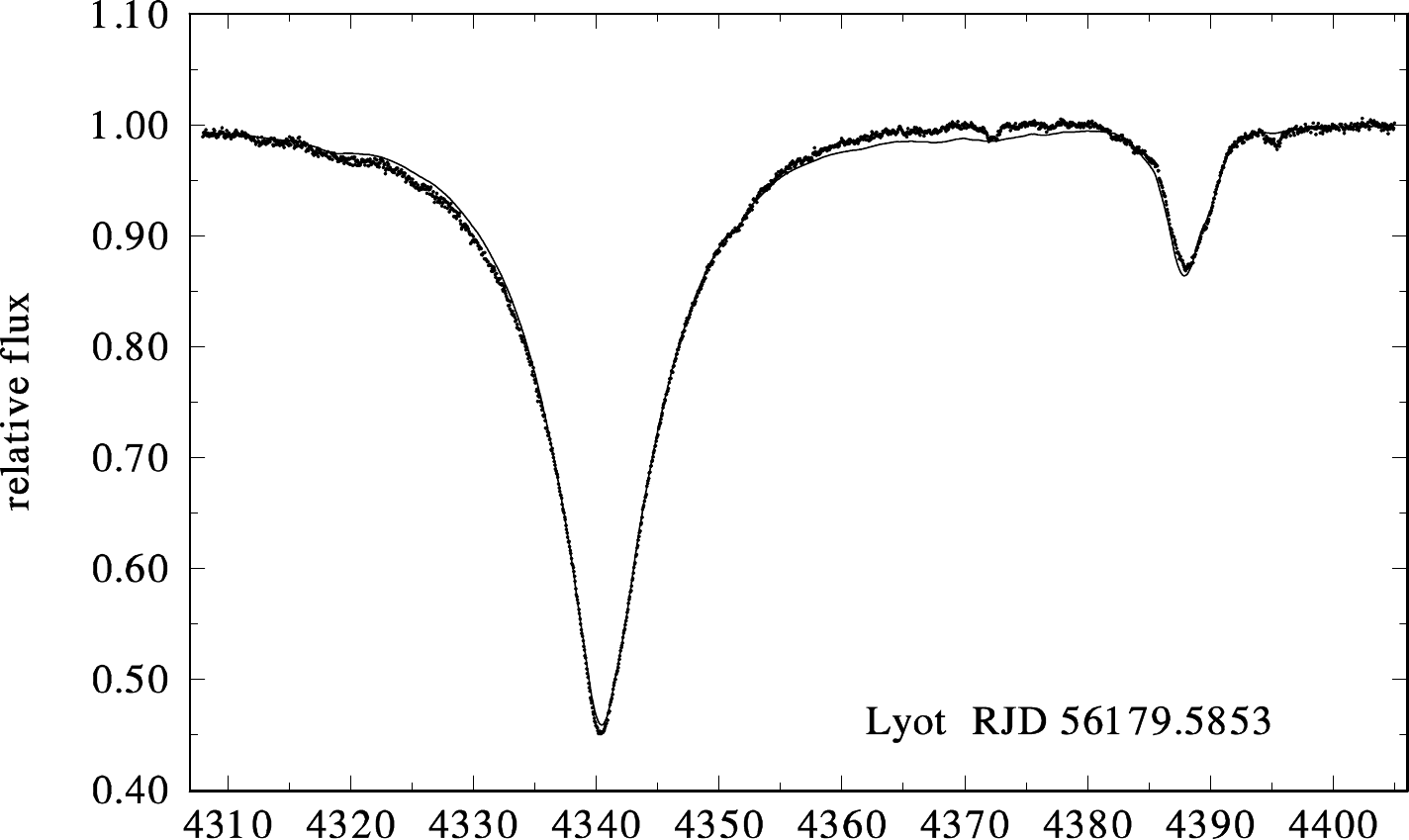}
\includegraphics[width=7.8cm]{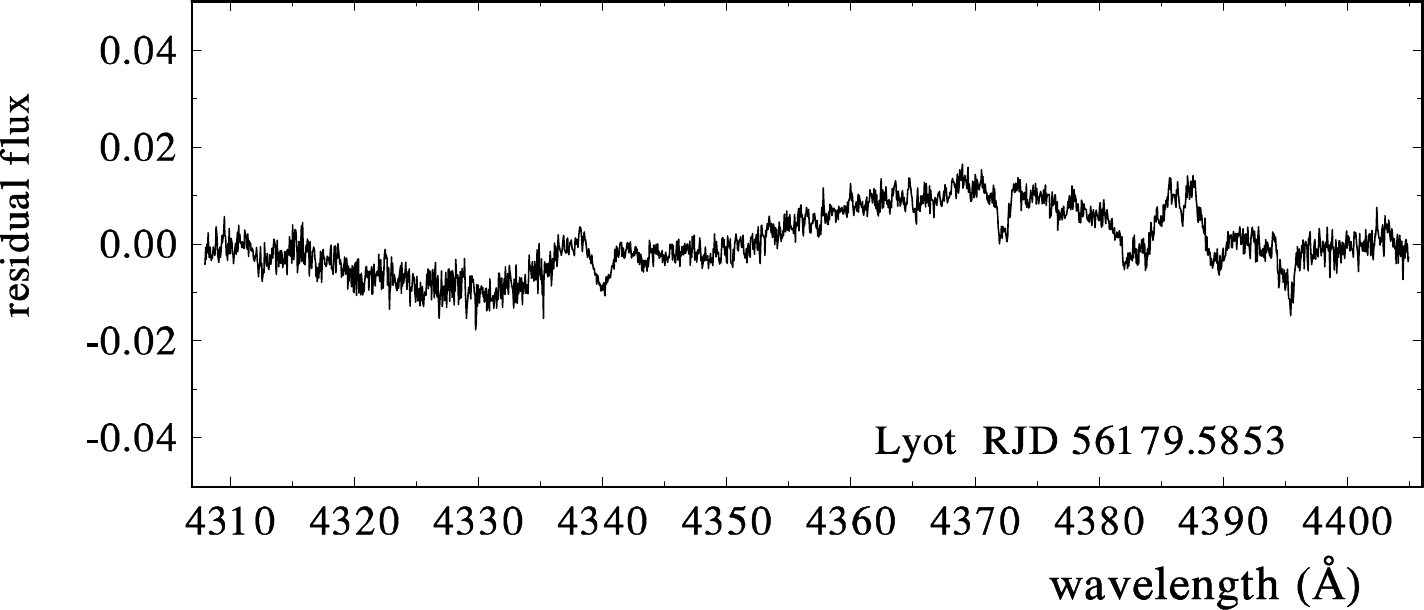}
\caption{Same as Fig.~\ref{synta} for the neighbourhood of
the H$\gamma$ line.}\label{syntb}
\end{figure}

\begin{figure}
\includegraphics[width=8.0cm]{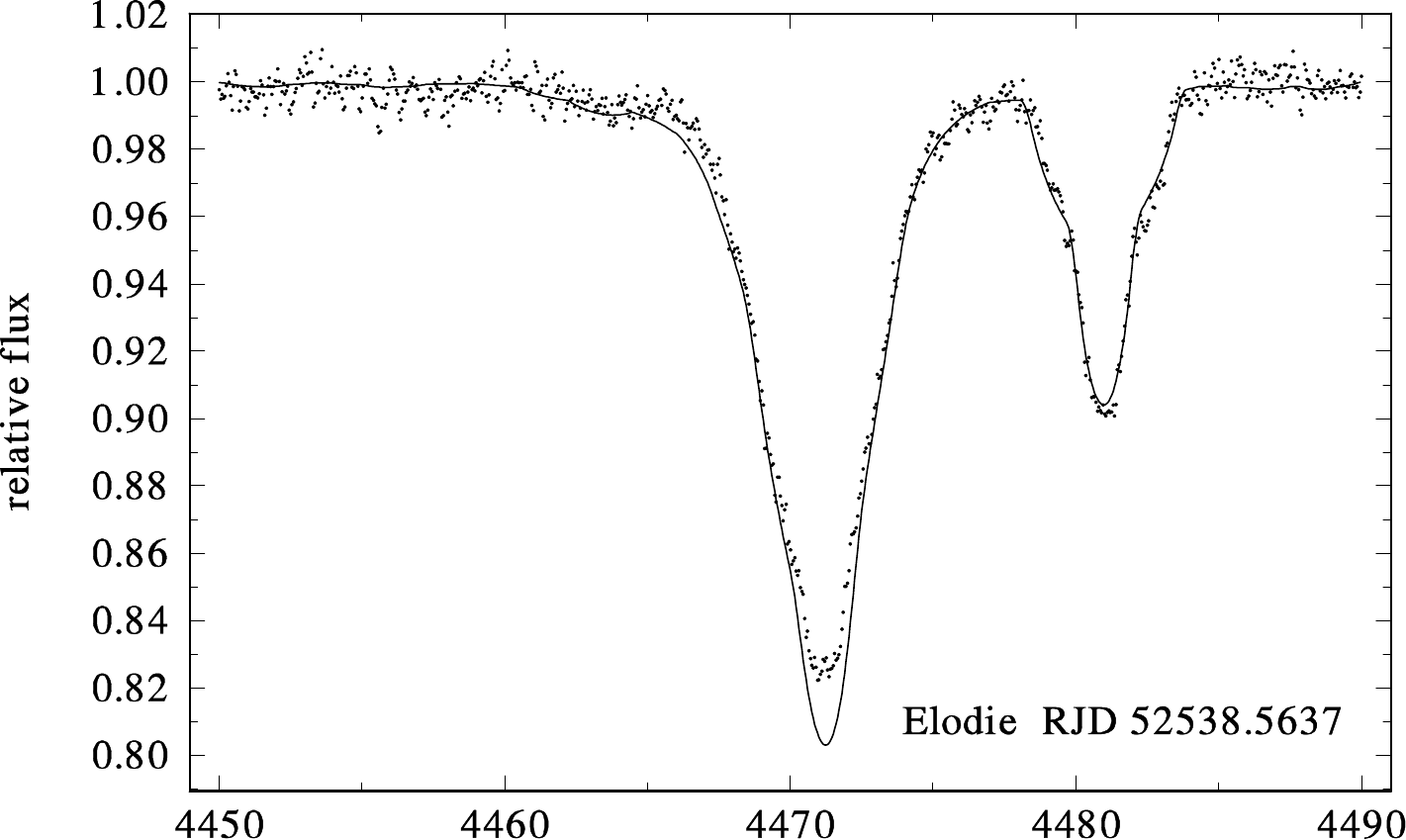}
\includegraphics[width=8.0cm]{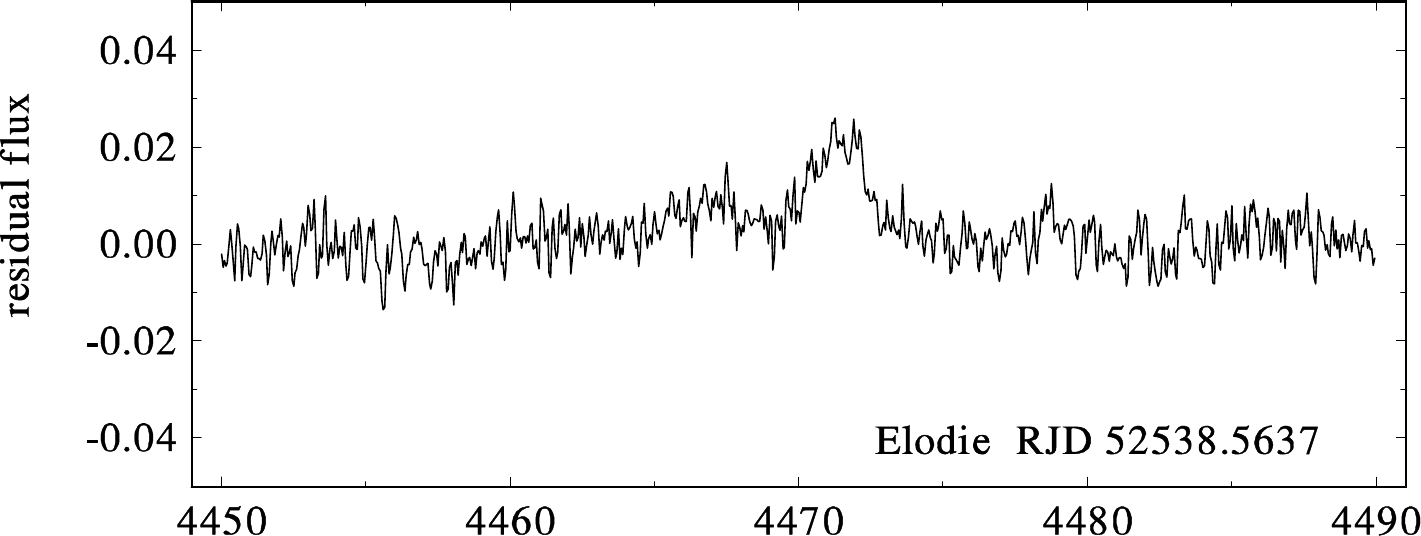}
\includegraphics[width=8.0cm]{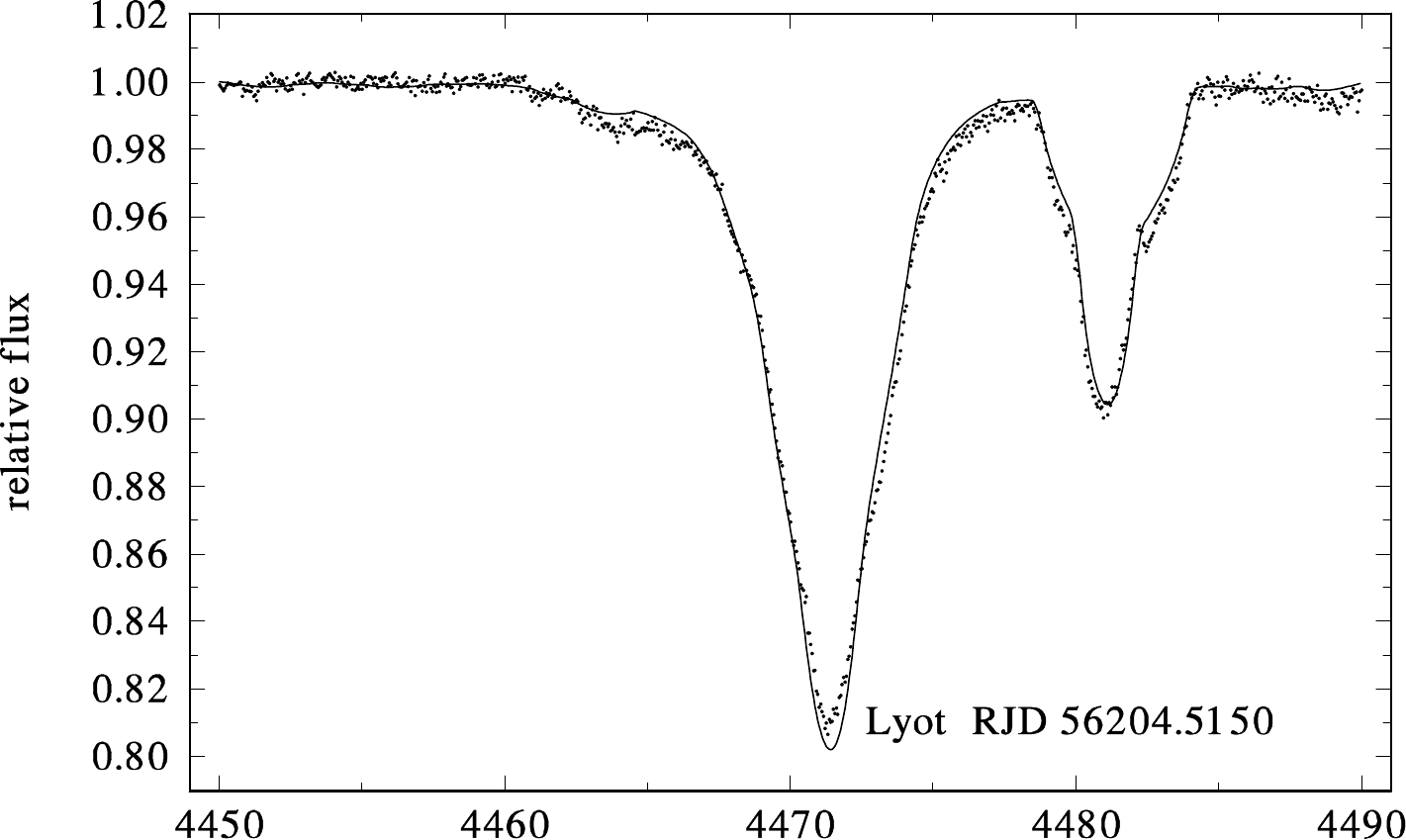}
\includegraphics[width=8.0cm]{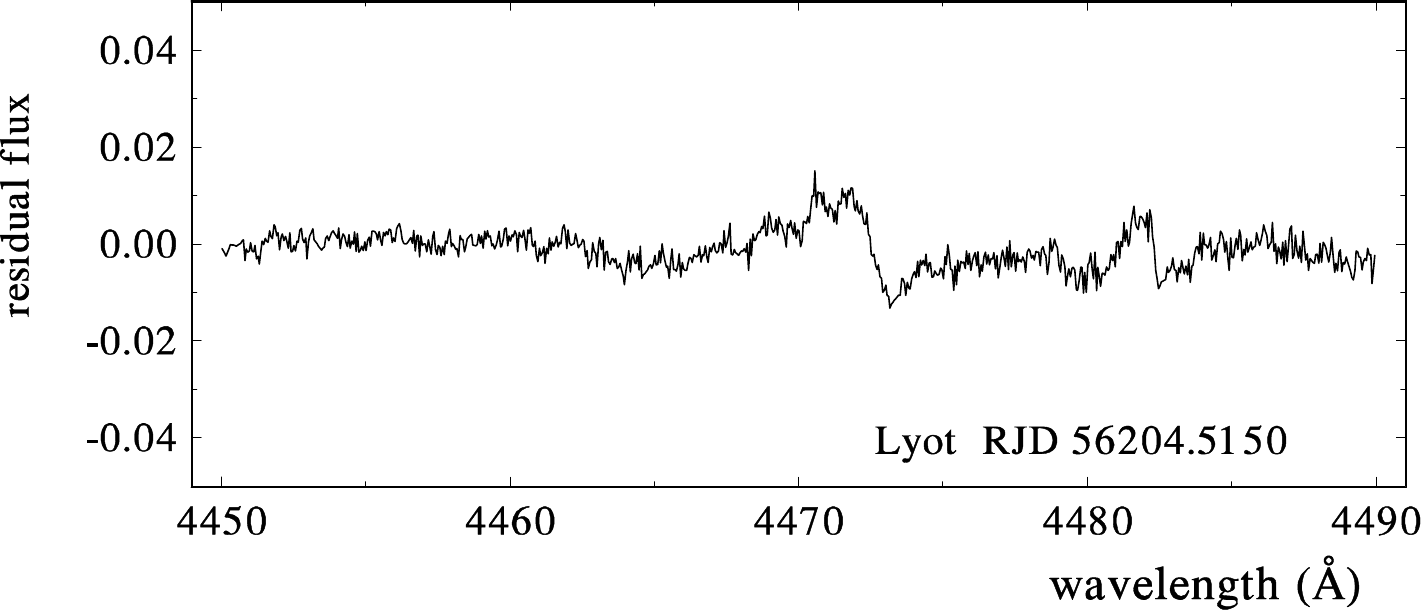}
\caption{Same as Fig.~\ref{synta} for the neighbourhood of
the \ion{H}{i}~4472~\AA\ and \ion{Mg}{ii}~4481~\AA\ lines.}\label{syntc}
\end{figure}

Using the Python program \pyt and two grids of synthetic spectra,
\citet{bstars} for \teff$>$15000~K, and \citet{pala2010} for \tef$<$15000~K,
we estimated the properties of the primary and tertiary, which are
summarised in Table~\ref{synpar}.
   Figures~\ref{synta}, \ref{syntb}, and \ref{syntc} show some examples of
the comparison of the observed and the best interpolated synthetic spectra in
detail for several different spectrograms and spectral regions.

\section{Details of the photometric data reduction and homogenisation}
\label{apc}

In an effort to bring at least all yellow-band photometric observations of
\vc on a comparable scale, we used several \ubv\ observations of \vc and
its comparison stars HR~96 and HD~567 secured with the 0.65~m reflector and
photoelectric photometer at Hvar. These observations
were reduced to the standard Johnson system with the help of
the program HEC22, based on non-linear transformation formulae \citep{hhj94}.
Extinction and its variations during observing nights were taken into account
during the data reduction.\footnote{The program suite with a detailed manual
is available at http://astro.troja.mff.cuni.cz/ftp/hec/PHOT/}
We then used the mean all-sky Hvar values of HD~96 and added them to the
magnitude differences $y-y_{\rm HR\,96}$ for both \vc and HD~567 Str\"omgren
APT observations of data set~12.

\begin{table*}
\begin{center}
\caption[]{Individual photometries of \vc and its comparison stars
transformed into the standard Johnson \ubv\ system}\label{jouubv}
\begin{tabular}{rrrcccccc}
\hline\hline\noalign{\smallskip}
Data set&  Star       &No. of  &$V$&$B$&$U$&\bv&\ub&Source \\
       &             &obs.    &(mag.)&(mag.)&(mag.)&(mag.)&(mag.)       \\
\noalign{\smallskip}\hline\noalign{\smallskip}
14A&    HR 96         &  23&5.748(10)&5.680(10)&5.369(21)&$-0.068$&$-0.311$&D\\
14A&   HD 567         &   8&7.215(06)&7.188(23)&6.789(07)&$-0.027$&$-0.399$&D\\
14A& V746 Cas         &   6&5.601(04)&5.486(06)&4.891(05)&$-0.115$&$-0.595$&D\\
\noalign{\smallskip}\hline\noalign{\smallskip}
14D&   HD 567         &  11&7.224(12)&7.194(22)&6.805(14)&$-0.029$&$-0.389$&D\\
14D& V746 Cas         &   6&5.606(05)&5.489(04)&4.901(05)&$-0.117$&$-0.588$&D\\
\noalign{\smallskip}\hline\noalign{\smallskip}
 11&    HR 96         & 157&5.737(05)&  --     &  --     &   --   &  --   &A\\
 11&   HD 567         &   8&7.206(09)&  --     &  --     &   --   &  --   &A\\
 11& V746 Cas         &   6&5.590(10)&  --     &  --     &   --   &  --   &A\\
\noalign{\smallskip}\hline\noalign{\smallskip}
 12& V746 Cas         &1635&5.606(01)&  --     &  --     &   --   &  --   &B\\
 12&   HD 567         & 847&7.235(01)&  --     &  --     &   --   &  --   &B\\
\noalign{\smallskip}\hline\noalign{\smallskip}
 13& V746 Cas         & 251&5.586(09)&5.475(11)&4.879(13)&$-0.111$&$-0.596$&E\\
\noalign{\smallskip}\hline\noalign{\smallskip}
\end{tabular}
\tablefoot{Column Data set}: \ \ the rows show
11. All-sky Hipparcos \hp\ observations, transformed to Johnson $V$ after
    \citet{hpvb};
12. Differential \uvby\ observations secured with the Four College Automatic
    Reflector relative to HR~96 = HD~2054. Also the check star
    HD~567 = BD$+51^\circ12$ was regularly observed.;
13. All-sky Geneva 7-C observations secured with the Mercator Telescope at
    LaPalma;
14A. Hvar all-sky photometry;
14D. Hvar differential photometry relative to HR~96.\\
Column Source: \ \  the rows show
A. \citet{esa97};
B. \citet{dukes2009} and this paper;
C. \citet{decatetal04};
D. this paper.
\end{center}
\end{table*}

Hipparcos \hp\ observations (data set 11) were transformed into the standard
Johnson $V$ magnitude after \citet{hpvb}, using the all-sky \bv\, and \ub\
indices of all three  stars derived at Hvar.

Finally, the Geneva 7-C all-sky observations (data set~13) were transformed
into the standard \ubv\ system using the transformation formul\ae\ devised
by \citet{hecboz01}.
All mean \ubv\ values for individual data sets thus obtained are summarised
in Table~\ref{jouubv} to illustrate the accuracy with which the conversion
into one system was possible.

\begin{figure}
\includegraphics[width=9.0cm]{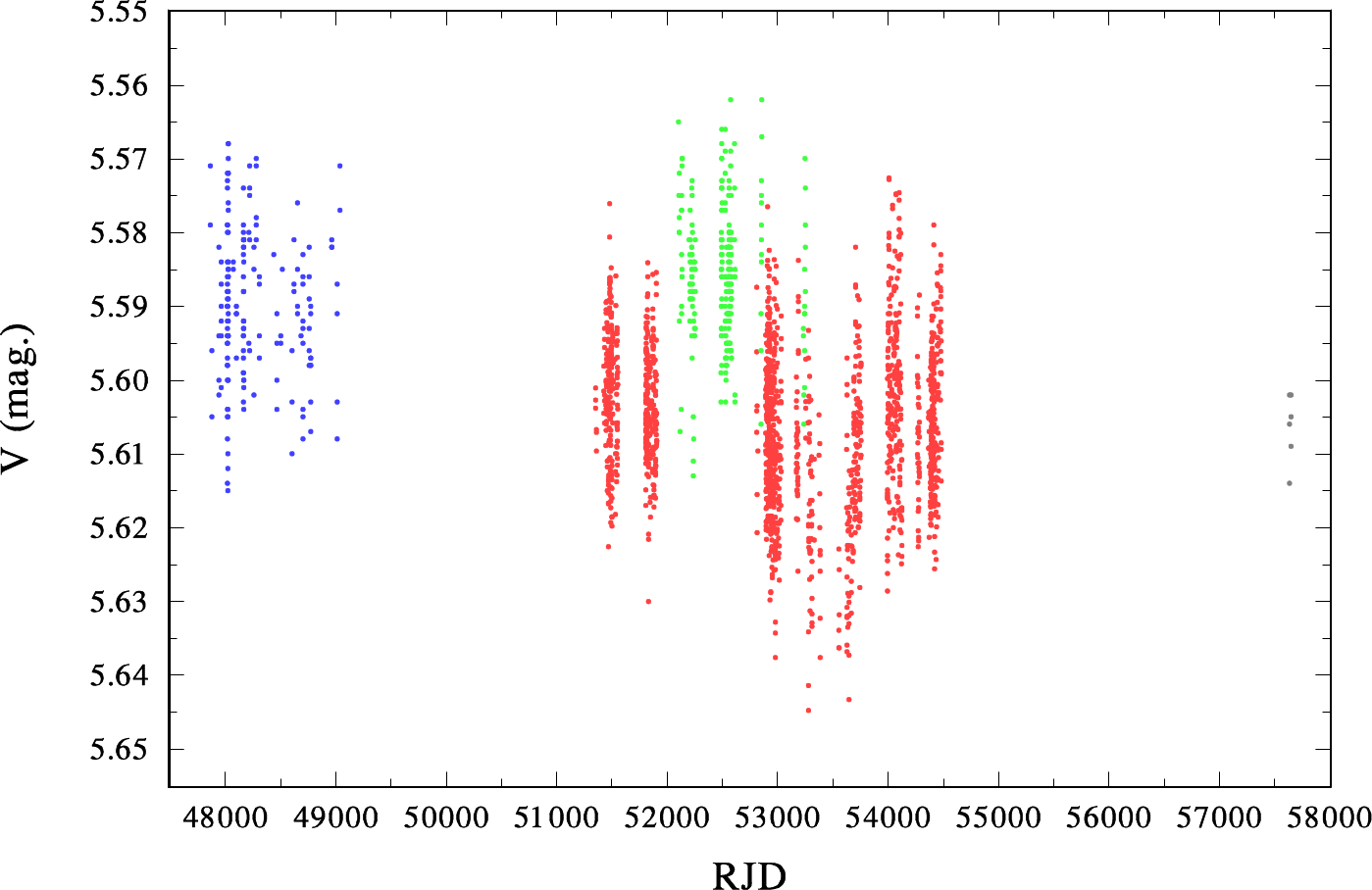}
\includegraphics[width=9.0cm]{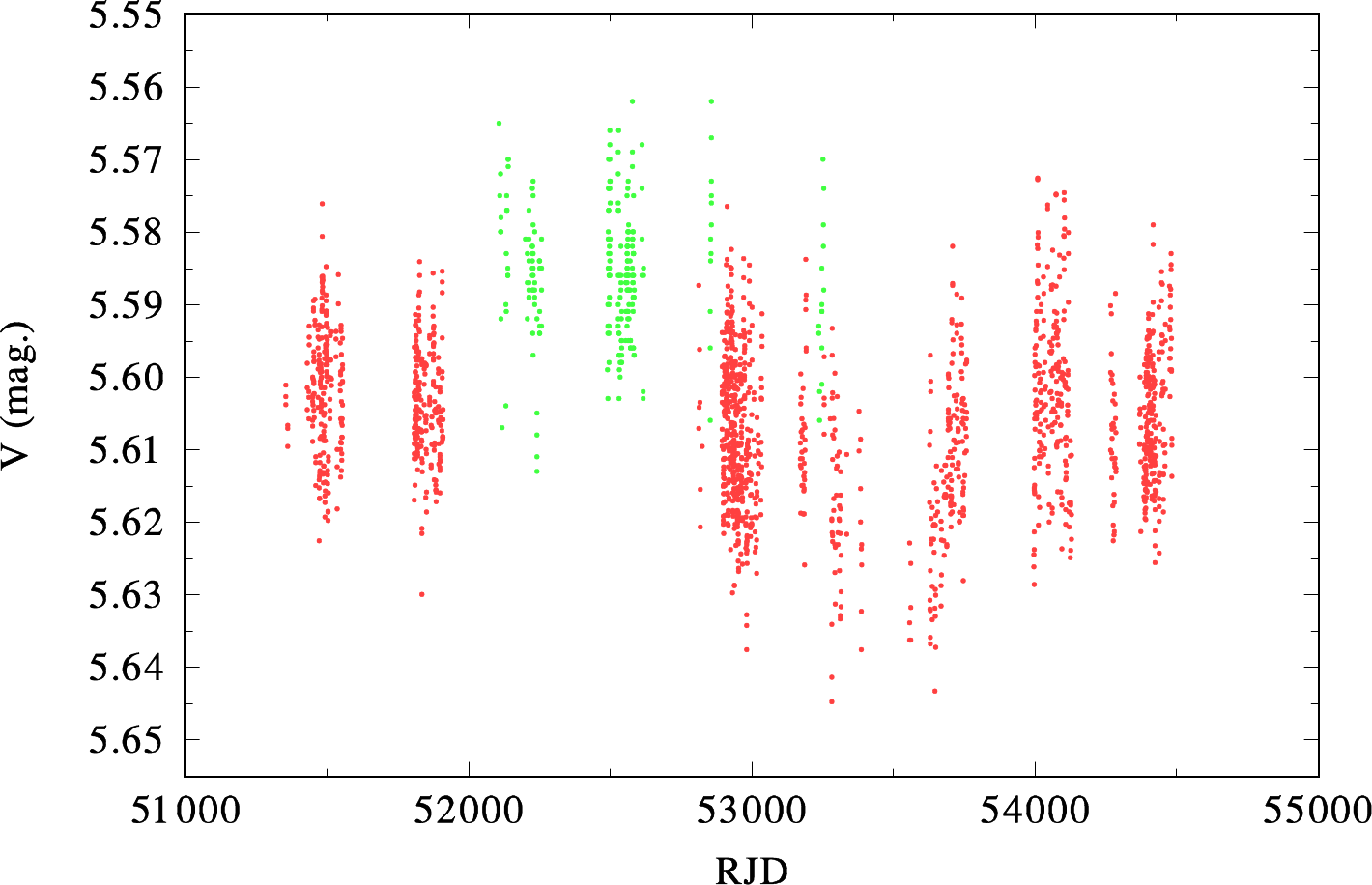}
\caption{Top: Time plot of $V$ photometry of \vc vs. time for the four
data sets. They are distinguished as follows:
blue: all-sky \hp\ magnitudes transformed to Johnson $V$;
red: differential Str\"omgren $y$ magnitude relative to the Hvar all-sky
$V$ magnitude of 5\m748 for HR~96;
green: Johnson  $V$ magnitude transformed from the all-sky Geneva 7-C
observations;
grey: standard differential Johnson $V$ magnitude relative to HR~96
secured at Hvar.
Bottom: Enlarged part of the plot for the time interval where
the $y$ and Geneva transformed $V$ magnitudes overlap.}\label{vtime}
\end{figure}

In Figure~\ref{vtime} we also plot all homogenised yellow-band observations
versus time. There are probably still some minor zero-point differences between
individual data sets (APT photometry in on the instrumental system, which,
however seems to be rather stable in time; Guinan, priv. com.). The plot in
the bottom panel of Fig.~\ref{vtime} is indicative of some small
secular variations, but such a statement would require verification through
differential observations relative to some other comparison than HR~96.

\begin{table}
\caption[]{Orbital solutions for the comparison star HR~96.}
\label{orbit-hr96}
\begin{center}
\begin{tabular}{rcccccl}
\hline\hline\noalign{\smallskip}
Element         & \citet{hube83}&Our new\\
\noalign{\smallskip}\hline\noalign{\smallskip}
$P$ (d)             &48.2905(42)  &48.2884(15)\\
$T_{\rm periastr.}$ &45248.12(60)$^\star$)&45248.29(35)\\
$T_{\rm super.c.}$  &45260.12      &45260.24 \\
$T_{\rm RV max.}$   &45250.43      &45250.58 \\
$e$             &0.384(29)       &0.376(25)  \\
$\omega$ (deg.) &320.0(6.8)         &321.0(2.9) \\
$K_1$ (\ks)     &30.1(1.0)       &29.02(74)  \\
$\gamma_{\rm DDO\,old}$ (\ks)& -- &$+9.37(56)$ \\
$\gamma_{\rm HRM}$ (\ks)    & -- &$+1.8(2.6)$ \\
$\gamma_{\rm DDO\,new}$ (\ks)& -- &$-0.23(39)$ \\
$\gamma_{\rm DAO}$ (\ks)    &$+$3.41(74)&$+$3.22(73)  \\
rms (\ks)           &not given        &2.10      \\
No. of RVs          &25         &36\\
\noalign{\smallskip}\hline\noalign{\smallskip}
\end{tabular}
\end{center}
\tablefoot{$^\star$) The epoch given by \citet{hube83}, 45240.60(60)
is an obvious misprint and cannot reproduce the phase plot
in his Fig.~1. We quote his period with the error derived also from all
36 RVs, which he fixed in the above reproduced solution based on the DAO
RVs only.\\
All epochs are in RJD;
rms is the rms of one observation of unit weight.}
\end{table}

The reason for the above statement is that the primary comparison HR~96 = HD~2054
is suspected to be a~CP star and it was found to be
a~spectroscopic binary with a 48\fd2905
period \citep{hube83}. Using the program \fotele,
we re-analysed available radial velocities, namely
2 RVs from the David Dunlap Observatory (DDO) prismatic 33~\Ame\ spectra
\citep{hube83}, 4 RVs from the DDO 40~\Ame\  grating-spectrograph spectra
\citep{hube70}, 5 RVs from Herstmonceux (HRM) Yapp-reflector prismatic
70-173~\Ame\ spectra \citep{palmer68}, and 25 RVs from the Dominion
Astrophysical Observatory (DAO) 15~\Ame grating spectra \citep{hube83}.
We weighted individual RVs by the weights inversely proportional to their
rms errors and then also applied similar external weights for individual
spectrographs. Unlike \citet{hube83}, we allowed for the determination
of individual systemic $\gamma$ velocities for the four individual
spectrographs. Our solution is compared with that of \citet{hube83} in
Table~\ref{orbit-hr96}

\begin{figure}
\includegraphics[width=9.0cm]{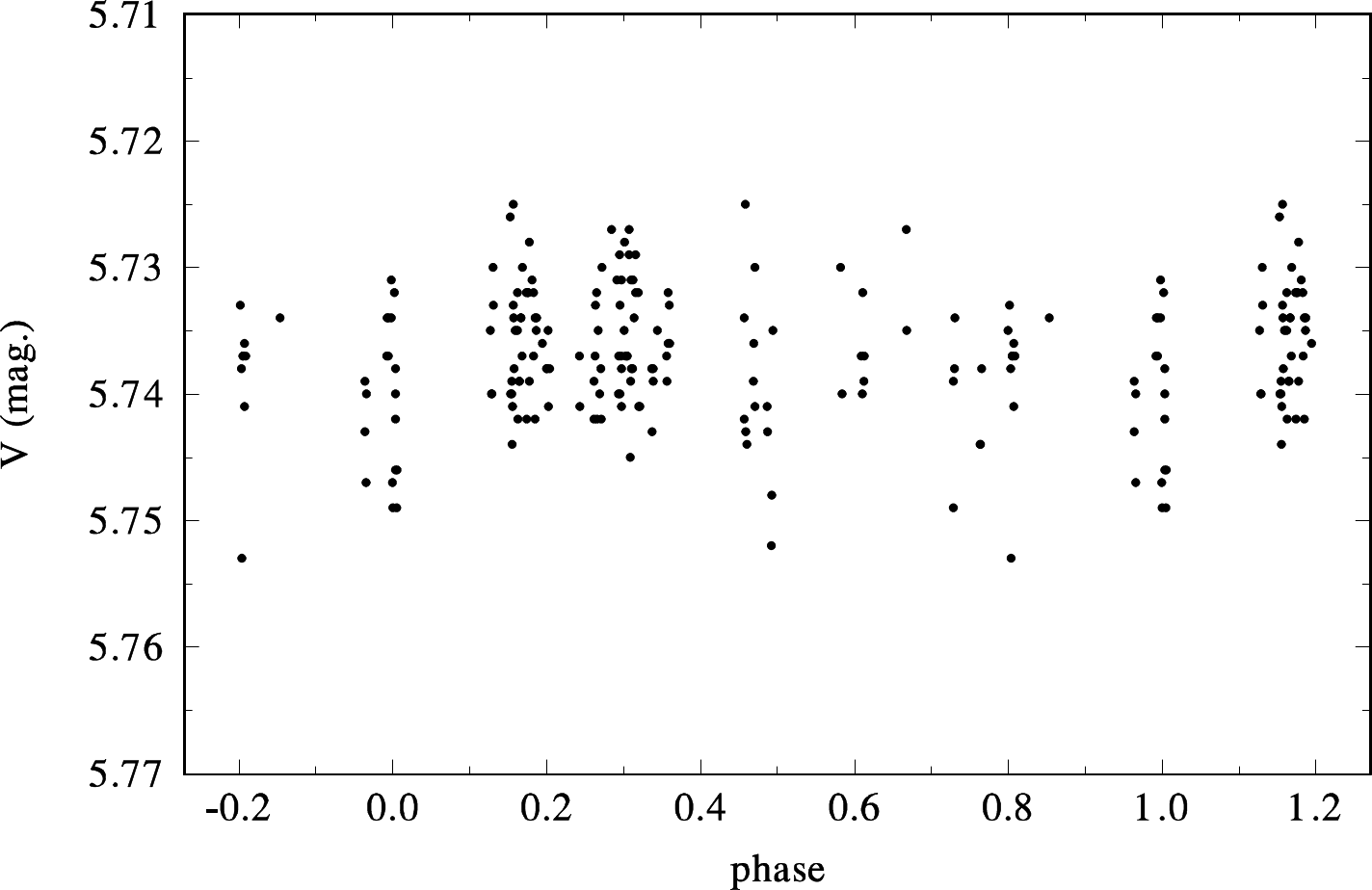}
\caption{Light curve of Hipparcos photometry of HR~96 transformed into
Johnson $V$ plotted for the ephemeris of the spectroscopic binary orbit
$T_{\rm super.conj.} = {\rm RJD}\,45260.24+48\fd2884$.}
\label{hr96hip}
\end{figure}

In Figure~\ref{hr96hip} we show a plot of the Hipparcos \hp\ photometry
transformed into Johnson $V$ magnitude versus phase of the 48\fd29 orbital period
of HR~96. A~small-amplitude variation
with a minimum near the superior conjunction of the binary is
detected. This can
naturally further complicate the period analysis of the photometry of \ve.
\end{appendix}
\end{document}